\documentclass[a4paper,11pt]{article}
\pdfoutput=1
\usepackage{jheppub}

\usepackage{amsmath} 
\usepackage{amsthm}
\usepackage{amssymb} 
\usepackage{dsfont}
\usepackage{yfonts}
\usepackage{hyperref}
\usepackage{floatrow,dcolumn,rotating}
\usepackage{array,xcolor,graphicx}
\usepackage{booktabs,multirow}
\usepackage[utf8]{inputenc}
\usepackage{mathtools}

\usepackage[normalem]{ulem}

\usepackage[all]{xy}
\usepackage{tikz}
\usetikzlibrary{arrows.meta}

\hypersetup{colorlinks=true,linkcolor=blue,citecolor=red,urlcolor=blue}

\newcommand{\be}{\begin{equation}}
\newcommand{\ee}{\end{equation}}
\newcommand{\bea}{\begin{eqnarray}}
\newcommand{\eea}{\end{eqnarray}}

\newcommand{\nn}{\nonumber\\}

\def\la{\langle}
\def\ra{\rangle}

\def\tr{{\mathrm{tr}}}

\def\grad{\nabla}

\def\sig{{\sigma}}
\def\le{\left}
\def\ri{\right}

\def\CA{\mathcal{A}}
\def\CB{\mathcal{B}}
\def\CC{\mathcal{C}}
\def\CD{\mathcal{D}}

\def\CH{\mathcal{H}}

\def\CJ{\mathcal{J}}

\def\CM{\mathcal{M}}
\def\CN{\mathcal{N}}

\def\CT{\mathcal{T}}
\def\CO{\mathcal{O}}

\def\CV{\mathcal{V}}
\def\CW{\mathcal{W}}

\title{Generalised global symmetries in holography:\\ magnetohydrodynamic waves in a strongly interacting plasma}
\author[a,c]{Sa\v{s}o Grozdanov}
\author[b,c]{and Napat Poovuttikul}
\affiliation[a]{Center for Theoretical Physics, Massachusetts Institute of Technology, \\ Cambridge, MA 02139, USA}
\affiliation[b]{University of Iceland, Science Institute, Dunhaga 3, IS-107, Reykjavik, Iceland}
\affiliation[c]{Instituut-Lorentz for Theoretical Physics, Leiden University,\\Niels Bohrweg 2, Leiden 2333 CA, The Netherlands}
\emailAdd{saso@mit.edu}
\emailAdd{nickpoovuttikul@hi.is}

\abstract{We begin the exploration of holographic duals to theories with generalised global (higher-form) symmetries. In particular, we focus on the case of magnetohydrodynamics (MHD) in strongly coupled plasmas by constructing and analysing a holographic dual to a recent, generalised global symmetry-based formulation of dissipative MHD. The simplest holographic dual to the effective theory of MHD that was proposed as a description of plasmas with any equation of state and transport coefficients contains dynamical graviton and two-form gauge field fluctuations in a magnetised black brane background. The dual field theory, which is closely related to the large-$N_c$, $\mathcal{N} = 4$ supersymmetric Yang-Mills theory at (infinitely) strong coupling, is, as we argue, in our setup coupled to a dynamical $U(1)$ gauge field with a renormalisation condition-dependent electromagnetic coupling. After constructing the holographic dictionary for gauge-gravity duals of field theories with higher-form symmetries, we compute the dual equation of state and transport coefficients, and for the first time analyse phenomenology of MHD waves in a strongly interacting, dense plasma with a (holographic) microscopic description. From weak to extremely strong magnetic fields, several predictions for the behaviour of Alfv\'{e}n and magnetosonic waves are discussed.}

\begin{document}

\maketitle
\flushbottom

\section{Introduction}\label{sec:Intro}

Magnetohydrodynamics (MHD) is a hydrodynamic theory of long-range excitations in plasmas (ionised gases) (see e.g. \cite{bellan2008fundamentals,freidberg2014,goedbloed2004principles,goedbloed2010advanced}), which has been applied to systems ranging from the physics of fusion reactors to astrophysical objects. In the modern language of hydrodynamics formulated as an effective field theory \cite{Dubovsky:2011sj, Endlich:2012vt, Grozdanov:2013dba,Nicolis:2013lma,Kovtun:2014hpa,Harder:2015nxa,Grozdanov:2015nea,Crossley:2015evo,Glorioso:2017fpd,Haehl:2015foa,Haehl:2015uoc,Torrieri:2016dko,Glorioso:2016gsa,Gao:2017bqf,Jensen:2017kzi}, MHD should describe the dynamics of infrared (IR) charge-neutral states in terms of massless effective degrees of freedom. These plasma ground states are characterised by an equation of state with a finite magnetic field. On the other hand, the electric field is suppressed due to the screening of electromagnetic interactions and is only induced on shorter length scales than the (thermodynamic) size of the system. In their standard form, the equations of motion that describe the evolution of plasmas are formulated as a combination of macroscopic fluid equations (continuity equation and the non-dissipative Euler, or dissipative Navier-Stokes equation), coupled to the microscopic electromagnetic Maxwell's equations. In ideal, non-dissipative form, the set of dynamical equations is
\begin{align}
\partial_t \rho + \vec \nabla \cdot \left(\rho\, \vec v\right) &= 0 \, , \label{MHD1} \\
\rho \left( \partial_t + \vec v \cdot \nabla \right) \vec v &= - \vec \nabla \,p + \vec J \times \vec B \, , \label{MHD2} \\ 
\partial_t \vec B &= \vec\nabla \times \left(\vec v \times \vec B \right)  , \label{MHD3} \\
\left( \partial_t + \vec v \cdot \nabla \right) \left( \frac{p}{\rho^\gamma} \right) &= 0 \, . \label{MHD4}
\end{align}
The magnetic field is constrained by
\begin{align}
\vec \nabla \cdot \vec B &= 0\, . \label{MHD5}
\end{align}
Eq. \eqref{MHD1} is the continuity equation and Eq. \eqref{MHD2} the Euler equation in the presence of the Lorentz force $\vec J \times \vec B$, with $\vec J$ given by the low-frequency limit of the Ampere's law ($\partial_t \vec E \to 0$) 
\begin{align}
\vec J = \frac{1}{\mu_0} \vec \nabla \times \vec B \, .
\end{align}  Eq. \eqref{MHD3} is the Faraday's induction law with the electric field fixed by the assumption of the ideal Ohm's law
\begin{align}\label{IdealOhm}
\vec E + \vec v \times \vec B  = 0 \, ,
\end{align}
which is derived by taking the conductivity in the (Lorentz transformed) Ohm's law $ \vec J / \sigma = \vec E + \vec v \times \vec B $ to infinity, i.e. $\sigma \to \infty$. The constraint equation \eqref{MHD5} is the magnetic Gauss's law. Since the ideal Ohm's law completely fixes $\vec E$, the electric Gauss's law plays no role in the equations of MHD. Eq. \eqref{MHD4} is the adiabatic equation of state relating density and pressure. Usually, one takes $\gamma = 5/3$. Altogether, Eqs. \eqref{MHD1}--\eqref{MHD4} give eight dynamical equations for eight unknown functions $\rho$, $p$, $\vec v$ and $\vec B$, subject to the magnetic field constraint \eqref{MHD5}. 

While the above equations are closed, solvable and have been successfully applied to a variety of phenomena in plasma physics, they are only applicable within the specific assumptions used to construct them. This means that they are only valid for electromagnetism controlled by Maxwell's equations in the limit of ideal Ohm's law (no possibility of strong-field pair production, etc.) and for the specific equation of state in Eq. \eqref{MHD4}. This equation of state encodes a separation between the fluid and the charge carrying sectors, for which the justification, beyond assuming weakly coupled Maxwell electromagnetism, also assumes very weak interactions between the fluid degrees of freedom and electromagnetism inside the plasma. Concretely, the latter statement is reflected in the equation of state permitting no dependence on the magnetic properties controlled by the charged sector. Furthermore, because of a lack of a symmetry principle behind the construction of ideal MHD, these equations are difficult to extend unambiguously to the most general, higher-order, dissipative theory in the gradient expansion (the Knudsen number expansion) \cite{Baier:2007ix,Bhattacharyya:2008jc,Romatschke:2009kr,Grozdanov:2015kqa}.\footnote{We note that in standard MHD, as formulated in Eqs. \eqref{MHD1}--\eqref{MHD4}, only the fluid sector has a well-defined and finite Knudsen number.} As such, the traditional formulation of MHD lacks generality and cannot be compatible with a variety of IR effective theories of plasmas that could (in principle) be derived from quantum field theory, in particular, in the presence of a strong magnetic field. 

These issues were addressed in a recent work \cite{Grozdanov:2016tdf}, where MHD was formulated by following the effective field theory philosophy behind the construction of relativistic hydrodynamics (see e.g. \cite{Kovtun:2012rj,Grozdanov:2015kqa}). Namely, MHD was formulated by only considering global conserved operators and writing them in terms of the most general hydrodynamic gradient expansion of the IR hydrodynamic fields \cite{Grozdanov:2016tdf}.\footnote{See also \cite{Schubring:2014iwa} and Ref. \cite{Hernandez:2017mch}, which includes a valuable comparison of various related past works, such as \cite{Huang:2011dc,Critelli:2014kra,Finazzo:2016mhm}. For a new treatment of charged fluids in an external electromagnetic field, see \cite{Kovtun:2016lfw,Hernandez:2017mch}. Of further interest is also a recently proposed field theory description of polarised fluids \cite{Montenegro:2017rbu}.} With such an expansion in hand, conservation equations then completely determine the temporal dynamics of a plasma with any equation of state. As in hydrodynamics, all of the details of the equation of state and transport coefficients are left to be determined by the microscopics of the underlying theory.  

The two relevant global symmetries describing the long-range dynamics of a plasma were argued to give the stress-energy tensor $T^{\mu\nu}$ and a conserved anti-symmetric two-form current $J^{\mu\nu}$ \cite{Grozdanov:2016tdf}:
\begin{align}
\nabla_\mu T^{\mu\nu} &= H^{\nu}_{\;\;\mu\sigma} J^{\mu\sigma} \,,  \label{EOM1}\\
\nabla_\mu J^{\mu\nu} & =0 \,.  \label{EOM2} 
\end{align} 
While $T^{\mu\nu}$ corresponds to conserved energy-momentum, $J^{\mu\nu}$ is the manifestation of a generalised global one-form $U(1)$ symmetry, which can be sourced (and gauged) by a two-form gauge field $b_{\mu\nu}$ \cite{Gaiotto:2014kfa}. $H^\nu_{~\mu\sigma}$ is a three-form field strength that can be turned on by an external two-form gauge field, $H = d b_{ext}$. This generalised global symmetry is a consequence of the absence of magnetic monopoles and directly corresponds to the conserved number of magnetic flux lines crossing a co-dimension two surface (in a four dimensional plasma). Normally, it is expressed in terms of the (topological) Bianchi identity
\begin{align}\label{BianchiId}
d F = 0 \, , 
\end{align}
where $F = d A$ and $A$ is the abelian electromagnetic field.\footnote{Note that in a theory with only electromagnetic fields, the number of electric flux lines crossing a two-surface is also conserved in four dimensions. For this reason, in absence of matter, the theory of electrodynamics has two one-form $U(1)$ symmetries. In terms of the photon field, the statement of the conservation of the electric one-form symmetry is analogous to its equation of motion: $\star\,d\star F = 0$.} In the language of a two-form current used in Eq. \eqref{EOM2}, 
\begin{align}\label{TwoFormJ}
J^{\mu\nu} = \frac{1}{2} \epsilon^{\mu\nu\rho\sigma} F_{\rho\sigma} \, .
\end{align}
The power in identifying Eq. \eqref{BianchiId} as a conservation equation of a global symmetry becomes apparent when one attempts to describe a phase of matter dominated by electromagnetic interactions, but without massless photons, i.e. the particles associated with $A$. In fact, this is precisely the situation in a plasma in which long-range electric forces are (Debye) screened and the photons are massive. Treating $J^{\mu\nu}$ as a globally conserved operator without invoking a massless gauge field $A$ can then be used directly to organise the infra-red dynamics of such states \cite{Grozdanov:2016tdf}. We note that in the language of generalised global symmetries, photons only become massless particles when the (one-form) symmetry is spontaneously broken---i.e. photons are the Goldstone bosons present in the broken symmetry phase.\footnote{We note that the order parameter that distinguishes between a broken and an unbroken magnetic one-form symmetry is an expectation value of the 't Hooft loop operator. When the symmetry is preserved, then the expectation value of the loop operator obeys the area law, $\la W_C \ra \sim \exp \left\{-T \, \text{Area}[C] \right\}$. On the other hand, in the symmetry broken phase with massless photons, the expectation value obeys the perimeter law, $\la W_C \ra \sim \exp \left\{-T \, \text{Perimeter}[C] \right\}$ \cite{Gaiotto:2014kfa}.} From this point of view, the Maxwell action is the effective Goldstone boson action that realises this symmetry non-linearly. 

The equations of motion \eqref{EOM1} and \eqref{EOM2} give seven dynamical equations (and one constraint). To solve them, we introduce the following hydrodynamical fields: a velocity field $u^\mu$, a temperature field $T$, a chemical potential $\mu$ that corresponds to the density of magnetic flux lines and a vector $h^\mu$, which can be though as a hydrodynamical realisation of a fluctuating magnetic field. The vectors are normalised as $u_\mu u^\mu = -1$, $h_\mu h^\mu = 1$, $u_\mu h^\mu = 0$, together resulting in $10 - 3 = 7$ degrees of freedom. A (directed) velocity flow of the plasma breaks the Lorentz symmetry from $SO(3,1)$ to $SO(3)$, which is further broken by the additional vector (magnetic field) to $SO(2)$.\footnote{Note that at zero temperature, in a plasma with a non-fluctuating temperature field, the symmetry is enhanced to $SO(1,1) \times SO(2)$ \cite{Grozdanov:2016tdf}.} The projector transverse to both $u^\mu$ and $h^\mu$ is defined as $\Delta^{\mu\nu} = g^{\mu\nu} + u^\mu u^\nu - h^\mu h^\nu$ and has a trace $\Delta^\mu_{~\mu} = 2$.

The constitutive relations for the conserved tensors with a positive local entropy production \cite{Glorioso:2016gsa} and charge conjugation symmetry can now be expanded to first order in derivatives as \cite{Grozdanov:2016tdf}
\begin{align}
T^{\mu\nu} & = (\varepsilon + p)\, u^{\mu}u^{\nu} + p \, g^{\mu\nu} - \mu\rho\, h^{\mu}h^{\nu} + \delta f \, \Delta^{\mu\nu} + \delta \tau \, h^{\mu}h^{\nu}  + 2 \, \ell^{(\mu}h^{\nu)} + t^{\mu\nu} \, ,\label{MHDstress-energy} \\
J^{\mu\nu} & = 2\rho \, u^{[\mu}h^{\nu]} + 2m^{[\mu}h^{\nu]} +  s^{\mu\nu} \, \label{MHDcurrent}, 
\end{align}
where
\begin{align}
\delta f & = -\zeta_{\perp} \Delta^{\mu\nu}\nabla_{\mu}u_{\nu} -\zeta_{\times}^{(1)} h^{\mu}h^\nu \grad_\mu u_\nu\, ,\label{visc1} \\
\delta \tau & = -\zeta_\times^{(2)}  \Delta^{\mu\nu}\grad_\mu u_\nu - \zeta_{\parallel} h^{\mu}h^{\nu} \nabla_{\mu} u_{\nu} \, ,\\ 
\ell^{\mu} & = -2\eta_{\parallel}\Delta^{\mu\sig}h^{\nu} \nabla_{(\sig}u_{\nu)} \,,\label{visc4} \\
t^{\mu\nu} & = -2\eta_{\perp}\le(\Delta^{\mu\rho}\Delta^{\nu\sig}- \frac{1}{2} \Delta^{\mu\nu}\Delta^{\rho\sig}\ri)\nabla_{(\rho}u_{\sig)} \, ,\\
m^{\mu} & = -2 r_{\perp}  \Delta^{\mu\beta}h^{\nu}  \left( T  \nabla_{[\beta}\le(\frac{h_{\nu]} \mu}{T}\ri)  + u_\sigma H^{\sigma}_{\;\;\beta\nu}\right)\, ,\label{mdef}\\
s^{\mu\nu} & = -2 r_{\parallel}  \Delta^{\mu\rho} \Delta^{\nu\sigma} \left( \mu\nabla_{[\rho} h_{\sigma]} + u_\lambda H^\lambda_{\;\;\rho\sigma} \right) \, . \label{sdef}
\end{align}
The frame choice which leads to this particular form of constitutive relations was specified in Ref. \cite{Grozdanov:2016tdf}. The thermodynamic relations between $\varepsilon$, $p$ and $\rho$, which need to be obeyed by the equation of state $p(T,\mu)$ are
\begin{align}
\varepsilon + p &= s T + \mu \rho \, , \label{ThermoRel1} \\
d p &= s \, dT+ \rho \, d\mu \, .
\end{align}
Furthermore, for the theory to be invariant under time-reversal, the Onsager relation implies that $\zeta_\times^{(1)} = \zeta_\times^{(2)}\equiv\zeta_\times $. Thus, first-order dissipative corrections to ideal MHD are controlled by seven transport coefficients: $\eta_\perp$, $\eta_\parallel$, $\zeta_\perp$, $\zeta_\parallel$, $\zeta_\times$, $r_\perp$ and $r_\parallel$. Each one can be computed from a set of Kubo formulae presented in \cite{Grozdanov:2016tdf,Hernandez:2017mch} and reviewed in Appendix \ref{appendix:kubo}. The transport coefficients should obey the following positive entropy production constraints: $\eta_\perp \geq 0 $, $\eta_\parallel \geq 0$, $r_\perp \geq 0 $, $r_\parallel \geq 0$, $\zeta_\perp \geq 0 $ and $\zeta_\perp \zeta_\parallel \geq \zeta_\times^2 $. In absence of charge conjugation symmetry, the theory has four additional transport coefficients, resulting in total in eleven transport coefficients \cite{Hernandez:2017mch}. The precise connection between the above formalism of MHD using the concept of generalised global symmetries and MHD expressed in terms of electromagnetic fields, which match in the limit of a small magnetic field (compared to the temperature of the plasma), was established in Ref. \cite{Hernandez:2017mch}.

Since the effective theory \cite{Grozdanov:2016tdf} makes no assumption regarding the microscopic details of the plasma, then, should such details somehow be computable from quantum field theory, or otherwise, the effective MHD can be used in solar plasma physics, fusion reactor physics, astrophysical plasma physics and even QCD quark-gluon plasma resulting from nuclear collisions. Of course, computing the microscopic properties of such systems is extremely difficult. In this work, we will resort to using holographic duality. By using standard holographic methods applicable to hydrodynamics \cite{Policastro:2001yc,Policastro:2002se,Policastro:2002tn}, our analysis will provide us with the required microscopic data of a strongly interacting toy model plasma needed to describe the phenomenology of MHD waves. 

In process, we will construct and develop holographic duality (the bulk/boundary dictionary) for field theories with generalised global (higher-form) symmetries. For this reason, this work should be thought of as not only a study of strongly interacting MHD but also as providing and executing for the first time the necessary systematic procedure for studying higher-form symmetries in holography. 

The paper is structured as follows: first, in Section \ref{sec:MatterEM}, we review important aspects of gauge theories with a matter sector coupled to dynamical $U(1)$, which can describe a plasma in the IR limit. In particular, we focus on the discussion of how to couple a strongly interacting field theory with a holographic dual to dynamical electromagnetism, all within a holographic setup. Then, in Section \ref{sec:Holography}, we explore this holographic setup in detail, develop the holographic dictionary for theories with higher-form symmetries and use it to compute the microscopic properties of the dual plasma, i.e. the equation of state and first-order transport coefficients. In Section \ref{sec:MHDWaves}, we then use this data to analyse the phenomenology of propagating MHD modes---Alfv\'{e}n and magnetosonic waves. Finally, we conclude with a discussion and a summary of the most important findings in Section \ref{sec:Discussion}. Three appendices are devoted to a derivation of the relevant Kubo formulae (Appendix \ref{appendix:kubo}), details regarding the derivation of horizon formulae for the transport coefficients (Appendix \ref{appendix:transport}) and a derivation of the magnetosonic dispersion relations (Appendix \ref{appendix:magnetosonicSpectrum}). 

Note added: We note that in addition to this paper on the holographic dual of MHD from the perspective of generalised global symmetries, a closely related work, i.e. Ref. \cite{Hofman:2017Something}, also studies various aspects of generalised global symmetries in gauge-gravity duality and holographic dual(s) of \cite{Grozdanov:2016tdf}. Although the two concurrent and complementary papers focus on different aspects of holography, there is overlap between our Sections \ref{sec:N4-II} and \ref{sec:Holography} and parts of Ref. \cite{Hofman:2017Something}. 

\section{Matter coupled to electromagnetic interactions}\label{sec:MatterEM}

A microscopic theory from which an effective description of a plasma can arise comprises of a matter sector that interacts through an electromagnetic $U(1)$ gauge field. In all theories that will be studied here, matter will only couple to electric flux lines. For this reason, the electric one-form symmetry will be explicitly broken. However, the magnetic one-form symmetry will remain a symmetry and $\partial_\mu J^{\mu\nu} = 0$, where $J = \star\, d A$ in a phase with spontaneously broken magnetic global one-form $U(1)$ symmetry. The simplest example of such a theory is quantum electrodynamics. In other theories, the matter sector may itself exhibit complicated physics with additional gauge interactions, such as in QCD. In this work, the theory that we will study contains an infinitely strongly coupled holographic matter sector (closely related to $\CN=4$ supersymmetric $SU(N_c)$ Yang-Mills) with infinite $N_c$. Because of the coupling between matter and dynamical electromagnetism, the holographic setup and the interpretation of results is somewhat subtle. For this reason, we begin our discussion by reviewing some relevant aspects of quantum field theory in a line of arguments similar to \cite{Fuini:2015hba}. 

\subsection{Quantum electrodynamics}\label{sec:QED}

The simplest example of a theory coupling matter to electromagnetism is quantum electrodynamics (QED). QED is a $U(1)$ gauge theory that contains a (massive) Dirac fermion $\psi$ (describing electrons and positrons) and a massless photon field $A_\mu$:\footnote{We use the mostly positive convention for the metric tensor, so that $\eta_{\mu\nu} = \{-1, +1,+1,+1\}$.} 
\begin{align}
S_{\scriptscriptstyle QED} = - \int d^4 x \left[ i \bar\psi \gamma^\mu D_\mu \psi  + m \bar \psi  \psi + \frac{1}{4 e^2} F_{\mu\nu} F^{\mu\nu}  \right].
\end{align}
$D_\mu$ is the gauge covariant derivative that couples $A_\mu$ to the fermion current (with the coupling $e$ scaled out from the interaction). For a detailed discussion of various properties of QED, see e.g. \cite{Weinberg:1995mt,Weinberg:1996kr,Peskin:1995ev}. 

The stress-energy tensor of the theory is
\begin{align}
T^{\mu\nu} = \frac{1}{2} \bar \psi i \left( \gamma^\mu D^\nu + \gamma^\nu D^\mu \right) \psi - \eta^{\mu\nu} \bar\psi \left(i\gamma^\lambda D_\lambda + m \right)\psi + \frac{1}{e^2} \left[ F^{\mu\lambda} F^\nu_{~\lambda} - \frac{1}{4} \eta^{\mu\nu} F^{\rho\sigma} F_{\rho\sigma} \right].
\end{align}
In the massless limit ($m=0$), the theory is classically scale invariant, which is reflected in the vanishing trace of the stress-energy tensor, $T^\mu_{~\mu} = 0$. Quantum mechanically, the theory does not remain scale invariant. The trace receives a correction proportional to the beta function of the electromagnetic coupling,
\begin{align}\label{QED_TraceAnomaly}
T^\mu_{~\mu} = - \frac{\beta(e)}{2 e^3} F_{\mu\nu} F^{\mu\nu}\,.
\end{align}
This is the anomalous breaking of scale invariance---the so-called trace anomaly. The running electromagnetic coupling $e(\mu)$ depends on the renormalisation group scale $\mu$.
To first order in perturbation theory, the beta function is
\begin{align}
\beta (e) = \mu \frac{d e}{d \mu }= \frac{e^3}{12 \pi^2}\,,
\end{align}
which, integrated on the interval $\mu \in \left[M , \Lambda \right]$, gives the running coupling
\begin{align}\label{CutOffDependentCouplingQED}
\frac{1}{e(\Lambda)^2} = \frac{1}{e(M)^2} - \frac{ \ln \left(\Lambda / M \right)}{6\pi^2}\, .
\end{align}
Here, $M$ is some IR renormalisation group scale at which the electric charge takes the renormalised physical value, $e_r = e(M)$, and $\Lambda$ is the UV cut-off. Note that at the Landau pole, $\Lambda = \Lambda_{EM}$, the left-hand-side of \eqref{CutOffDependentCouplingQED} vanishes. On the other hand, the expectation value of the stress-energy tensor is a physical quantity and therefore cannot depend on $\mu$ . This statement is encoded in the following identity, which leads to the Callan-Symanzik equation:
\begin{align}\label{TMuIndep}
\mu \frac{d}{d \mu} \left\langle T^{\mu\nu} \right\rangle = 0 \, .
\end{align}

Since we are interested in neutral IR plasma states in QED that can be described by an effective theory of MHD, we can consider the (ground state) expectation value of the photon field to produce a non-zero magnetic field and a vanishing (screened) electric field,
\begin{align}
\left\langle A_\mu \right\rangle = \frac{1}{2}\CB \left( x^1 \delta^2_{~\mu} - x^2 \delta^1_{~\mu}\right).
\end{align}
$\CB$ is the magnitude of the ``background" magnetic field pointing in the $x^3 = z$ direction. The IR spectrum of the theory has a gapped-out photon, i.e. long-range charge neutrality, which allows us to neglect quantum fluctuations of $A_\mu$. For such a plasma state, Eq. \eqref{QED_TraceAnomaly} yields
\begin{align}\label{TTraceQED}
\left\langle T^{\mu}_{~\mu} \right\rangle = - \frac{\beta(e)}{e^3} \CB^2 = - \frac{1}{12\pi^2} \CB^2 + \CO\left(e^2\right).
\end{align}
Furthermore, the expectation value of the stress-energy tensor can be conveniently split into the matter (containing matter-light interactions) and the purely electromagnetic parts,
\begin{align}
\left\langle T^{\mu\nu} \right\rangle &= \left\langle T^{\mu\nu}_{{\scriptscriptstyle matter} }  (\mu) \right\rangle +  \frac{1}{e (\mu)^2 } \left[ F^{\mu\lambda} F^\nu_{~\lambda} - \frac{1}{4} \eta^{\mu\nu} F^{\rho\sigma} F_{\rho\sigma} \right] \nn
&= \left\langle T^{\mu\nu}_{{\scriptscriptstyle matter}}  (\Lambda / M ) \right\rangle + \left( \frac{1}{e_r^2} - \frac{ \ln \left(\Lambda / M \right)}{6\pi^2} \right) \frac{\CB^2}{2}   \times
{\scriptsize
 \begin{bmatrix}
     1 & 0 & 0 & 0 \\
     0 & 1 & 0 & 0 \\
     0 & 0 & 1 & 0 \\
     0 & 0 & 0 & -1
   \end{bmatrix} },
\end{align}
where in the second line, we chose to evaluate the expectation value at the UV cut-off $\mu = \Lambda$. Note that because $\left\langle T^{\mu\nu}\right\rangle$ is $\mu$-independent (cf. Eq. \eqref{TMuIndep}), this choice does not influence the final value of $\left\langle T^{\mu\nu} \right\rangle$. 

\subsection{Strongly interacting holographic matter coupled to dynamical electromagnetism}\label{sec:N4}

We now turn our attention to the holographic strongly interacting theory that will be investigated in the remainder of this paper. Throughout our discussion, it will prove useful to think of the matter sector as that of the best understood holographic example---the conformal $\CN=4$ supersymmetric Yang-Mills theory (SYM) with an infinite number of colours $N_c$ and an infinite 't Hooft coupling $\lambda$. However, as will become clear below, the theory dual to our holographic setup will not be precisely the $\CN=4$ SYM theory coupled to a $U(1)$ gauge field, but rather its deformation, of which the microscopic definition will not be investigated in detail. Instead, the model studied here should be considered as a bottom-up construction---the simplest dual of a strongly coupled plasma, which can be described with magnetohydrodynamics in the infrared limit.

The field content of $\CN = 4$ SYM is four Weyl fermions, three complex scalars and a vector field, all transforming under the adjoint representation of $SU(N_c)$. The theory also has an $SU(4)_R$ R-symmetry owing to its extended supersymmetry. The adjoint fields together represent the matter content of a hypothetical plasma, which further requires the fields to be (minimally) coupled to an electromagnetic $U(1)$ gauge group (with $e$ the electromagnetic coupling). In $\CN = 4$ SYM, this can be achieved by gauging the $U(1)_R$ subgroup of $SU(4)_R$. Under $U(1)_R$, the Weyl fermions transform with the charges $\{+3,-1,-1,-1\}/\sqrt{3}$ and the complex scalars all have charge $+2/\sqrt{3}$ (for details regarding the choice of the normalisation, see \cite{Fuini:2015hba}). Such a system can be considered as a strongly coupled toy model for a QCD plasma in which the quarks interact with photons as well as with the $SU(3)$ vector gluons. 

A crucial fact about $\CN = 4$ SYM is that the $R$-current of $\CN = 4$ becomes anomalous in the presence of electromagnetism. For this reason, the $U(1)_R$, which is gauged, is also anomalous and thus the theory has to be deformed in some way to reestablish its self-consistency. As pointed out in \cite{Fuini:2015hba}, one way to do this is by adding a set of spectator fermions that only interact electromagnetically and ``absorb" the anomaly. We will assume that the gauge anomaly can be cancelled by some deformation of the theory so that the quantum expectation value of the $U(1)_R$ R-current $J_R^\mu $ remains conserved, $\nabla_\mu \la J_R^\mu \ra = 0$. We can then write the total bare action of the $SU(N_c) \times U(1)$ gauge theory as
\begin{align}
S_{{\scriptscriptstyle plasma}} = S_{\scriptscriptstyle matter} + \int d^4 x \, A_\mu J^\mu_R \, - \frac{1}{4 e^2} \int d^4 x \, F_{\mu\nu}  F^{\mu\nu} \,,
\end{align}
where $A_\mu$ is the dynamical electromagnetic gauge field and $F = d A$. The expectation value of the conserved operator $J^\mu_R$ contains a trace over the colour index of the adjoint matter field and therefore scales as $N_c^2$. Since it is coupled to a single photon, the Maxwell part of the total plasma action $S_{{\scriptscriptstyle plasma}}$ contains no powers of $N_c$.

As in the QED plasma, we will consider the photons to be gapped out from the IR spectrum so that $A_\mu$ will only produce a (classical) magnetic field  
\begin{align}
\la A_\mu \ra = \frac{1}{2}\CB \left( x^1 \delta^2_{~\mu} - x^2 \delta^1_{~\mu}\right).
\end{align}
In order to maintain the neutrality of the plasma, we will set the electric $U(1)_R$ chemical potential to zero, $\mu_R = \left\langle A_0 \right\rangle= 0$.\footnote{For a discussion of supersymmetric gauge theories with non-zero R-charge densities, see e.g. \cite{Yamada:2006rx,Cherman:2013rla}} For this reason, the electric one-form (or vector) conserved $U(1)_R$ R-current will play no role in the hydrodynamic IR limit of the theory, so $\la J^{\mu}_R \rangle = 0$.

The plasma has a conserved stress-energy tensor to which both the matter (along with its interaction with the electromagnetic field) and the purely electromagnetic sectors contribute,
\begin{align}\label{TmunuSecQFT}
\left\langle T^{\mu\nu} \right\rangle &= \left\langle T^{\mu\nu}_{{\scriptscriptstyle matter} }  (\Lambda/M) \right\rangle +  \frac{1}{e (\Lambda/M)^2 } \left[ \la F^{\mu\lambda} F^\nu_{~\lambda} \ra - \frac{1}{4} \eta^{\mu\nu} \la F^{\rho\sigma} F_{\rho\sigma}\ra \right] .
\end{align}
The trace of the superconformal theory again experiences an anomaly proportional to the beta function of the electromagnetic coupling (cf. Eq. \eqref{TTraceQED}), which in $\CN = 4$ theory turns out to be one-loop exact in the presence of a background electromagnetic field and follows from a special case of the NSVZ beta function due to the fact that the $U(1)_R$ sector has a remaining $\CN = 1$ supersymmetry (see Refs. \cite{Freedman:1998tz,Anselmi:1997ys,Fuini:2015hba}),\footnote{Note that as $N_c \to \infty$, $N_c^2 - 1 \approx N_c^2$.}
\begin{align}\label{N4TraceAnomaly}
\la T^\mu_{~\mu} \ra = - \frac{\beta(e)}{e^3} \CB^2 = - \frac{N_c^2}{4 \pi^2} \CB^2 \,.
\end{align}
The beta function for the inverse electromagnetic coupling is then
\begin{align}\label{NSVZN4}
\beta\left(1/e^2\right) = \mu \frac{d e^{-2}}{d\mu} =  - \frac{N_c^2}{2 \pi^2} \left[ \frac{1}{6} \sum_{\alpha=1}^4 \left(q_{\text f}^\alpha\right)^2  + \frac{1}{12} \sum_{a=1}^3 \left(q_{\text s}^a\right)^2 \right]= - \frac{N_c^2}{2 \pi^2} \, ,
\end{align}
with the fermionic and the scalar R-charges being $q_{\text f}^\alpha = \{+3,-1,-1,-1\} / \sqrt{3}$ and $q_{\text s}^\alpha = \{2,2,2\}/\sqrt{3}$, respectively. In analogy with Eq. \eqref{CutOffDependentCouplingQED} in QED, by integrating the beta function equation, we find
\begin{align}\label{CutOffDependentCouplingN4}
\frac{1}{e^2(\Lambda) } = \frac{1}{e^2 (M) } - \frac{N_c^2}{2\pi^2} \ln \left(\Lambda / M \right).
\end{align}
It is essential to stress that even though our holographic theory will not be exactly dual to the $\CN = 4$ SYM theory, it will give us the same trace anomaly and thus the same electromagnetic beta function. Since the NSVZ beta function \eqref{NSVZN4} is only sensitive to the matter content, this match can be interpreted as our working with a theory with the $U(1)$-gauged matter content and R-charges of $\CN = 4$ but with a deformed Lagrangian and possibly additional matter that is ungauged under the $U(1)$.

Beyond the stress-energy tensor of the theory discussed thus far, the only other (generalised) global symmetry of interest to describing a plasma state is the higher-form $U(1)$ symmetry that corresponds to the conserved number of magnetic flux lines crossing a two-surface. The symmetry results in a conserved two-form current $\la J^{\mu\nu} \ra \neq 0$ and was discussed in Section \ref{sec:Intro}. The generating functional of the field theory that can be used to study MHD of a magnetised plasma in which the two globally conserved operators are $T^{\mu\nu}$ and $J^{\mu\nu}$ is therefore
\begin{align}\label{GenFunTJ}
W\left[g_{\mu\nu}, b_{\mu\nu} \right] = \left\la \exp\left[i \int d^4x \sqrt{-g} \left( \frac{1}{2} T^{\mu\nu}g_{\mu\nu} + J^{\mu\nu} b_{\mu\nu}\right) \right] \right\ra .
\end{align}
The remainder of this paper is devoted to constructing and analysing its holographic bulk dual.

\subsection{Holographic dual}\label{sec:N4-II}

The simplest holographic dual of a strongly interacting state with the generating functional \eqref{GenFunTJ} is one that contains a five-dimensional bulk with a dynamical graviton (metric tensor $G_{ab}$) described by the Einstein-Hilbert action, a negative cosmological constant and a two-form bulk gauge field $B_{ab}$:\footnote{Throughout this paper, we use Greek and Latin letters to denote the boundary and bulk theory indices, respectively.}
\begin{align}\label{BulkAction}
S = \frac{1}{2\kappa_5^2} \int d^5x \sqrt{-G} \left( R + \frac{12}{L^2} -\frac{1}{3 e_H^2} H_{abc}H^{abc} \right) .
\end{align}
In standard (Dirichlet) quantisation, the two fields asymptote to $g_{\mu\nu}$ and $b_{\mu\nu}$ at the boundary and source $T^{\mu\nu}$ and $J^{\mu\nu}$. Furthermore, $H$ is the three-form defined as $H = d B$. In component notation, $B = \frac{1}{2} B_{ab} \, dx^a \wedge dx^b$ and $H = \frac{1}{6} H_{abc} \, dx^a \wedge dx^b \wedge dx^c$. The two-form gauge field action is the bulk Maxwell Lagrangian $F \wedge \star\, F$ written in terms of the five-dimensional Hodge dual three-form $H = \star\, F$, giving the Lagrangian term $H \wedge \star \, H$. In most of our work, we will set $e_H = L = 1$. Because the two bulk theories are related by dualisation, the background solution to the equations of motion derived from \eqref{BulkAction} give rise to the same magnetised black brane solution known from the Einstein-Maxwell theory \cite{D'Hoker:2009mmn}. 

In the absence of the two-form term, the action \eqref{BulkAction} arises from a consistent truncation of type IIB string theory on $S^5$ and is upon identification of the Newton's constant $\kappa_5 = 2 \pi / N_c$ dual to pure $\CN = 4$ SYM at infinite $N_c$ and infinite 't Hooft coupling $\lambda$. For reasons discussed above, the full dual of the action \eqref{BulkAction} is unknown and we are not aware of a mechanism for deriving this action from a consistent truncation of ten-dimensional type IIB supergravity. Nevertheless, for purposes of comparing the sizes of matter and electromagnetic contributions to the total operator expectation values, it will prove useful to keep the definition of $\kappa_5$ in terms of the number of colours $N_c$ of the hypothetical dual deformed $\CN = 4$ SYM coupled to dynamical electromagnetism. 

To show further evidence that the action \eqref{BulkAction} is a sensible dual of a strongly coupled MHD plasma, it is useful to elucidate the connection between Eq. \eqref{BulkAction} and the Einstein-Maxwell theory. To put an uncharged holographic theory in an external magnetic field, one normally adds the Maxwell action $F \wedge \star\, F$ with $F = d A$ to the Einstein-Hilbert bulk action. If one imposes Dirichlet boundary conditions on the bulk one-form $A_a$, then $A_a$ sources the R-current $J^\mu_R$ at the boundary, $\int d^4 x\, J^\mu_R \delta A_\mu$, and thus the electromagnetic field $A_\mu$ is external and non-dynamical. The investigation of the physics of such a setup with an external magnetic field was initiated in \cite{D'Hoker:2009mmn} and studied in numerous subsequent works, including recent \cite{Fuini:2015hba,Janiszewski:2015ura,Critelli:2014kra,Finazzo:2016mhm,Ammon:2017ded}. The semi-classical generating functional of the field theory dual to the Einstein-Maxwell bulk action with Dirichlet boundary conditions corresponds to
\begin{align}\label{ZExtField}
Z_0 [ A_\mu] = \int \CD \Phi \, \exp \left\{  i S_0(\Phi) + i\int d^4 x A_\mu J^\mu_R (\Phi) \right\},
\end{align} 
where $S_0$ is the strongly coupled field theory action that depends on a set of fields $\Phi$, which we collectively denote as $\Phi$. For uncharged solutions of the Einstein-Maxwell theory, $S_0$ is the $\CN = 4$ SYM action and $A_\mu$ is an external gauge field that sources the $U(1)_R$ current. The bounday gauge field $A_\mu$ can be made dynamical by performing a Legendre transform of \eqref{ZExtField} and adding a kinetic term for $A_\mu$ (see also Refs. \cite{Grozdanov:2016tdf,Hernandez:2017mch}):
\begin{equation}\label{ZExtField2}
\begin{aligned}
Z[j_{ext}^\mu] &= \int \CD A\int  \CD\Phi \,\exp \left\{{i S_0(\Phi) + i \int d^4x \left(A_\mu J^\mu_R(\Phi)-\frac{1}{4e^2}F_{\mu\nu}F^{\mu\nu} + A_\mu j^\mu_{ext}\right) }\right\},\\
&= \int \CD A\, Z_0[ A_\mu] \,\exp \left\{i\int d^4x \left(-\frac{1}{4e^2}F_{\mu\nu}F^{\mu\nu} + A_\mu j^\mu_{ext} \right)  \right\},
\end{aligned}
\end{equation}
where $j^\mu_{ext}$ is the external current which sources the dynamical $U(1)$ gauge field $A_\mu$. To describe a stable plasma state, which is charge neutral in equilibrium, we must impose $J^\mu_R + j^\mu_{ext} = 0$. The variation $\int d^4x\, A_\mu \delta j^\mu_{ext}$ can then be used to obtain correlation functions of the dynamical vector field. Now, since $j^\mu_{ext}$ is conserved, one can express it through an anti-symmetric two-form $b_{\mu\nu}$ as $\epsilon^{\mu\nu\rho\sigma} \partial_\nu b_{\rho\sigma}$, which, upon integration by parts, yields a dualised $\int d^4 x \, J^{\mu\nu} b_{\mu\nu}$, where $J^{\mu\nu}$ is the anti-symmetric current from Eq. \eqref{TwoFormJ}. Furthermore, as we will explicitly see in Section \ref{sec:Holography}, the gravitational dual formulation of a theory with a two-form source $b_{\mu\nu}$ and a corresponding conserved two-form current $J^{\mu\nu}$ allows us to interpret the kinetic Maxwell term in \eqref{ZExtField2} as a double-trace deformation $\int d^4 x \, J_{\mu\nu} J^{\mu\nu}$ of a CFT (with a broken scale invariance). The necessity of imposing double-trace deformations to ensure that the $U(1)$ boundary gauge field be dynamical will thus require us to impose mixed boundary conditions \cite{Witten:2001ua} on the two-form gauge field.\footnote{See also Refs. \cite{Pomoni:2008de,Heemskerk:2010hk,Faulkner:2010jy,Grozdanov:2011aa} and references therein.} 

Instead of imposing Dirichlet boundary conditions on the bulk $A_a$, one can work in alternative quantisation and impose Neumann boundary conditions. Such a choice exchanges the interpretation of the normalisable and the non-normalisable mode in $A_a$. From the dual field theory point of view, this can be interpreted as the Legendre transform of the boundary theory, as in Eq. \eqref{ZExtField2}, leading to the variation $\int d^4 x \, A_\mu \delta J^\mu_R$. Physically, this means that in alternative quantisation, an external current sources a dynamical (boundary) vector field.\footnote{See Refs. \cite{Witten:2003ya,Marolf:2006nd,Jokela:2013hta} for discussions regarding the exchange of boundary conditions and (emergent) dynamical gauge fields in lower-dimensional theories.} The two boundary theories, one with Dirichlet and one with Neumann boundary conditions, are normally related by a double-trace deformed RG flow. In our case, we require the boundary double-trace deformation $\int d A \wedge \star\, dA$, or its Hodge dualised $\int d^4 x \, J_{\mu\nu} J^{\mu\nu}$, to be explicitly present regardless of the choice of the quantisation.

From the point of view of the quantum bulk theory, as in a lower-dimensional theory \cite{Faulkner:2012gt}, the Einstein-Maxwell bulk (quantum) path integral runs over the metric and the Maxwell field $A_a$. Alternatively, one can write the path integral over the fields strength $F_{ab}$, but at the expense of ensuring the Bianchi identity $dF = 0$ by introducing a Lagrange multiplier $B_{ab}$: 
\begin{equation}
Z \supset \int \CD F_{ab} \, \CD B_{ab}  \,\, \text{exp} \left\{ i \, \frac{N_c^2}{8\pi^2} \int d^5 x \sqrt{-G} \, \left( F_{ab}F^{ab} +e_H^{-1} B_{ab}  \epsilon^{abcde}\nabla_c F_{de} \right)\right\} .
\end{equation}
Since the second (Bianchi identity) term vanishes for any classical field solution, it has no influence on the saddle point of the path integral. However, it does generate a non-zero contribution to the boundary action, i.e. $\left(N_c^2 / 8 \pi^2 e_H \right) \int d^4x  \, \epsilon^{\mu\nu\rho\sigma} F_{\rho\sigma} B_{\mu\nu}$, which is precisely the source term $\int d^4x \, J^{\mu\nu} b_{\mu\nu}$ once we identify $B_{\mu\nu} \sim b_{\mu\nu}$ and $J^{\mu\nu} \sim \epsilon^{\mu\nu\rho\sigma}F_{\rho\sigma}$. The precise dictionary between the bulk and boundary quantities will be discussed in Section \ref{sec:holoRenorm}. By varying the action with respect to $F_{ab}$, one obtains the equation of motion 
\begin{equation}\label{relnFandBbulk}
F^{ab} = e_H^{-1} \epsilon^{abcde}\nabla_c B_{de} \, .
\end{equation}
Then, the field strength $F_{ab}$ can be integrated out in the saddle point approximation which gives the two-form gauge field Lagrangian term from Eq. \eqref{BulkAction}. Furthermore, in the language of the Einstein-Maxwell theory, by using Eq. \eqref{relnFandBbulk}, one finds the relation between the one-form R-current $J^\mu_R$ and the $B_{ab}$ field:
\begin{equation}
\la J^\mu_R \ra =- \frac{N_c^2}{2 \pi^2} \lim_{u\to 0} F^{u\mu} = -\frac{N_c^2}{2 \pi^2  e_H}\lim_{u\to 0}\epsilon^{\mu\nu\rho\sigma}\partial_\nu B_{\rho\sigma} \, ,
\end{equation}
where $u$ is the radial coordinate and $u = 0$ the boundary of the bulk spacetime. Thus, imposing Dirichlet boundary conditions on $B_{ab}$ (in this case, necessarily with an additional double-trace deformation) corresponds to treating $J^\mu_R$ as a source, which is the same as performing alternative quantisation discussed above. This is again consistent with the interpretation that the dual field theory of \eqref{BulkAction} contains dynamical photons. Furthermore, as we will see from a detailed holographic renormalisation in Section \ref{sec:holoRenorm}, the (double-trace) boundary counter-terms, which are required to keep the on-shell action finite, will give us precisely the Maxwell theory for $A_\mu$ (dual of $b_{\mu\nu}$) on the boundary, including a renormalised electromagnetic coupling $e_r$, as in QED.\footnote{We note that the way the Maxwell Lagrangian arises on the boundary is equivalent to the way holographic matter can be coupled to dynamical gravity on a cut-off brane \cite{Gubser:1999vj}. There too, a holographic counter-term gives rise to the Einstein-Hilbert action at the cut-off brane (the boundary) of a more intricately foliated bulk. As shown by Gubser in \cite{Gubser:1999vj}, such a theory can result in a radiation (CFT)-dominated FRW universe at the boundary with the stress-energy tensor of the $\CN =4 $ SYM driving the expansion.} All further details of this holographic setup will be presented in Section \ref{sec:Holography}. 

\section{Holographic analysis of theories with generalised global symmetries: equation of state and transport coefficients}\label{sec:Holography}
In this section, we study the relevant details of the simplest holographic theory with Einstein gravity coupled to a higher-form (two-form) bulk field, cf. Eq. \eqref{BulkAction}, which can source a two-form current associated with the $U(1)$ one-form generalised global symmetry in the boundary theory. In other words, we construct the holographic dictionary for theories with higher-form symmetries. As our main goal is to study the phenomenology of MHD waves in a strongly coupled plasma using the dispersion relations of \cite{Grozdanov:2016tdf}, we will use holography only to provide us with the necessary microscopic data: the equation of state and the transport coefficients. 

In Section \ref{sec:ActionAndBrane}, we will begin by discussing details of the magnetic brane solution \cite{D'Hoker:2009mmn,D'Hoker:2009bc} supported by the bulk action introduced in Section \ref{sec:N4-II}. In Section \ref{sec:holoRenorm}, we will consider holographic renormalisation of the theory in question and show how the bulk gives rise to a dual theory coupled to dynamical electromagnetism (as in Section \ref{sec:MatterEM}). In particular, we will derive the expectation values of the stress-energy tensor $\la T^{\mu\nu} \ra$ and the two-form $\la J^{\mu\nu}\ra$ and show that they satisfy the Ward identities \eqref{EOM1} and \eqref{EOM2}. We will also recover and match all expected renormalisation group properties, such as the beta function of the electromagnetic coupling, from the point of view of the bulk calculation. In Section \ref{sec:EOS}, we will then compute and analyse thermal and magnetic properties of the equation of state of the dual plasma. Finally, in Section \ref{transport-maintext}, we will derive the membrane paradigm formulae for the seven transport coefficients required to describe first-order dissipative MHD \cite{Grozdanov:2016tdf} and compute them.\footnote{These horizon formulae are analogous to the expression for shear viscosity in $\CN=4$ theory \cite{Iqbal:2008by}. For more recent discussions of other transport coefficients that can be computed directly from horizon data, see e.g. \cite{Gubser:2008sz,Saremi:2011ab,Donos:2014cya,Banks:2015wha,Davison:2015taa,Gursoy:2014boa,Grozdanov:2016ala}.} Further details regarding the horizon formulae for the transport coefficients can be found in Appendix \ref{appendix:transport}.

\subsection{Holographic action and the magnetic brane}\label{sec:ActionAndBrane}

A holographic action dual to a plasma state with a low-energy limit that can be described by MHD was stated in Eq. \eqref{BulkAction}. Including the boundary Gibbons-Hawking term and the (relevant) holographic counter-term, the full action is 
\begin{equation}\label{HoloAction}
\begin{aligned}
S &= \frac{N_c^2}{8 \pi^2} \Bigg[\int d^5x \sqrt{-G}\, \left( R+\frac{12}{L^2} -\frac{1}{3e_H^2} H_{abc}H^{abc} \right) \\
&\qquad + \int_{\partial M} d^4x \sqrt{-\gamma} \left( 2\, \tr K - 6 + \frac{1}{e_H^2} \CH_{\mu\nu}\CH^{\mu\nu}\ln \CC\right)\Bigg],
\end{aligned}
\end{equation}
where $\tr K$ is the trace of the extrinsic curvature of the boundary ($\partial M$) defined by an outward normal vector $n^a$. For convenience, we set both the AdS radius $L=1$ and $e_H=1$. The two-form $\CH_{\mu\nu}$ is defined as a projection of the three-form field strength in the direction normal the boundary, $\CH_{\mu\nu} = n^a H_{a\mu\nu}$. $\CC$ is a dimensionless number that needs to be adjusted to fix the renormalisation condition, of which the details will be discussed below. The equations of motion that follow are
\begin{align}
R_{ab} + 4G_{ab} - \left( H_{acd} H_b^{\;\;cd}-\frac{2}{9} G_{ab}H_{cde} H^{cde}\right) &= 0 \,  , \label{EOMTheory1}\\
\frac{1}{\sqrt{-G}} \partial_a\left( \sqrt{-G} H^{abc}\right) &= 0 \, \label{EOMTheory2} .
\end{align}

Since the theory \eqref{HoloAction} is S-dual to the Einstein-Maxwell theory, we can express the magnetised black brane solution of \cite{D'Hoker:2009mmn} by dualising the Maxwell terms and writting
\begin{equation}
\begin{aligned}
ds^2 &= G_{ab}dx^a dx^b = r_h^2\left(-F(u) dt^2 + \frac{e^{2\CV(u)}}{v}(dx^2+dy^2) +
\frac{e^{2\CW(u)}}{w}dz^2\right) + \frac{du^2}{4u^3 F(u)}\, , \\ 
H &=  \frac{B r_h^2 e^{-2\CV+\CW}}{2u^{3/2}\sqrt{w}} \,dt\wedge dz\wedge du \, .
  \end{aligned}
  \label{StefanAnsatz}
\end{equation}
The equations of motion \eqref{EOMTheory1} for this ansatz reduce to three second-order ordinary differential equations (ODE's) for $\{F,\,\CV,\,\CW\}$ and one additional first-order constraint. The equation of motion derived from the variation of the two-form \eqref{EOMTheory2} is automatically satisfied. The equations are equivalent to those derived from the Einstein-Maxwell theory \cite{D'Hoker:2009mmn} upon identification of the Maxwell bulk two-form $F$ with $F = \mathcal{B} \, dx \wedge dy$, where $\mathcal{B} = Br_h^2/v $.\footnote{The metric ansatz is chosen to have the form used in \cite{Janiszewski:2015ura}. It can be obtained from the ansatz $ds^2 = -U(r)dt^2 + e^{2V(r)}(dx^2+dy^2) + e^{2W(r)}dz^2+ dr^2/U(r)$ used in \cite{D'Hoker:2009mmn} by performing a coordinate transformation $r=r_h/\sqrt{u}$ and shifting $V$ and $W$ by constant $-\ln v$ and $-\ln{w}$, respectively, which are chosen so that the near-boundary expansion has the form $ds^2 = (1/u) \, \eta_{\mu\nu}dx^\mu dx^\nu + du^2 / (4u^2)$.
} 

The undetermined functions $F$, $\CV$ and $\CW$ are obtained numerically by using the shooting method. We first expand the background fields near the horizon as 
\begin{equation}
\begin{aligned}
F &= f_{1}^h (1-u) + f_{2}^h(1-u)^2+\CO(1-u)^3 \,,\\
\CV &= v_{0}^h + v^h_{1}(1-u) + \CO(1-u)^2 \,,\\
\CW &= w^h_{0} + w^h_{1}(1-u) + \CO(1-u)^2 \, ,
\end{aligned}
\end{equation}
where the constants $\{f^h_i,\,v^h_i,\,w^h_i\}$ can be written in terms of $\{f_1^h,\,v^h_0,\,w^h_0\}$ after solving the equations of motion order-by-order near the horizon. The scaling symmetry of our background ansatz then allows us to rescale $dx$ and $dy$ so that $v_0^h = w_0^h = 0$. Next, we match the numerical solutions generated by shooting from the horizon towards the boundary, where the analytical near-boundary expansions of the metric functions are
\begin{equation}
\begin{aligned}
F &=\frac{1}{u}\left( 1 +f_1^b \sqrt{u}+ \frac{f_1^b}{4} u + f_4^b u^2 +\CO(u^{3/2})+
\left(\frac{\CB^2}{3}+\CO(\sqrt{u}) \right) u^2\ln u\right) ,\\
e^{2\CV} &= \frac{1}{u}\left(v + v f_1^b \sqrt{u} + \frac{v (f^b_1)^2}{4}u + v^b_4 u^2 + \CO(u^{3/2})
- \left(\frac{\CB^2}{6}+\CO(\sqrt{u}) \right)u^2\ln u\right),\\
e^{2\CW} &=\frac{1}{u}\left(w + w f_1^b \sqrt{u} + \frac{w (f^b_1)^2}{4}u - \frac{2wv^b_4}{v} u^2 + \CO(u^{3/2})
- \left( \frac{w\CB^2}{3}+\CO(\sqrt{u}) \right)u^2\ln u\right).\\
\end{aligned}
\label{eqn:nearBndExpansion}
\end{equation}
As before, one can solve for the coefficients $\{f^b_i,\,v^b_i,\,w^b_i\}$ in terms of $\{f^b_1,\,f^b_4,\,v^b_4\}$. Furthermore, $f_1^b$ can be removed by residual diffeomorphism freedom of the metric ansatz \cite{Janiszewski:2015ura}. For a given value of $B = v \CB/r_h^2$, we can therefore generate a numerical background by shooting from the initial conditions of the functions set by the near-horizon expansion with $\{f^h_1,v^h_0,w^h_0 \}= \{\hat f ,0,0\}$. The numerical value of $\hat f$ is chosen so that the near-boundary expansion has $f^b_1 = 0$. The near-boundary behaviour of this function then determines the properties of the dual field theory. Note that the theory is governed by a one-parameter family of such numerical solutions characterised by the dimensionless ratio $T/\sqrt{\mathcal{B}}$, where $T = f^h_1r_h / 2\pi $ is the Hawking temperature (see Eq. \eqref{TempEnt}). In practice, this ratio can be tuned by changing the parameter $B$ of the background ansatz. The numerical solver encounters stiffness problems when $B \approx \sqrt{3}$, i.e. where the temperature is close to zero. All of our numerical results will therefore stop near $T / \sqrt{\CB} = 0$. In this work, we do not attempt an independent analysis of the theory at $T = 0$.

\subsection{Holographic renormalisation and the bulk/boundary dictionary}\label{sec:holoRenorm}

The next step in analysing the dual of \eqref{HoloAction} is a systematic holographic renormalisation. In this section, we derive the one-point functions of the stress-energy tensor $\la T_{\mu\nu} \ra $ and the two-form current $\la J_{\mu\nu}\ra$, and show that they satisfy the Ward identities of magnetohydrodynamics \eqref{EOM1} and \eqref{EOM2} \cite{Grozdanov:2016tdf}, which in terms of operator expectation values take the form
\begin{align}\label{WardIden1}
\grad_\nu \la T^{\mu\nu}\ra = \tilde H^{\mu}_{~\lambda\sigma} \la J^{\lambda\sigma}\ra \, , &&  \grad_\mu \la J^{\mu \nu}\ra =0 \,,
\end{align}
where $\tilde H = db$ is the field strength of the background gauge field $b$ in field theory. The precise definition of these quantities will become clear below. Since we are only interested in the expansion of MHD to first order in the gradient expansion around a flat (boundary) background, it will be sufficient to only work with terms that contain no more than two derivatives along the boundary directions. The procedure for obtaining holographic renormalisation will closely follow Refs. \cite{deHaro:2000vlm,Taylor:2000xw}.\footnote{This part of the calculation was performed by using the Mathematica package xAct \cite{xAct}.} 

We begin by writing the bulk metric in the Fefferman-Graham coordinates \cite{deHaro:2000vlm}
\begin{equation}\label{FGcoord}
ds^2_\text{FG} =G_{ab}\, dx^a dx^b= \frac{d\rho^2}{4\rho^2} + \gamma_{\mu\nu}(\rho,x)dx^\mu dx^\nu = \frac{d\rho^2}{4\rho^2} + \frac{1}{\rho} g_{\mu\nu}(\rho,x) dx^\mu dx^\nu \, ,
\end{equation}
so that near the boundary, $\rho \approx 0$, the metric $g_{\mu\nu}$ can be expanded as  
\begin{equation}\label{nearBoundaryFGmetric}
g_{\mu\nu}(\rho,x) = g^{(0)}_{\mu\nu}(x) + \rho g^{(1)}_{\mu\nu}(x) + \rho^2 \left(g^{(2)}_{\mu\nu}(x) + \tilde h_{\mu\nu}(x) \ln\rho \right)+\CO(\rho^3) \, .
\end{equation}
Note that Greek (boundary) indices in a tensor $A^{\mu\nu}$ are raised with the metric $g^{\mu\nu}_{(0)}$, which satisfies $g^{(0)}_{\mu\nu}g^{\mu\nu}_{(0)} = 4$. There are two types of covariant derivatives that we will use: $\grad_\mu^{(g)}$ and $\grad_\mu$. Firstly, $\grad_\mu^{(g)}$ and $\grad^\mu_{(g)} \equiv g^{\mu\nu} \grad_\mu^{(g)}$ are defined with respect to the metric $g_{\mu\nu}(\rho,x)$, while $\grad_\mu$ and $\grad^\mu\equiv g^{\mu\nu}_{(0)} \grad_\mu$ are defined through the metric $g^{(0)}_{\mu\nu}(x)$. The Ricci tensors of $g_{\mu\nu}$ and $g^{(0)}_{\mu\nu}$ are denoted by $R^{(g)}_{\mu\nu}$ and $R^{(0)}_{\mu\nu}$, respectively. 

The components of bulk two-form gauge field $B_{ab}$ in the boundary field theory directions can similarly be expanded near the boundary as  
\begin{equation}\label{nearBoundaryFGgaugefield}
B_{\mu\nu}(\rho,x) = B^{(0)}_{\mu\nu}(x) + B^{(1)}_{\mu\nu}(x) \ln \rho + \CO(\rho) \, .
\end{equation}
In the boundary directions, the three-form field strength is defined as $ H_{\mu\nu\sigma} = \partial_\mu B_{\nu\sigma} + \partial_\nu B_{\sigma\mu} + \partial_\sigma B_{\mu\nu}$, with the near-boundary expansion $ H_{\mu\nu\sigma}(\rho,x) = H^{(0)}_{\mu\nu\sigma}(x) + H^{(1)}_{\mu\nu\sigma}(x) \ln \rho + \CO(\rho)$.  Each $ H^{(n)}$ is defined in terms of $B^{(n)}$, i.e. in the same way at each order. Note that both quantities $B^{(0)}_{\mu\nu}$ and $B^{(1)}_{\mu\nu}$ are related to the two-form gauge field source of the boundary theory, $ \int d^4x \sqrt{-g}\, J^{\mu\nu}\delta b_{\mu\nu}$. The variation of the regularised bulk on-shell contribution from the $H^2$ term, evaluated at the boundary cut-off $\rho = \rho_\Lambda$, is  
\begin{equation}
\delta S_{\scriptscriptstyle on-shell} = -\frac{N_c^2 }{4 \pi^2} \int d^4x \sqrt{-g} \, \CH^{\mu\nu} \left(\delta B_{\mu\nu}^{(0)}  + \delta B^{(1)}_{\mu\nu} \ln \, \CC^2\rho_\Lambda \right) .
\end{equation}
This expression makes it clear that the boundary source should be identified with the linear combination of $B^{(0)}_{\mu\nu}$ and $B^{(1)}_{\mu\nu} \ln \,\CC^2\rho_\Lambda$ in the parenthesis. Thus, $\mathcal{H}^{\mu\nu}$ sets the expectation value of $J^{\mu\nu}$ in the boundary theory. However, due to the fact that, by definition, $J^{\mu\nu} = \frac{1}{2} \epsilon^{\mu\nu\lambda\sigma}F_{\lambda\sigma}$, of which the expectation value contains no colour trace, we need to identify the combination of $B^{(0)}_{\mu\nu}$ and $B^{(1)}_{\mu\nu}$ with the field theory source $b_{\mu\nu}$ by including a factor proportional to $1/N_c^2$, i.e., 
\begin{equation}\label{eq:B0toSourceb}
B^{(0)}_{\mu\nu} +  B^{(1)}_{\mu\nu}\ln\, \CC^2\rho_\Lambda  = \frac{4 \pi^2 }{N_c^2 } b_{\mu\nu} \, .
\end{equation}
In holography, such boundary conditions are knows as {\em mixed} boundary conditions. They arise in the presence of double-trace deformations \cite{Witten:2001ua}, which is precisely how the logarithmically running $H^2$ term in the renormalised on-shell action should be interpreted. From the point of view of the boundary field theory, as we will see below, this term is a consequence of dynamical boundary electromagnetism---it is the boundary Maxwell action. Now, since the source $b_{\mu\nu}$ is a physical quantity, it cannot depend on the cut-off scale $\rho_\Lambda$. Hence, the renormalisation group equation
\begin{align}\label{RGEqB}
\frac{db_{\mu\nu}}{d \rho_\Lambda}  = 0\,
\end{align}
prompts us to set the value of $\CC \sim 1 / \sqrt{\rho_\Lambda}$, which makes the on-shell action formally finite in the limit of $\rho_\Lambda \to 0$. Of course, we need to scale $\CC\to\infty$ so that the product $\CC^2\rho_\Lambda$ remains finite. As we will see below, the proportionality constant in the relation between $\CC^2$ and $\rho_\Lambda$ sets the value of the renormalised electromagnetic coupling, and corresponds to the choice of the renormalisation group condition. This procedure replaces the necessity to keep the cut-off scale of the theory explicit in our final results and replaces the need to explicitly choose the Landau pole scale in favour of choosing the renormalisation group scale, or the electromagnetic coupling. With these boundary conditions in hand, the expectation value of $J^{\mu\nu}$ can then be obtained by taking a variational derivative of the on-shell action with respect to the source $b_{\mu\nu}$. 

The Ward identities \eqref{WardIden1} can be obtained by solving the equations of motion \eqref{EOMTheory1} and \eqref{EOMTheory2} \cite{Taylor:2000xw}. In the Fefferman-Graham coordinates \eqref{FGcoord}, these equations (together with the trace of \eqref{EOMTheory1}) become 
\begin{align}
\frac{1}{2}\tr\left[g^{-1}g'' \right] - \frac{1}{4}\tr\left[g^{-1}g'g^{-1}g' \right] +\frac{1}{3}\rho^2\tr\left[ g^{-1}B'g^{-1}B' \right]-\frac{1}{18}\rho \,\tr[ g^{-1} H^2] &=0 \, ,\label{FGeqn1}  \\
\frac{1}{2}\left(\grad_\mu^{(g)}\tr g' - \grad^\nu_{(g)}g'_{\mu\nu}\right) - \rho^2  H_{\mu\alpha\beta}\left(g^{-1}B'g^{-1}\right)^{\alpha\beta} &=0 \, , \label{FGeqn2} \\
\rho\left( 2g''_{\mu\nu} - 2(g'g^{-1}g')_{\mu\nu} + g'_{\mu\nu} \tr[g^{-1}g']\right) + R^{(g)}_{\mu\nu}-2g'_{\mu\nu} - g_{\mu\nu} \tr[g^{-1}g'] & \nonumber \\
\quad +  8\rho^3\left[ (B'g^{-1}B')_{\mu\nu}-\frac{1}{3}g_{\mu\nu}\tr\left[g^{-1}B'g^{-1}B' \right]\right]+\rho^2 \left[  H^2_{\mu\nu}-\frac{2}{9}g_{\mu\nu}\tr[g^{-1} H^2] \right] & = 0 \,,\label{FGeqn3}\\
\frac{d}{d\rho}\left( 2\rho \left( g^{-1}B'g^{-1} \right)^{\mu\nu} \right)+\frac{1}{2}\grad^\lambda_{(g)}\left( g^{\mu\alpha}g^{\nu\beta}H_{\lambda\alpha\beta} \right)& = 0 \,, \label{FGeqn4} \\
\grad_\nu \left( g^{-1}B'g^{-1}\right)^{\mu\nu} &= 0\, , \label{FGeqn5}
\end{align}
where $g^{-1}$ denotes the matrix inverse of $g$ (in components, this is $g^{\mu\nu}$) and where
\begin{align}
\tr[ g^{-1}B' g^{-1}B'] = -B'_{\mu_1\mu_2}B'_{\nu_1\nu_2}g^{\mu_1\nu_1}g^{\mu_2\nu_2}\,,&&  H^2_{\mu\nu} = H_{\mu\lambda_1\lambda_2}H_{\nu\sigma_1\sigma_2} g^{\lambda_1\sigma_1}g^{\lambda_2\sigma_2} \, .
\end{align}
Expanding equations \eqref{FGeqn1}--\eqref{FGeqn5} around small $\rho$, we find that
\begin{align}
g^{(1)}_{\mu\nu} = \frac{1}{2}\left( R^{(0)}_{\mu\nu}-\frac{1}{6}g^{(0)}_{\mu\nu} R^{(0)}\right),&& (g^{(1)})^\mu_{~\mu} = \frac{1}{6} R \, .
\end{align}
Since $g^{(1)}_{\mu\nu}$ is proportional to second derivatives of the boundary metric, and we are only keeping track of terms up to second order in boundary derivatives, we can ignore terms with $g^{(1)}_{\mu\nu}$. The remaining equations of motion can thus be written as 
\begin{align}
(g^{(2)})^\mu_{~\mu} -\frac{1}{3} B^{(1)}_{\mu\nu}B^{(1)\mu\nu}=0\,, \qquad \tilde h^\mu_{~\mu} = 0 \, , \qquad \grad_\nu B^{(1)\mu\nu} &= 0 \, ,\label{FGrel1}\\
- H_{\mu\nu\lambda}^{(0)}B^{(1)\nu\lambda} + \grad^\nu_{(0)}\left(  g^{(0)}_{\mu\nu}  (g^{(2)})^\lambda_{~\lambda} - g_{\mu\nu}^{(2)}-\frac{1}{2}\tilde h_{\mu\nu}  \right) &=  0 \, ,\label{FGrel2}\\
 \tilde h_{\mu\nu} + \frac{1}{2}\left(4 B^{(1)}_{\mu\lambda} (B^{(1)})_\nu^{~\lambda}- g_{\mu\nu}^{(0)} B^{(1)}_{\lambda\sigma}B^{(1)\lambda\sigma}  \right) &=0 \, . \label{FGrel3}
\end{align}

The expectation values of the stress-energy tensor and the two-form current follow from the generating functional \eqref{GenFunTJ}:
\begin{align}\label{Def1ptFunction}
\la T^{\mu\nu} \ra = -\frac{2i}{\sqrt{-g^{(0)}}} \frac{\delta \ln W}{\delta g^{(0)}_{\mu\nu}}\, ,&& \la J^{\mu\nu} \ra = - \frac{i}{\sqrt{-g^{(0)} }} \frac{\delta \ln W}{\delta b_{\mu\nu}} \, .
\end{align}
In holography, $W$ is computed from the (on-shell) action \eqref{HoloAction}, giving us\footnote{Note that in order to raise indices of the boundary theory expectation values, one needs to use the induced metric $\gamma_{\mu\nu}$.} 
\begin{align}
\la T_{\mu\nu}\ra =& -\frac{N_c^2}{4 \pi^2} \lim_{\epsilon \to 0}\frac{r_h^2}{\epsilon}\left(K_{\mu\nu} - \gamma_{\mu\nu} K - 3 \gamma_{\mu\nu}+ \frac{1}{2}R[\gamma]_{\mu\nu}  \right. \nn
&\left. -\frac{1}{4}\gamma_{\mu\nu}R[\gamma]- \left(\CH_{\mu\lambda}\CH_\nu^{\;\;\lambda}-\frac{1}{4}\gamma_{\mu\nu}\CH_{\alpha\beta}\CH^{\alpha\beta}\right)\ln (\CC^2\rho)\right) \biggr|_{\rho=\rho_\Lambda} \label{defT} \, , \\
\la J_{\mu\nu}\ra =& -\lim_{\epsilon \to 0}\,\CH_{\mu\nu} \big|_{\rho=\rho_\Lambda} \, .\label{defJ}
\end{align}
Note that while the expectation value of $T^{\mu\nu}$ scales as $N_c^2$, the expectation value of $J^{\mu\nu}$ is of order $\CO(1)$. 

By using Eq. \eqref{FGrel1} and the fact that $\CH_{\mu\nu} = n^\rho H_{\rho\mu\nu} = -2 B^{(1)}_{\mu\nu}+\CO(\rho)$, we find that the boundary two-form current is conserved:
\begin{align}\label{JConserHolRG}
\grad_{(0)}^\mu \la J_{\mu\nu} \ra=  2  \grad^\mu B_{\mu\nu}^{(1)}= 0 \,.
\end{align}
Using the definition \eqref{TwoFormJ}, which gives $\la J_{\mu\nu} \ra = \frac{1}{2} \epsilon_{\mu\nu\rho\sigma} \la F^{\rho\sigma} \ra$ and connects Eq. \eqref{JConserHolRG} with the Bianchi identity, we find that $\star \, B^{(1)}$ sets the expectation value of the Maxwell field strength $ \la F \ra$. Furthermore, the (regularised) stress-energy tensor \eqref{defT} becomes
\begin{equation}\label{holographicStress}
\la T_{\mu\nu}\ra =\lim_{\rho_\Lambda \to 1/(L\Lambda)^2}\frac{N_c^2}{2 \pi^2 }\left(g^{(2)}_{\mu\nu} - g^{(0)}_{\mu\nu}(g^{(2)})^\lambda_{~\lambda}  +\frac{1}{2}\tilde h_{\mu\nu} +\tilde h_{\mu\nu}\ln\left(\CC^2 \rho\right)+ \CO(\rho,\partial^2)\right) \biggr|_{\rho=\rho_\Lambda} .
\end{equation}
It is useful to write $\rho_\Lambda = 1/(L\Lambda)^2$, where $\Lambda$ is the UV cut-off energy of the theory and $L$ is the AdS radius which we set to be $L=1$. As discussed in Section \ref{sec:MatterEM}, the choice of the constant $\CC$ must now be made in order to fix the renormalisation condition, which will render the renormalised expectation value $\la T_{\mu\nu}\ra$ physical and finite in the formal limit of $\Lambda \to \infty$. This again implies that $\CC^2 \rho_\Lambda$ has to be finite and invariant under the change of the cut-off scale, which is consistent with the renormalisation group-invariant condition for the dual field theory source $b_{\mu \nu}$ in Eq. \eqref{RGEqB}. It will prove useful to introduce a renormalisation group-invariant energy scale $M_\star = \Lambda/\CC$, which is the energy scale associated with the Landau pole. Furthermore, we also introduce the combination $1/e_r^2 = \ln (\Lambda L/\CC)$, which, as we shall see shortly, plays the role of the renormalised electromagnetic coupling. 

To see how the constant $\CC$ in Eq. \eqref{holographicStress} is related to our discussion in Section \ref{sec:MatterEM}, we write the last term by introducing a mass scale $M$:
\begin{equation}\label{splitting}
\frac{N_c^2}{\pi^2}\tilde h_{\mu\nu}\ln(\Lambda L/\CC) = \frac{N_c^2}{\pi^2} \tilde h_{\mu\nu} \ln(\Lambda/M) + \tilde h_{\mu\nu}\left( \frac{2}{e_r^2} - \frac{N_c^2 }{\pi^2} \ln(\Lambda/M) \right) . 
\end{equation}
What can be seen from Eq. \eqref{splitting} is that this splitting precisely reproduces the way the logarithmic divergence enters into the stress-energy tensor from two different pieces of the Lagrangian: the matter content (with its coupling to the photons) and the electromagnetic (Maxwell) part: 
\begin{align}
\la T_{\mu\nu} \ra  = \la T^{\scriptscriptstyle matter}_{\mu\nu}\ra + \la T^{\scriptscriptstyle EM}_{\mu\nu}\ra \,,
\end{align}
with the two terms being
\begin{align}
\la T^{\scriptscriptstyle matter}_{\mu\nu}\ra &= \frac{N_c^2}{2\pi^2}\left(g^{(2)}_{\mu\nu} - g^{(0)}_{\mu\nu}(g^{(2)})^\lambda_{~\lambda}  +\frac{1}{2}\tilde h_{\mu\nu} \right)  - \frac{N_c^2}{\pi^2} \tilde h_{\mu\nu} \ln(\Lambda/M)  \, , \\
\la T^{\scriptscriptstyle EM}_{\mu\nu}\ra &=  -\left(\frac{2}{e_r^2} - \frac{N_c^2}{\pi^2}  \ln \left(\Lambda / M \right)\right) \tilde h_{\mu\nu} \, .
\end{align}
Finally, we note that the electromagnetic $\la T^{\scriptscriptstyle EM}_{\mu\nu}\ra$ would follow precisely from the Maxwell boundary action, which induces a double-trace deformation into the boundary field theory (see discussion below Eq. \eqref{eq:B0toSourceb}) 
\begin{align}
S_{\scriptscriptstyle EM} = -\frac{1}{4 e(\Lambda/M)^2} \int d^4x \sqrt{-g} F_{\mu\nu}F^{\mu\nu} \, , \qquad \frac{1}{e(\Lambda/M)^2} = \left( \frac{1}{e_r^2} - \frac{N_c^2}{2\pi^2} \ln (\Lambda/M) \right)\, ,
\end{align}
upon using Eq. \eqref{FGrel3} and the fact that the bulk $\star \, B^{(1)}$ determines $\la F_{\mu\nu}\ra $: 
\begin{align}
\la T^{\scriptscriptstyle EM}_{\mu\nu}\ra &=\frac{1}{e(\Lambda/M)^2}\left(\la F_{\mu\alpha}F_\nu^{\;\;\alpha}\ra - \frac{1}{4} \eta_{\mu\nu} \la F_{\alpha\beta}F^{\alpha\beta} \ra \right)  \nn
&= \frac{1}{e(\Lambda/M)^2}\left(\la F_{\mu\alpha} \ra \la F_\nu^{\;\;\alpha}\ra - \frac{1}{4} \eta_{\mu\nu} \la F_{\alpha\beta} \ra\la F^{\alpha\beta} \ra \right) \, ,
\end{align}
where the last equality follows from the fact that quantum fluctuations of the photon field are suppressed in the boundary QFT. Our holographic calculation thus fully reproduces Eq. \eqref{TmunuSecQFT}, which followed from the field theory discussion in Section \ref{sec:N4}. Furthermore, the running electromagnetic coupling constant matches the one found from field theory (cf. Eq. \eqref{CutOffDependentCouplingN4}) \cite{Fuini:2015hba}. Hence, our holographic setup appears to contain the $U(1)$-gauged matter content of the $\CN = 4$ SYM theory. In terms of bulk quantities, the renormalised stress-energy tensor and the two-form current are
\begin{align}
\la T_{\mu\nu}\ra &=  \frac{N_c^2}{2\pi^2} \left(g^{(2)}_{\mu\nu} - g^{(0)}_{\mu\nu}(g^{(2)})^\lambda_{~\lambda}  +\frac{1}{2}\tilde h_{\mu\nu} \right)  - \frac{2}{e_r^2} \tilde h_{\mu\nu}  \, , \label{THolFin} \\
\la J_{\mu\nu} \ra & =
2B^{(1)}_{\mu\nu} \, \label{JHolFin} ,
\end{align}
where, as in Section \ref{sec:MatterEM}, $e_r$ is the renormalised coupling which needs to be set by experimental input---the renormalisation condition. In practice, this constant is fixed by choosing the value of $\CC$ in \eqref{defT}. For the same reasons as in any QFT with the Landau pole, there is therefore an inherent ambiguity in holographic results, which has to be fixed by external physically-motivated input. Here, instead of simply choosing the Landau pole scale, which would have rendered all our results explicitly dependent on the UV cut-off scale of the theory, we underwent a renormalisation group analysis of the theory and traded the cut-off scale for the renormalisation group scale $M$, which set the more physically relevant electric charge $e_r$. As a result, the stress-energy tensor in Eq. \eqref{THolFin} and all other physical quantities are formally independent of the cut-off scale $\Lambda$.

We conclude this section by noting that the relation \eqref{FGrel2} and a relation between $\tilde H$, $H^{(0)}$ and $H^{(1)}$ implies that the Ward identity for the stress-energy tensor satisfies Eq. \eqref{EOM1}, or in terms of our holographic notation, $\grad_\nu \la T^{\mu\nu}\ra = \tilde H^{\mu}_{~\lambda\sigma} \la J^{\lambda\sigma}\ra$, as in Eq. \eqref{WardIden1}.

\subsection{The equation of state}\label{sec:EOS}

To find the equation of state of our theory, we can use the renormalised stress-energy tensor \eqref{THolFin} and the two-form current \eqref{JHolFin} computed in the previous section. The results are then expressed in terms of the near-boundary expansions \eqref{eqn:nearBndExpansion}, which can be read off from the numerical background. Upon changing the radial coordinate from the Fefferman-Graham $\rho$ to $u$ used in Section \ref{sec:ActionAndBrane}, the logarithmic term in the near-boundary expansion becomes shifted by
\begin{equation}
	 \tilde h_{\mu \nu} \ln \rho  = \tilde h_{\mu \nu}  \ln u + \tilde h_{\mu \nu} \ln (r_h/L) \, .
\end{equation}
Hence, in order to extract $g^{(2)}_{\mu \nu}$ in the Fefferman-Graham coordinates from the near-boundary expansion in the $u$ coordinate, one has to take into account the fact that the term proportional to $u^2$ is a combination of $g^{(2)}_{\mu \nu}$ and $\tilde h_{\mu \nu} \ln (r_h/L)$. This effectively changes the value of the renormalised electromagnetic coupling and the resulting stress-energy tensor in equilibrium, written in terms of variables in \eqref{eqn:nearBndExpansion}, are 
\begin{align}
\left\langle T^{tt} \right\rangle &=  \frac{N_c^2}{2\pi^2} \left[ - \frac{3}{4} f^b_4 r_h^4  + \frac{\CB^2}{8 \pi\bar\alpha} \right], \label{THol1} \\
\left\langle T^{xx} \right\rangle = \left\langle T^{yy} \right\rangle &=  \frac{N_c^2}{2\pi^2}\left[ \left( - \frac{1}{4} f^b_4 + \frac{v^b_4}{v} \right)r_h^4- \frac{\CB^2}{4}  + \frac{\CB^2}{8 \pi \bar\alpha}  \right] ,\label{THol2} \\
\left\langle T^{zz} \right\rangle &=  \frac{N_c^2}{2\pi^2} \left[ \left(- \frac{1}{4} f^b_4 - 2 \frac{v^b_4}{v} \right)r_h^4   - \frac{\CB^2}{8 \pi \bar \alpha} \right],\label{THol3}
\end{align}
where we have used the (renormalised) fine-structure constant of the electromagnetic coupling in the plasma
\begin{equation}
\frac{1}{4\pi \alpha} = \ \frac{1}{e_r^2} +  \ln \left(\frac{r_h}{L}\right) =  \ln \left( M_\star r_h\right) .
\end{equation}
The argument of the logarithm is nothing but the energy scale of the Landau pole $M_\star$ (introduced below Eq. \eqref{holographicStress}) measured in the units of energy set by $1/r_h$. For convenience, we will rescale $\alpha$ by $N_c^2 / 2 \pi^2$ (or $| \beta(1/e^2)|$):
\begin{align}
\bar\alpha =  \frac{N_c^2}{2\pi^2}\alpha \, .
\end{align}
The coupling $\bar\alpha$ (or alternatively, the dimensionless ratio between the Landau pole scale $M_\star$ and the energy scale set by $1/r_h$) has to be fixed by experimental observations as in any other quantum field theory, which is not easy in an unrealistic toy model. 

In studying strongly coupled MHD, it is phenomenologically relevant to not only consider the matter and light-matter interactions, but to also include large electromagnetic self-interactions encoded in the Maxwell action. However, since we are working with a holographic large-$N_c$ matter sector and a single photon, it is unnatural to expect a Maxwell term of the same order. The choice that we make here is to set the rescaled constant $\bar\alpha$ to the physically motivated $\bar\alpha = 1/137$. There are several ways to think about this choice: one is imagining that our plasma contains magnetic properties, which have non-trivial scalings with $N_c$, while another interpretation may assume that the bulk studied here could remain a valid dual of a theory with a reasonably small $N_c$. Of course, by considering only a classical bulk theory, we are restricting the strict validity of any computed observable to the limit of $N_c \to \infty$. As soon as one moves towards finite $N_c$, it becomes crucial to estimate the size of subleading $1/N_c^2$ corrections (topological expansion in the string coupling $g_s$)---an endeavour in holography (and string theory) which to date has been largely neglected and will continue being neglected in this work.\footnote{For some discussions of $1/N_c^2$ corrections to the thermodynamic free energy (the equilibrium partition function) and hydrodynamic long-time tails, see \cite{Denef:2009yy,Denef:2009kn,CaronHuot:2009iq,Arnold:2016dbb,Castro:2017mfj}.} A less problematic limit is that of the infinite 't Hooft coupling, which is also implied by the choice of our action.\footnote{For recent discussions of coupling-dependent holography, see \cite{Stricker:2013lma,Waeber:2015oka,Grozdanov:2016vgg,Grozdanov:2016zjj,Grozdanov:2016fkt} and references therein.} Perhaps the best interpretation is one of an ``agnostic choice" led by our having to fix a free parameter to some value. We will return to a more careful investigation of the dependence of our results on this choice in Section \ref{sec:alphaDependence}.

The expectation values of the stress-energy tensor expressed in \eqref{THol1}--\eqref{THol3} are related to the MHD stress-energy tensor in Eq. \eqref{MHDstress-energy} by
\begin{align}
\la T^{tt}\ra = \varepsilon \, ,&& \la T^{xx}\ra = p\, ,&& \la T^{zz}\ra = p-\mu\rho \,.
\end{align}
We note that, as required in a conformal field theory with a trace anomaly induced by electromagnetic interactions, the trace of stress-energy tensor is non-zero. The holographic two-form current,  
\begin{equation}
\la J^{tz}\ra  = \CB = \frac{Br_h^2}{v}\,,
\end{equation}
is related to the equilibrium magnetic flux line density appearing in the MHD equation \eqref{MHDcurrent} as
$
\la J^{tz} \ra = \rho.
$
Temperature and entropy can be expressed in terms of the background geometry as
\begin{align}\label{TempEnt}
T = \frac{1}{2 \pi } f^h_1r_h \,,&& s =  \frac{N_c^2}{2\pi^2} \left( \frac{\pi r_h^3}{v\sqrt{w}} \right)  ,
\end{align}
and are therefore independent of the renormalised electromagnetic charge. The chemical potential, which is conjugate to the density of magnetic flux lines, can be computed by using the thermodynamic identity $\varepsilon + p = s T + \mu \rho$ (cf. \eqref{ThermoRel1}):
\begin{align}
\mu = \frac{ \la T^{xx} \ra - \la T^{zz} \ra }{\la J^{tz} \ra} =  \frac{N_c^2}{2\pi^2} \left( \frac{3v^b_4}{B}-\frac{B}{4 v} + \frac{B}{4\pi v \bar\alpha} \right)r_h^2\,.
\end{align}
Note that with our choice of the bulk theory scalings, $\rho \sim \CO(1)$ and $\mu \sim \CO(N_c^2)$. Furthermore, while $T \sim\CO(1)$, $p$, $\varepsilon$ and $s$ all scale as $\CO(N_c^2)$.

\begin{figure}[tbh]
\center
\includegraphics[width=0.49\textwidth]{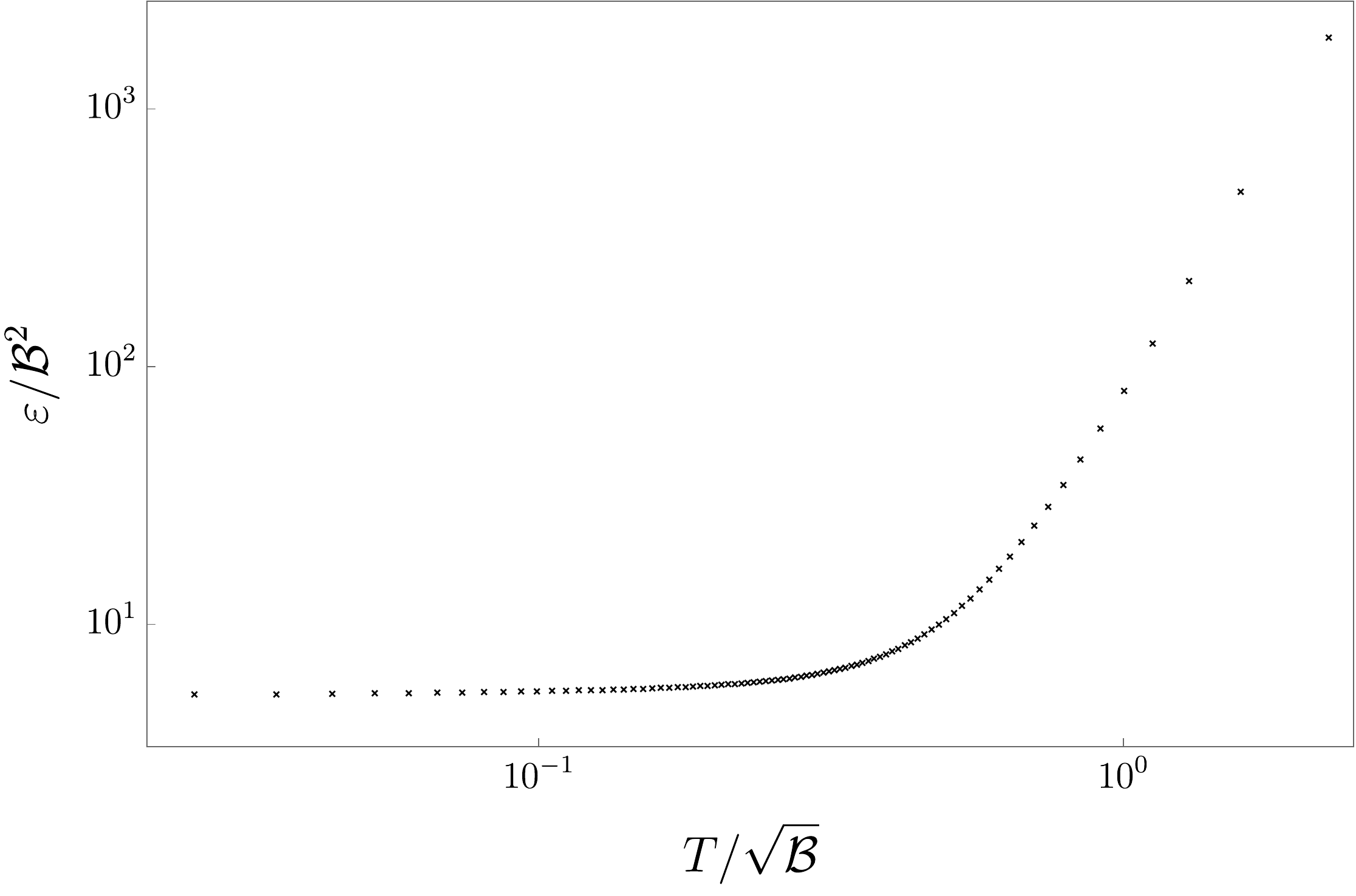}
\includegraphics[width=0.49 \textwidth]{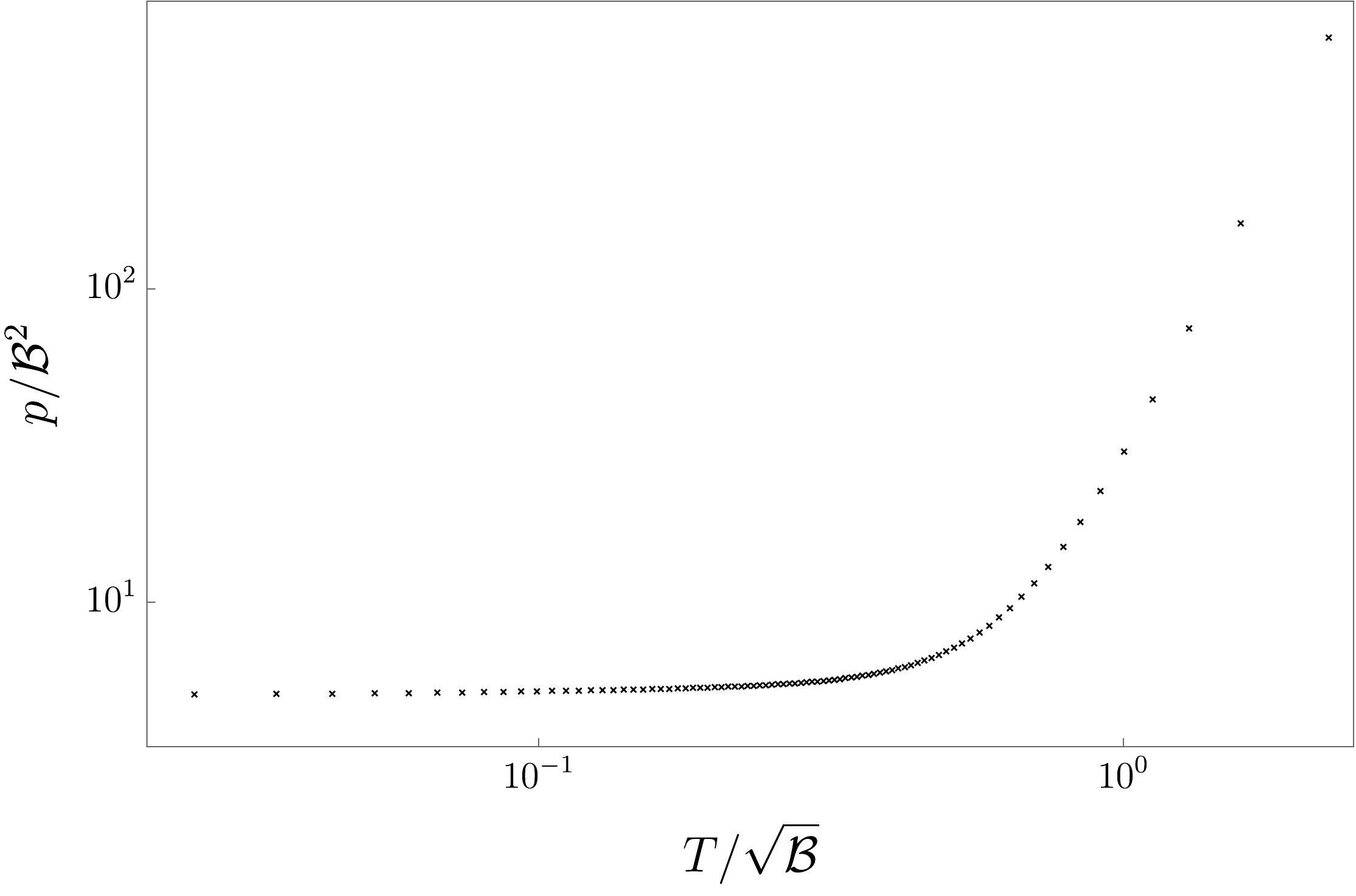}
\includegraphics[width=0.49\textwidth]{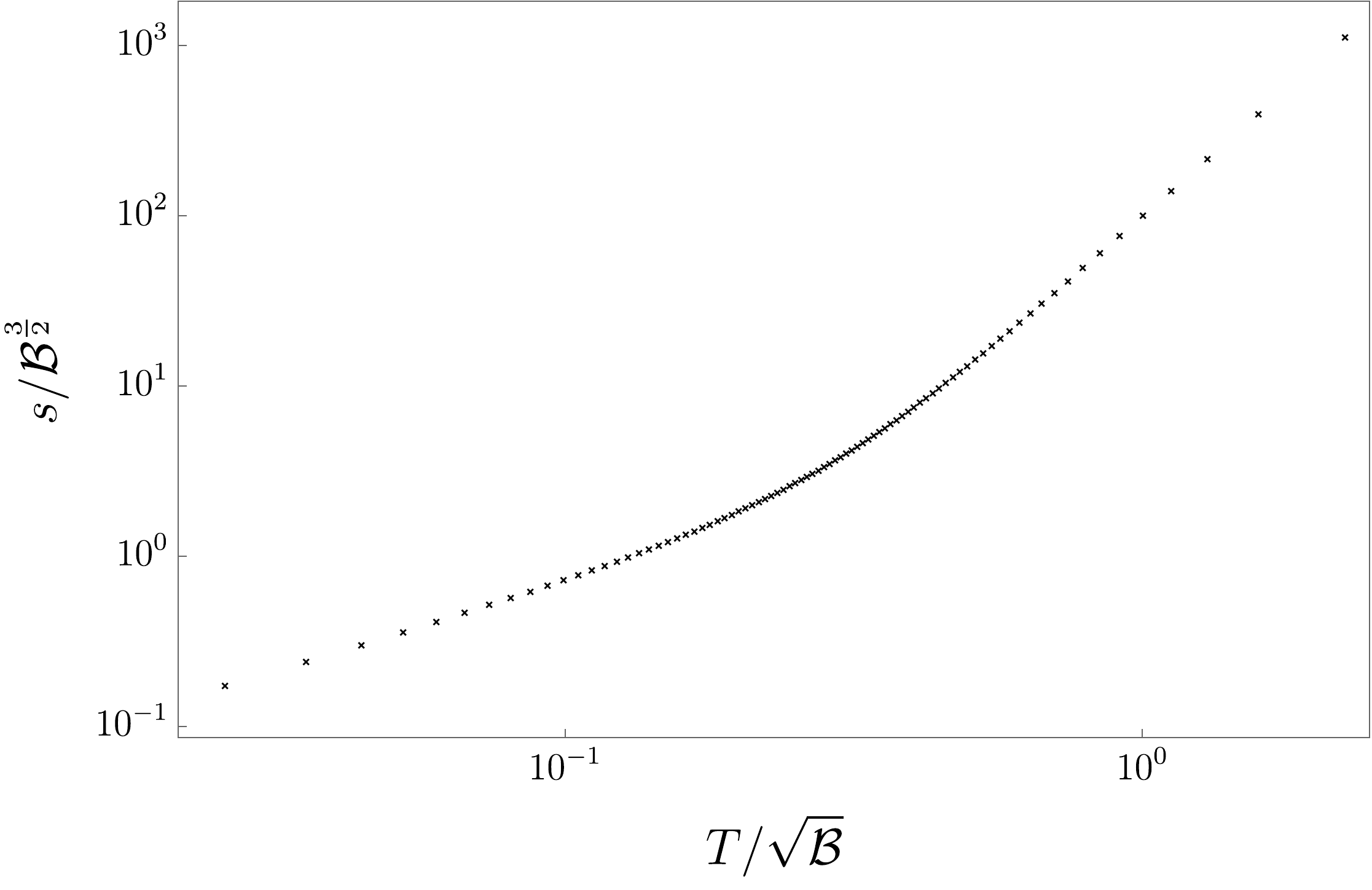}
\includegraphics[width=0.49 \textwidth]{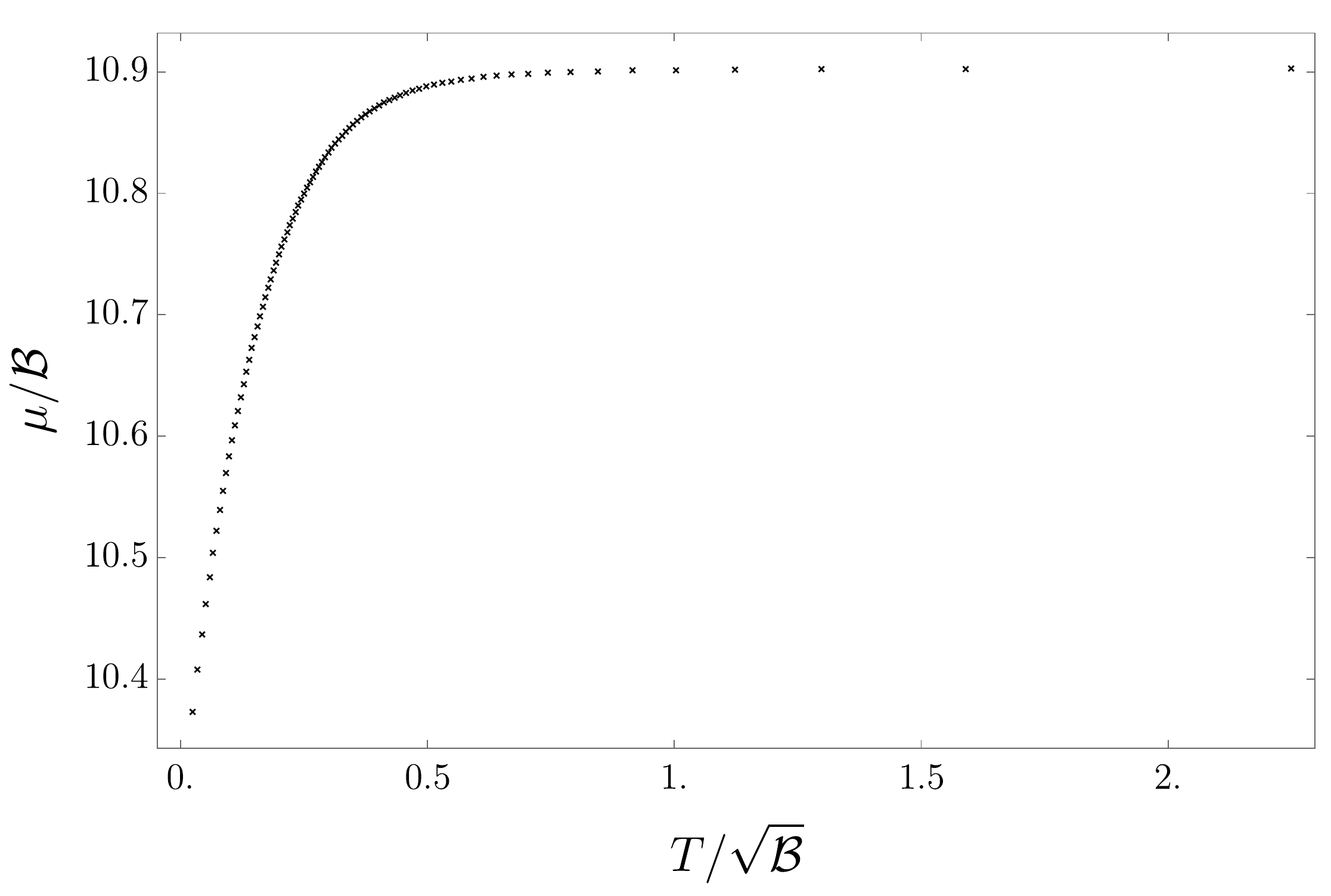}
\caption{Dimensionless energy density $\varepsilon/\CB^2$ (top-left), pressure
  $p/\CB^2$ (top-right), entropy density $s/\CB^{3/2}$ (bottom-left) and chemical potential $\mu/\CB$ (bottom-right), in units of $N_c^2 / (2\pi^2) $, plotted as a function of the dimensionless parameter $T/\sqrt{\CB}$. The first three plots use logarithmic scales on both axes.}
\label{fig:Thermo}
\end{figure}

Using the above relations, we can perform two consistency checks on our holographic setup and numerical calculations of the background. First, the value of the pressure computed from the stress-energy tensor component $\la T^{xx}\ra = p$ can be compared with the value of the Euclidean on-shell action, $p = -i (\beta V_3)^{-1}S_{\scriptscriptstyle on-shell}$, where $\beta = 1 / T$ and $V_3$ is the spatial volume of the theory. Secondly, we can compute $\varepsilon + p - \mu \rho$ from the stress-energy tensor evaluated near the boundary and by using the thermodynamic relation \eqref{ThermoRel1}, check whether its values agree with $s T$ computed purely from horizon quantities. Both calculations show consistency of our setup in that we find $\la T^{xx}\ra =  -i (\beta V_3)^{-1}S_{\scriptscriptstyle on-shell}$ and $\la T^{tt}\ra + \la T^{zz}\ra = sT$, within numerical precision.

We can now plot various thermodynamic quantities in a dimensionless manner by dividing them by appropriate powers of $\CB$. The natural dimensionless parameter with respect to which we present our numerical results is $T / \sqrt{\CB}$. The results for the energy density, pressure, entropy density and chemical potential are shown in Figure \ref{fig:Thermo}. The theory has two distinct regimes: the low- and the high-temperature regimes, or alternatively, the strong and weak magnetic field regimes, respectively. The high-temperature regime $T / \sqrt{\CB} \gg 1$ is one to which MHD has been historically applied and to which the formulation of MHD, which assumes a weak-field separation between fluid and charge degrees of freedom can be applied. The claim presented in the Ref. \cite{Grozdanov:2016tdf} is that within the dual formulation, however, MHD applies for all values of $T / \sqrt{\CB}$ provided that the state remains in the hydrodynamic regime. The profiles of the thermodynamic functions in Figure \ref{fig:Thermo} show a smooth crossover between the two regimes, which occurs around 
\begin{align}\label{Crossover}
T/\sqrt{\CB} \approx 0.5-0.7 \,.
\end{align} 
By using numerical fits, the equation of state in the two limits behaves as expected on dimensional grounds \cite{Grozdanov:2016tdf}. We present our numerical results in Table \ref{table:EOS}.

\begin{table}[tbh]
\begin{center}
    \begin{tabular}{| l | l | l |}
    \hline
     & weak field ($T / \sqrt{\CB} \gg 1$) & strong field ($T / \sqrt{\CB} \ll 1$) \\ \hline
    $\varepsilon~~$ & $\frac{N_c^2}{2\pi^2} \left( 74.1 \times T^4 \right)$  & $\frac{N_c^2}{2\pi^2}\left( 5.62\times \CB^2 \right)$  \\ \hline
    $p~~$ & $\frac{N_c^2}{2\pi^2} \left( 25.3 \times T^4 \right)$ & $\frac{N_c^2}{2\pi^2} \left( 5.32 \times \CB^2\right) $  \\ \hline
    $s~~$ & $\frac{N_c^2}{2\pi^2} \left( 99.4 \times T^3 \right)$ & $\frac{N_c^2}{2\pi^2}\left( 7.41 \times\CB\, T \right)$ \\ \hline
    $\mu~~$ & $\frac{N_c^2}{2\pi^2} \left( 10.9\times \CB \right)$& $\frac{N_c^2}{2\pi^2}\left( 2.88 \times \CB \right)$  \\ \hline
    \end{tabular}
\end{center}
\caption{Approximate asymptotic behaviour of the equation of state in weak- and strong-field limits for $\bar\alpha = 1/137$.}
\label{table:EOS}
\end{table}

In the limit of $\CB \to 0$, the weak-field result approximately limits to the equation of state of a strongly coupled, thermal $\CN = 4$ plasma, dual to a five dimensional AdS-Schwarzschild black brane with $p_{\scriptscriptstyle \CN = 4} = \frac{1}{8} N_c^2 \pi^2 T^4 $; i.e. $ \lim_{\CB\to 0 } p_{\scriptscriptstyle weak} \approx 1.28 \times N_c^2 T^4$ and  $ p_{\scriptscriptstyle \CN = 4} \approx 1.23 \times N_c^2 T^4$. We also note that the value of the pressure at low temperature strongly depends on the renormalised (re-scaled) fine structure constant $\bar \alpha$, which we set to $\bar\alpha = 1/137$.

\subsection{Transport coefficients}\label{transport-maintext}

Next, we compute the seven transport coefficients, $\eta_\perp$, $\eta_\parallel$, $r_\perp$, $r_\parallel$, $\zeta_\perp$, $\zeta_\parallel$ and $\zeta_\times$, by using the Kubo formulae derived in \cite{Grozdanov:2016tdf,Hernandez:2017mch} and reviewed in Appendix \ref{appendix:kubo}. The procedure only requires us to turn on time-dependent fluctuations of the background fields without any spatial dependence, $G_{ab} \to G_{ab} + \delta G_{ab}(t)$ and $B_{ab} \to B_{ab} + \delta B_{ab}(t)$. The perturbations asymptote to the boundary sources $\delta g^{(0)}_{\mu\nu}$ and $\delta b^{(0)}_{\mu\nu}$ of the dual stress-energy tensor and the two-form current. In the absence of spatial dependence, the fluctuations decouple into five separate channels, from which the seven transport coefficients are computed, with each channel containing one independent dynamical second-order equation. The sets of decoupled fluctuations responsible for their respective transport coefficients are 
\begin{equation}
\begin{aligned}
\eta_{\perp}  &: \quad \delta G_{xy} \,,\\
\eta_{\parallel}  &: \quad \delta G_{xz}, \,\delta B_{tx},\, \delta B_{xu} \,,\\
\zeta_\perp, \, \zeta_\parallel,\zeta_\times &: \quad \delta G_{tt}, \,\delta G_{xx},\,\delta G_{yy},\,\delta
G_{zz}, \,\delta B_{tz},\, \delta G_{tu},\, \delta B_{zu},\, \delta G_{uu} \,,\\
r_{\perp} &: \quad \delta B_{xz}, \,\delta G_{tx},\, \delta G_{xu} \,,\\
r_{\parallel}  &: \quad \delta B_{xy}  \,,  
\end{aligned}
\label{eqn:fluc-channel}
\end{equation}
with only one of the three bulk viscosities being independent. Each one of the transport coefficients can then be related to a membrane paradigm-type formula and can be expressed in terms of a simple expression. We summarise these relations here and discuss their derivation below:
\begin{equation}
\label{horizonFormulae}
\begin{aligned}
\eta_\perp &= \frac{N_c^2}{2\pi^2} \left( \frac{r_h^3}{4v\sqrt{w}}\right) = \frac{1}{4\pi} s \,,\\
\eta_\parallel &= \frac{N_c^2}{2\pi^2} \left( \frac{r_h^3}{4w^{3/2}} \right) = \frac{1}{4\pi} \frac{v}{w} s \,,\\
r_\perp &= \frac{2\pi^2}{N_c^2} \left( \frac{\sqrt{w}}{r_h} \right) \left(\frac{ \mathfrak{b}^{(-)}_{xz}(1)}{ \mathfrak{b}^{(-)}_{xz}(0)}\right)^2, \\
r_\parallel &=  \frac{2\pi^2}{N_c^2} \left(  \frac{v}{r_h\sqrt{w}} \right) ,\\
\zeta_\perp =\frac{1}{4}\zeta_\parallel = -\frac{1}{2}\zeta_\times  &= \frac{N_c^2}{2\pi^2} \left( \frac{r_h^3}{12 v\sqrt{w}} \left( \frac{6+B^2}{6-B^2}\right)^2
  \left[ \frac{ \mathfrak{Z}^{(-)}(1)}{ \mathfrak{Z}^{(-)}(0)} \right]^2 \right) ,
\end{aligned}
\end{equation}
where $\mathfrak{b}^{(-)}$ and $\mathfrak{Z}^{(-)}$ are the time-independent solutions of the fluctuations $\delta B_{xz}$ and $Z_s = \delta G^x_{~x} + \delta G^y_{~y} - (2\CV'/\CW') \delta G^z_{~z}$, respectively. The arguments denote that the functions are evaluated either at the horizon, $u=1$, or the boundary, $u=0$. Note that the value at the boundary is set by the Dirichlet boundary conditions.

What we see is that the ratio of the transverse shear viscosity (w.r.t. the background magnetic field) to entropy density is universal, resulting in $\eta_\perp/s = 1/4\pi$. Furthermore, the expressions for $\eta_\parallel$ and $r_\parallel$ only depend on the background quantities $v$ and $w$, while $\zeta_\perp$, $\zeta_\parallel$ and $r_\perp$ also depend on the fluctuations of the fields.\footnote{For a holographic derivation of bulk viscosity in neutral relativistic hydrodynamics, see \cite{Gubser:2008sz}.}

In order to derive the horizon formulae, we use the Wronskian method (see e.g. \cite{Davison:2015taa}). Here, we will only explicitly show the derivation of the transverse resistivity $r_\perp$. The other formulae from Eq. \eqref{horizonFormulae} are derived in Appendix \ref{appendix:transport}. First, we combine the equations of motion for the relevant fluctuations, $\delta B_{xz}$, $\delta G_{tx}$, and $\delta G_{xu}$, into a single second-order differential equation by eliminating the metric fluctuations,
\begin{equation}\label{eomForRperp}
\delta B_{xz}''+\left(\frac{3}{2u}+ \frac{F'}{F} -\CW'  \right)\delta B_{xz}' +
\left(\frac{\omega^2}{4r_h^2 u^3F^2} -\frac{B^2 e^{-4\CV}}{u^3F} \right)\delta B_{xz} = 0 \,.
\end{equation}
Since we are only computing first-order transport coefficients, it is sufficient to solve Eq. \eqref{eomForRperp} to linear order in $\omega$. To find the solution, we assume that there exists a time-independent solution $\mathfrak{b}_{xz}^{(-)}(u)$, which asymptotes to a constant both at the boundary and the horizon. At the boundary, this asymptotic value is related to the source of the two-form background gauge field, i.e. $\mathfrak{b}_{xz}^{(-)}(u\to 0) = \delta B_{xz}^{(0)}$. The time-dependent information is contained in the second solution, linearly independent from $\mathfrak{b}^{(-)}_{xz}$. We refer to this solution as $\mathfrak{b}_{xz}^{(+)}$. It can be expressed as an integral over the Wronskian $W_R$ of \eqref{eomForRperp}:
\begin{align}
\mathfrak{b}_{xz}^{(+)}(u) = \mathfrak{b}_{xz}^{(-)}(u) \int^1_u du'
\frac{ W_R (u') }{ \left( \mathfrak{b}_{xz}^{(-)}(u')\right)^2} \,,
\end{align}
where
\begin{align}
W_R (u) = \exp\left[ -\int_u^1 du' \left(   \frac{3}{2u'}+\frac{F' (u')}{F(u')} - \CW' (u')\right) \right] = \frac{1}{u^{3/2}F e^{-\CW}} \, .
\end{align}
The near-boundary and the near-horizon expansions of $\mathfrak{b}_{xz}^{(+)}$ are
\begin{equation}
\mathfrak{b}_{xz}^{(+)} = \begin{dcases}
\sqrt{w}\left[\mathfrak{b}_{xz}^{(-)}(0)\right]^{-1}\ln u +\CO(\sqrt{u}) \,,& ~~~~\text{for}\;\; u\approx 0 \, ,\\
-r_h\left[2\pi T\mathfrak{b}_{xz}^{(-)}(1)\right]^{-1}\ln (1-u) +\CO(1-u) \,,& ~~~~\text{for}\;\; u\approx 1 \,. 
\end{dcases}
\label{eqn:asymptbxz}
\end{equation}
Finally, $\delta B_{xz}(\omega,u)$ is then the following linear combination of the two solutions: 
\begin{equation}
\delta B_{xz}(\omega,u) = \mathfrak{b}_{xz}^{(-)}(u) + \alpha(\omega)\mathfrak{b}_{xz}^{(+)}(u)+\CO(\omega^2) \,.
\end{equation} 

The coefficient $\alpha(\omega)$ can be determined by imposing a regular ingoing boundary condition at the horizon, which corresponds to computing a retarded dual correlator \cite{Son:2002sd,Herzog:2002pc}:
\begin{align}
\delta B_{xz} (u) = (1- u)^{- \frac{i\omega}{4\pi T}} \tilde B_{xz}\,.
\end{align}
The function $\tilde B_{xz}(u)$ is regular at the horizon. This choice of the boundary condition implies that near the horizon, $\delta B_{xz}$ behaves as
\begin{equation}\label{ingoingBxz}
\delta B_{xz}(u) = \mathfrak{b}_{xz}^{(-)}(u) + \alpha(\omega)
\mathfrak{b}_{xz}^{(+)}(u)  + \ldots = \mathfrak{b}_{xz}^{(-)}(1) \left(1-\frac{i\omega}{4\pi
    T}\ln(1-u) \right) +\ldots \,.
\end{equation}
Comparing Eq. \eqref{ingoingBxz} with the asymptotic behaviour of $\mathfrak{b}_{xz}^{(+)}$ in \eqref{eqn:asymptbxz}, we find $\alpha = (i\omega/2 r_h)\left[\mathfrak{b}_{xz}^{(-)}(1)\right]^2$. Thus, the near-boundary expression for $\delta B_{xz}$ becomes
\begin{equation}
\delta B_{xz}(u) = \mathfrak{b}_{xz}^{(-)}(0) \left(1+\frac{i\omega}{2r_h}\sqrt{w}
  \left[ \frac{ \mathfrak{b}_{xz}^{(-)}(1)}{ \mathfrak{b}_{xz}^{(-)}(0)} \right]^2\ln u\right) +\CO(u) \,.
\end{equation}
By substituting this expression into the expectation value \eqref{defJ} of the two-form current $\la J^{\mu\nu}\ra$, we obtain 
\begin{equation}
\la \delta J^{xz}\ra = \lim_{u\to 0}\left( 2u^{3/2}\sqrt{F}\delta B_{xz}'(u) \right)= \frac{2\pi^2}{N_c^2}\left( 2i\omega r_h^{-1}\sqrt{w} \left[ \frac{ \mathfrak{b}_{xz}^{(-)}(1)}{ \mathfrak{b}_{xz}^{(-)}(0) }\right]^2 \right)\delta b_{xz} + \CO(\omega^2) \, .
\end{equation}
The expression on the right-hand-side of the second equation is obtained by using the relation between $\delta B_{xz}^{(0)}$, $\delta B_{xz}^{(1)}$ and $\delta b_{xz}$ in \eqref{eq:B0toSourceb}. Note that the dependence on the electromagnetic coupling enters into the one-point function at order $\omega^2$ and, thus, $\bar\alpha$ plays no role in the holographic formula for the resistivity; $r_\perp$ and other first-order transport coefficients are independent of the renormalised electromagnetic coupling. Finally, using the Kubo formula for $r_\perp$, which is derived and presented in Eq. \eqref{practicalKubo} of Appendix \ref{appendix:kubo}, we recover the expression presented in Eq. \eqref{horizonFormulae}. All of the six remaining transport coefficients can be obtained by following the same procedure. We refer the reader to Appendix \ref{appendix:transport} for their detailed derivations.  

\begin{figure}[tbh]
\center
\includegraphics[width=0.482\textwidth]{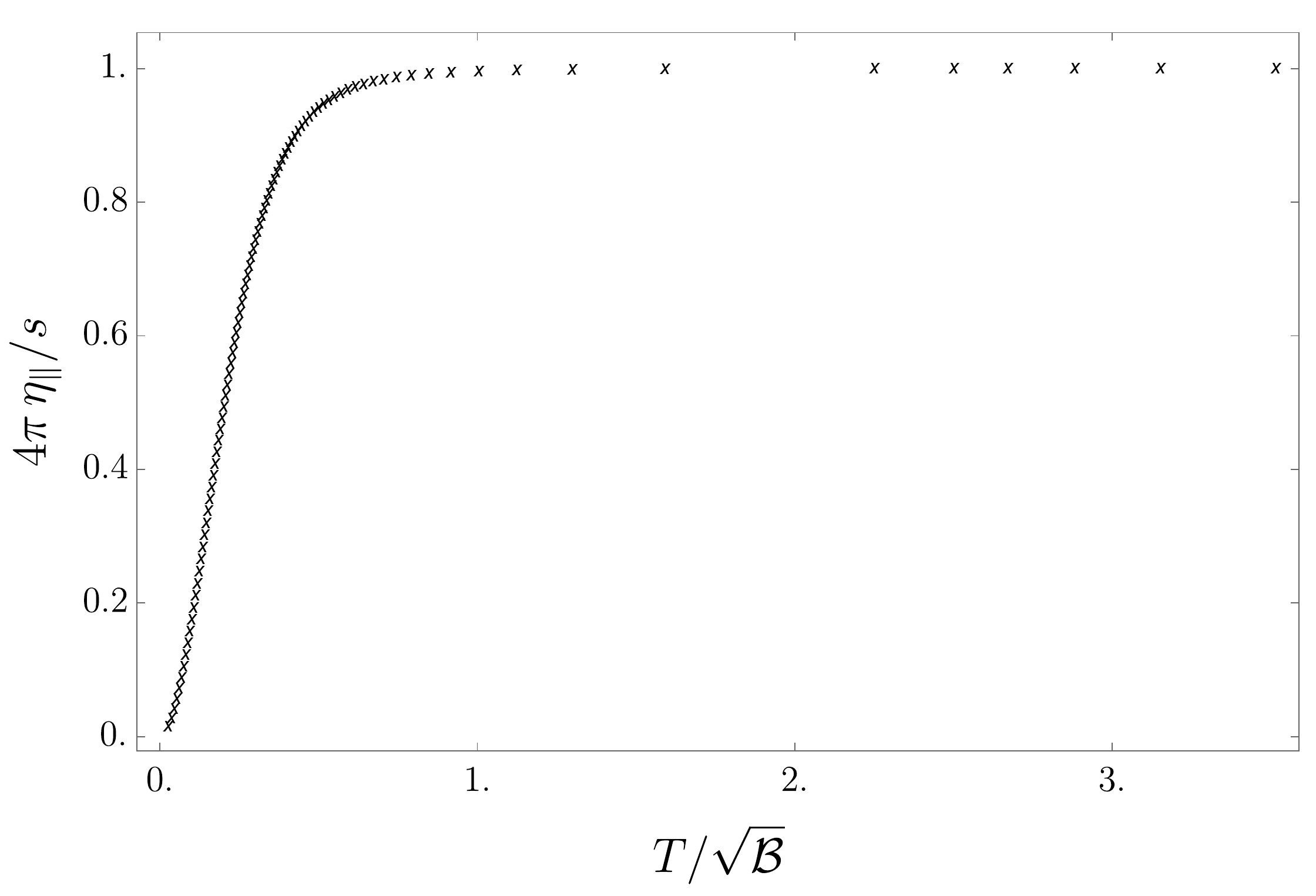}
\includegraphics[width=0.482\textwidth]{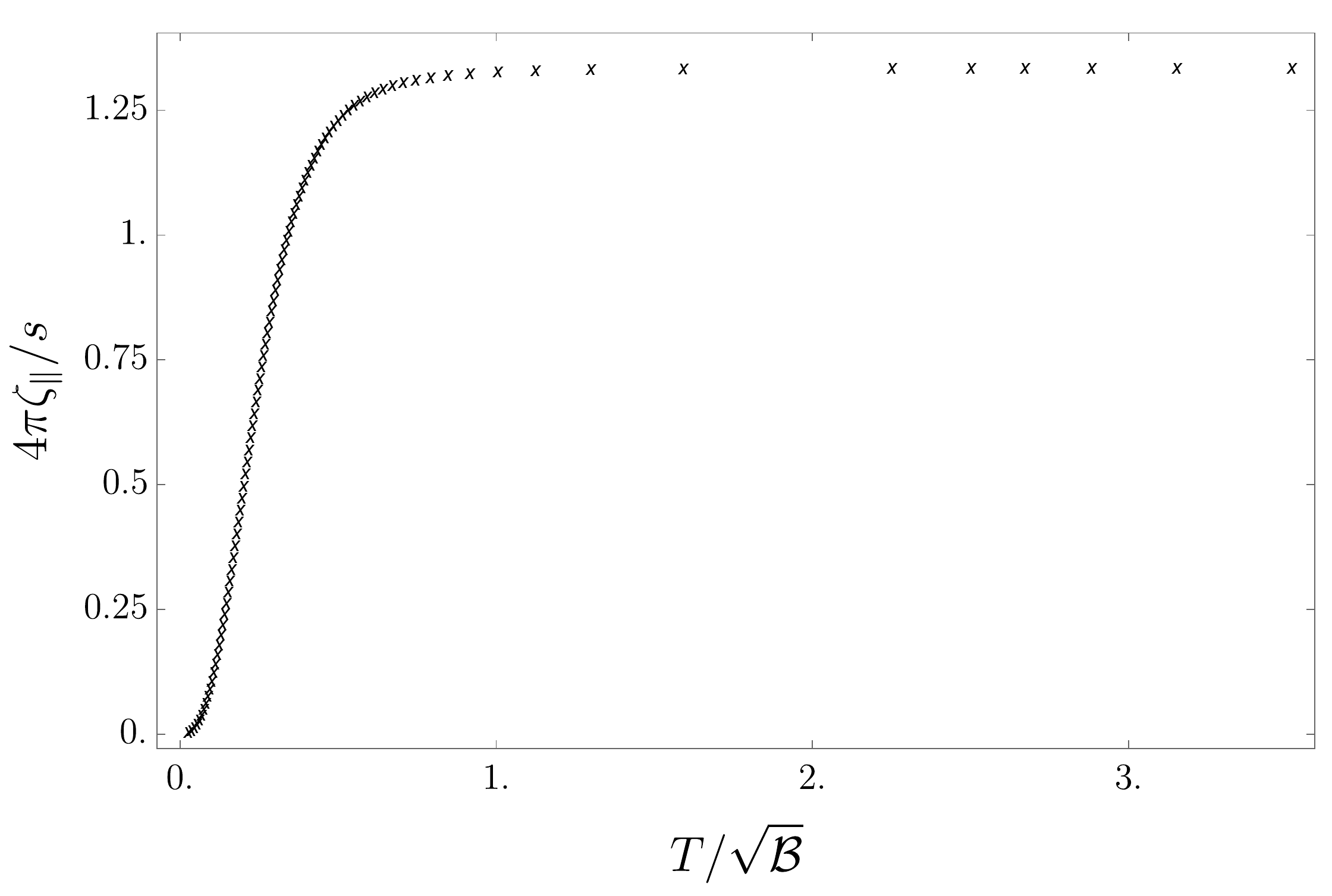}
\includegraphics[width=0.482\textwidth]{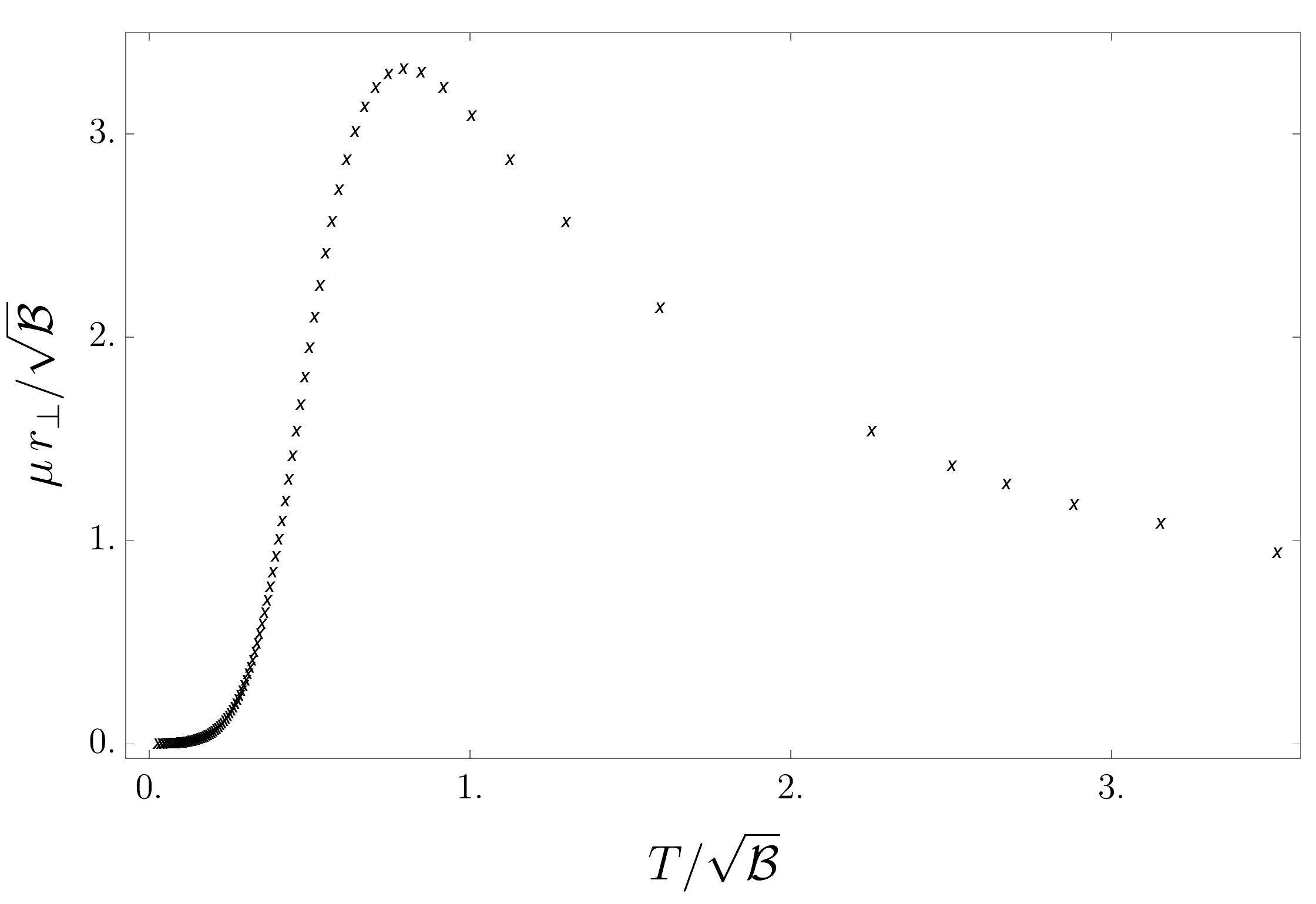}
\includegraphics[width=0.482\textwidth]{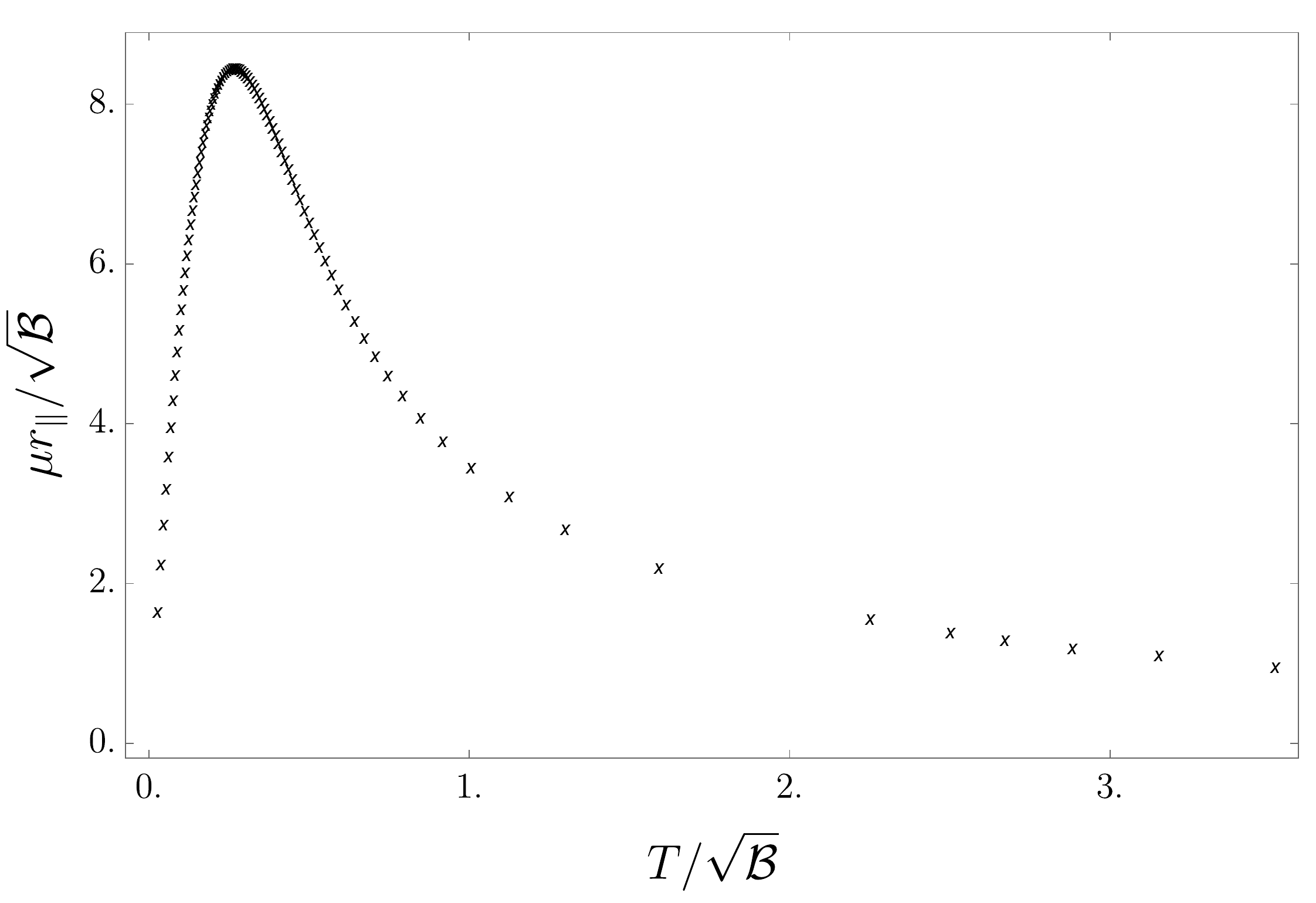}
\caption{The plots of (dimensionless) first-order transport coefficients as a function of $T / \sqrt{\CB}$.}
\label{fig:Transport}
\end{figure}

The plots of the (dimensionless) transport coefficients $\eta_{\parallel}$, $\zeta_{\parallel}$, $r_{\perp}$ and $r_\parallel$ as a function of $T/\sqrt{\CB}$ are presented in Figure \ref{fig:Transport}. The remaining three viscosities can easily be inferred from Eq. \eqref{horizonFormulae}. In particular, $\eta_{\perp} / s = 1 / (4\pi)$, $\zeta_\perp = \zeta_\parallel / 4$ and $\zeta_\times = - \zeta_\parallel / 2$. We note that all transport coefficients satisfy the positive entropy production bounds discussed in Section \ref{sec:Intro}. It is interesting that the bulk viscosity inequality $\zeta_\perp \zeta_\parallel \geq \zeta_\times^2$ is saturated, i.e. $\zeta_\perp \zeta_\parallel = \zeta_\times^2$ in the plasma studied here for all parameters of the theory.

We can now investigate the behaviour of the transport coefficients in the two extreme limits of $T / \sqrt{\CB} \to 0$ and $T/ \sqrt{\CB} \to \infty$, i.e. the strong- and the weak-field regimes, respectively. The leading-order power-law scalings (which we assume) and the coefficients follow from numerical fits. The results are presented in Table \ref{table:TC}.

\begin{table}[tbh]
\begin{center}
    \begin{tabular}{| l | l | l |}
    \hline
     & weak field ($T / \sqrt{\CB} \gg 1$) & strong field ($T / \sqrt{\CB} \ll 1$) \\ \hline
    $\eta_\perp~~$ & $\frac{s}{4\pi}$  &  $\frac{s}{4\pi}$ \\ \hline
    $\eta_\parallel~~$ & $1.00\times \frac{s}{4\pi}$ & $\frac{s}{4\pi} \left( 21.32\times \frac{T^2}{\CB} \right)$  \\ \hline
    $\zeta_\perp~~$ & $0.33\times \frac{s}{4\pi}$ &$\frac{s}{4\pi} \left( 16.34\times  \frac{T^3}{\CB^{3/2}} \right)  $   \\ \hline
    $\zeta_\parallel~~$ & $1.33\times \frac{s}{4\pi}$ & $\frac{s}{4\pi} \left( 65.37\times  \frac{T^3}{\CB^{3/2}} \right)  $\\ \hline
    $\zeta_\times~~$ &$-0.66 \times \frac{s}{4\pi}$ & $ - \frac{s}{4\pi} \left( 32.69 \times  \frac{T^3}{\CB^{3/2}} \right)  $  \\ \hline
    $r_\perp~~$ &  $\frac{\CB}{\mu}  \left( 3.37\times  \frac{1}{T} \right)$ & $\frac{\sqrt{\CB}}{\mu}\left(4.7\times \frac{T^3}{\CB^{3/2}} \right) $ \\ \hline
    $r_\parallel~~$ & $\frac{\CB}{\mu} \left( 3.37\times  \frac{1}{T}\right) $& $ \frac{\sqrt{\CB}}{\mu} \left(  62.3\times \frac{T}{\sqrt{\CB}}\right) $ \\  \hline
    \end{tabular}
\end{center}
\caption{Approximate asymptotic behaviour of all first-order transport coefficients in weak- and strong-field limits. The temperature-dependent scaling of the shear viscosities at low temperature agrees with what was reported in Ref. \cite{Critelli:2014kra}.}
\label{table:TC}
\end{table}

Since the entropy density $s$ vanishes in the limit of zero temperature, all first-order transport coefficients vanish in the strong-field limit of $T\to0$. Furthermore, as we will see, all (first-order) dissipative effects also vanish in the $T\to 0$ limit. These observations are consistent with predictions of \cite{Grozdanov:2016tdf} based on symmetry arguments.

In the regime of a weak magnetic field, $T\gg \sqrt{\CB}$, we find that both shear viscosities $\eta_\perp$ and $\eta_\parallel$ converge to $\eta_\perp = \eta_\parallel = s/(4\pi)$ as $\CB/T^2 \to 0$. On the other hand, the longitudinal bulk viscosity limits to $\zeta_\parallel \to 4 \eta / 3$, which is consistent with the fact that as $\CB/T^2 \to 0$, the evolution of the plasma should be governed by uncharged relativistic conformal hydrodynamics (see e.g. \cite{Kovtun:2012rj} or Appendix \ref{appendix:transport}). Indeed, both resistivities, $r_\perp$ and $r_\parallel$, also tend to zero in the limit. 

We also note that the weak-field behaviour of $r_\perp$ and $r_\parallel$ is consistent with the assumption used to construct standard (ideal) MHD, whereby conductivity is taken to infinity, $\sigma \approx 1 / r \to \infty$, and whereby one adds corrections proportional to $1/\sigma$.\footnote{See Ref. \cite{Grozdanov:2016tdf} for a discussion regarding the subtleties in relating resistivities to conductivities.} In other words, small weak-field resistivities are compatible with the assumption of ideal Ohm's law, which gives rise to Eq. \eqref{IdealOhm} (see also our discussion around this equation in Section \ref{sec:Intro}.). Furthermore, note that in standard MHD, only one resistivity (conductivity) is typically added to include dissipative corrections. What we see is that in our theory, the two resistivities take similar values in the weak-field limit in which standard MHD applies. However, in the strong-field limit, they assume drastically different values, including a different scaling with $T/\sqrt{\CB}$. This observation therefore further points to the important role of anisotropic effects in MHD \cite{Grozdanov:2016tdf} and the necessity for using the formulation of \cite{Grozdanov:2016tdf,Hernandez:2017mch} as one moves from the weak- to the strong-field regime.

The fact that $r_\perp$ and $r_\parallel$ tend to zero both in the limits of $T / \sqrt{\CB} \to 0$ and $T \sqrt{\CB} \to \infty$, along with the positivity of the entropy production bounds $r_\perp \geq 0$ and $r_\parallel \geq 0$ \cite{Grozdanov:2016tdf}, implies that there always exists a maximum value of the resistivities at some intermediate $T / \sqrt{\CB}$. It would be interesting to find the sizes of these maxima in experimentally realisable systems and probe the regimes of the ``least conductive" plasmas. Finally, it would be interesting to further investigate the connection between maximal $r$ and various discussions of lower bounds on conductivities, e.g. \cite{PhysRevLett.111.125004,Grozdanov:2015qia,Lucas:2017ggp}.

\section{Magnetohydrodynamic waves in a strongly coupled plasma}\label{sec:MHDWaves}

We are now ready to use the information obtained from the holographic analysis of Section \ref{sec:Holography} to study dissipative dispersion relations of magnetohydrodynamic waves in a toy model of a strongly coupled plasma. We will use the theory of MHD \cite{Grozdanov:2016tdf}, which is a phenomenological effective theory, and supplement it with microscopic details---the equation of state and transport coefficients---of the holographic setup investigated above. We will be particularly interested in the dependence of the MHD modes on the angle between momentum and magnetic field, as well as the ratio between temperature and the strength of the magnetic field. The 't Hooft coupling of interactions in the matter sector is not tuneable in our model, however, the electromagnetic coupling is. In all sections, except in Section \ref{sec:alphaDependence}, it will be set to $\alpha =  2 \pi^2 / 137 N_c^2$. 

Before presenting the numerical results, we review the relevant facts about MHD modes. For a detailed derivation of these results, see Ref. \cite{Grozdanov:2016tdf} and for a discussion of the general procedure, see Refs. \cite{Kadanoff,Kovtun:2012rj}. First, we write the hydrodynamic variables $u^\mu$, $h^\mu$, $T$ and $\mu$ in terms of oscillating modes perturbed around their near-equilibrium values, e.g. $u^\mu \to (1,0,0,0) + \delta u^\mu \, e^{-i \omega t + i k x \sin\theta + i k z \cos\theta}$, so that $\theta \in [0, \pi/2]$ measures the angle between the equilibrium magnetic field pointing in the $z$-direction and the wave momentum $k$ in the $x$--$z$ plane. The dispersion relations $\omega(k)$ are then derived from the equations of MHD, i.e. Eqs. \eqref{EOM1} and \eqref{EOM2}, with the external $H_{\mu\nu\rho} = 0$. The solutions depend on the angle $\theta$, temperature $T$ and the strength of the magnetic field (or the chemical potential of the magnetic flux number density), parametrised in our solutions by $\CB$. Any dimensionless quantity will only depend on the single dimensionless ratio $T / \sqrt{\CB}$. The resulting modes can be decomposed into two channels---odd and even under the reflection of $y \to -y$. The first channel is the transverse Alfv\'{e}n channel. The second is the magnetosonic channel with two branches of solutions: slow and fast magnetosonic waves.  

The linearised MHD equations of motion \eqref{EOM1} and \eqref{EOM2} need to be expanded in the hydrodynamic regime in powers of small $\omega / \Lambda_h \ll 1 $ and $k/ \Lambda_h \ll 1$, where $\Lambda_h$ is the UV cut-off of the effective theory. In standard MHD, where $T \gg \sqrt{\CB}$, then $\Lambda_h \approx T$, whereas in the strong-field regime of $T \ll \sqrt{\CB}$, the cut-off can be set by the magnetic field, then $\Lambda_h \approx \sqrt{\CB}$. As argued in \cite{Grozdanov:2016tdf}, hydrodynamics may exist all the way to $T \to 0$, even when $\delta T = 0$. Such an expansion, performed to some order, gives rise to a polynomial equation in $\omega$ and $k$. For example, in the Alfv\'{e}n channel, within first-order dissipative MHD,
\begin{equation}\label{AlfvenDispersion}
\begin{aligned}
-\omega^2 + \left(\frac{\mu\rho \cos^2\theta }{\varepsilon+p}\right)k^2 - i \left[ \left( \frac{\mu r_\perp}{\rho} + \frac{\eta_\parallel}{\varepsilon+p} \right)\cos^2\theta + \left( \frac{\mu r_\parallel}{\rho} + \frac{\eta_\perp}{\varepsilon+p} \right)\sin^2\theta  \right] \omega k^2  & \\
+ \frac{\mu }{2 \rho (\varepsilon+p)}\left( r_\perp \cos^2 \theta + 2 r_\parallel \sin^2\theta \right)\left( \eta_\perp \sin^2\theta + \eta_\parallel \cos^2\theta \right)k^4 &= 0 \, .
\end{aligned}
\end{equation}
The two solutions of the quadratic equation for $\omega$ are given by 
\begin{align}\label{AlfvenSolFull}
\omega = -\frac{i}{2}(\CD_{A,+})k^2 \pm \frac{k}{2}\sqrt{\CV_A^2 \cos^2\theta - (\CD_{A,-})^2 k^2} \,,
\end{align}
where $\CD_{A,+}$ and $\CD_{A,-}$ are 
\begin{equation}
\begin{aligned}
\CD_{A,\pm} &= \left( \frac{\mu r_\perp}{\rho} \pm \frac{\eta_\parallel}{\varepsilon+p} \right)\cos^2\theta + \left( \frac{\mu r_\parallel}{\rho} \pm \frac{\eta_\perp}{\varepsilon+p} \right)\sin^2\theta \,.
\end{aligned}
\end{equation}
One can now series expand $\omega(k) = \CD_0 k + \CD_1 k^2$, or alternatively, plug this ansatz in Eq. \eqref{AlfvenDispersion} and solve it order-by-oder in $k$. What we find is the Alfv\'{e}n wave dispersion relation \cite{Grozdanov:2016tdf}:
\begin{equation}\label{AlfvenDispersion2}
\omega = \pm \CV_A k\cos\theta  - \frac{i}{2} \left(  \frac{1}{\varepsilon+p}\left( \eta_\perp\sin^2\theta  + \eta_\parallel \cos^2\theta \right)+ \frac{\mu}{\rho} \left(r_\perp \cos^2\theta + r_\parallel \sin^2\theta \right)\right)k^2 \, ,
\end{equation}
where the speed is given by $\CV_A^2 = \mu\rho/(\varepsilon+p)$. The dispersion relation appears to be well-defined for any angle $\theta \in [0, \pi/2]$ between momentum and equilibrium magnetic field. In particular, if we were to take the $\theta \to \pi/2$ limit, \eqref{AlfvenDispersion2} would yield two diffusive modes, both with dispersion relation
\begin{align}\label{AlfDiffK0First}
\omega = - \frac{i}{2} \left( \frac{\eta_\perp}{\varepsilon + p } + \frac{\mu r_\parallel}{\rho} \right) k^2\, ,
\end{align}
which are, however, unphysical and only result from an incorrect order of limits of $k$ and $\theta$.  

\begin{figure}[tbh]
\center
\includegraphics[width=.49\textwidth]{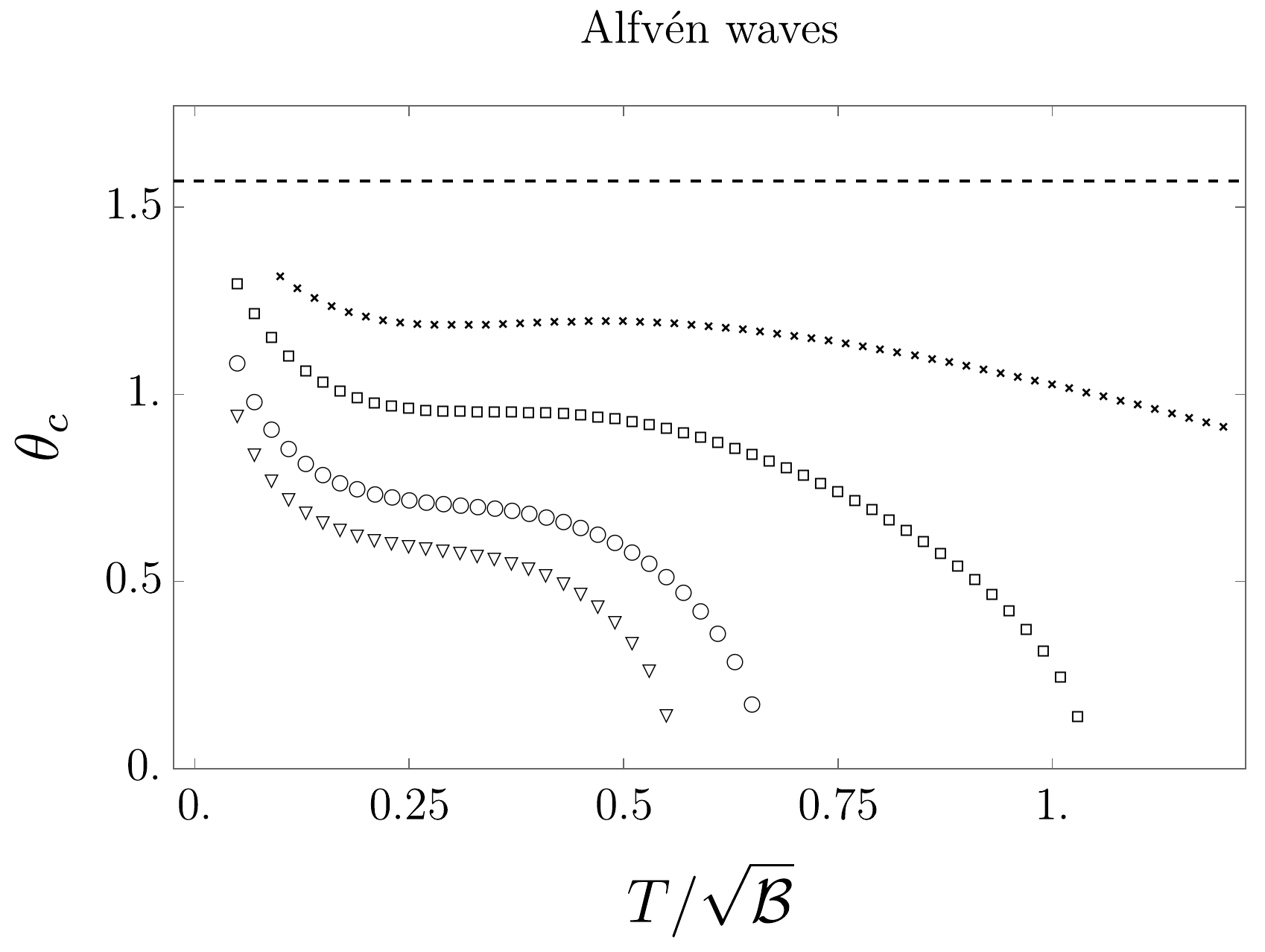}
\includegraphics[width=.49\textwidth]{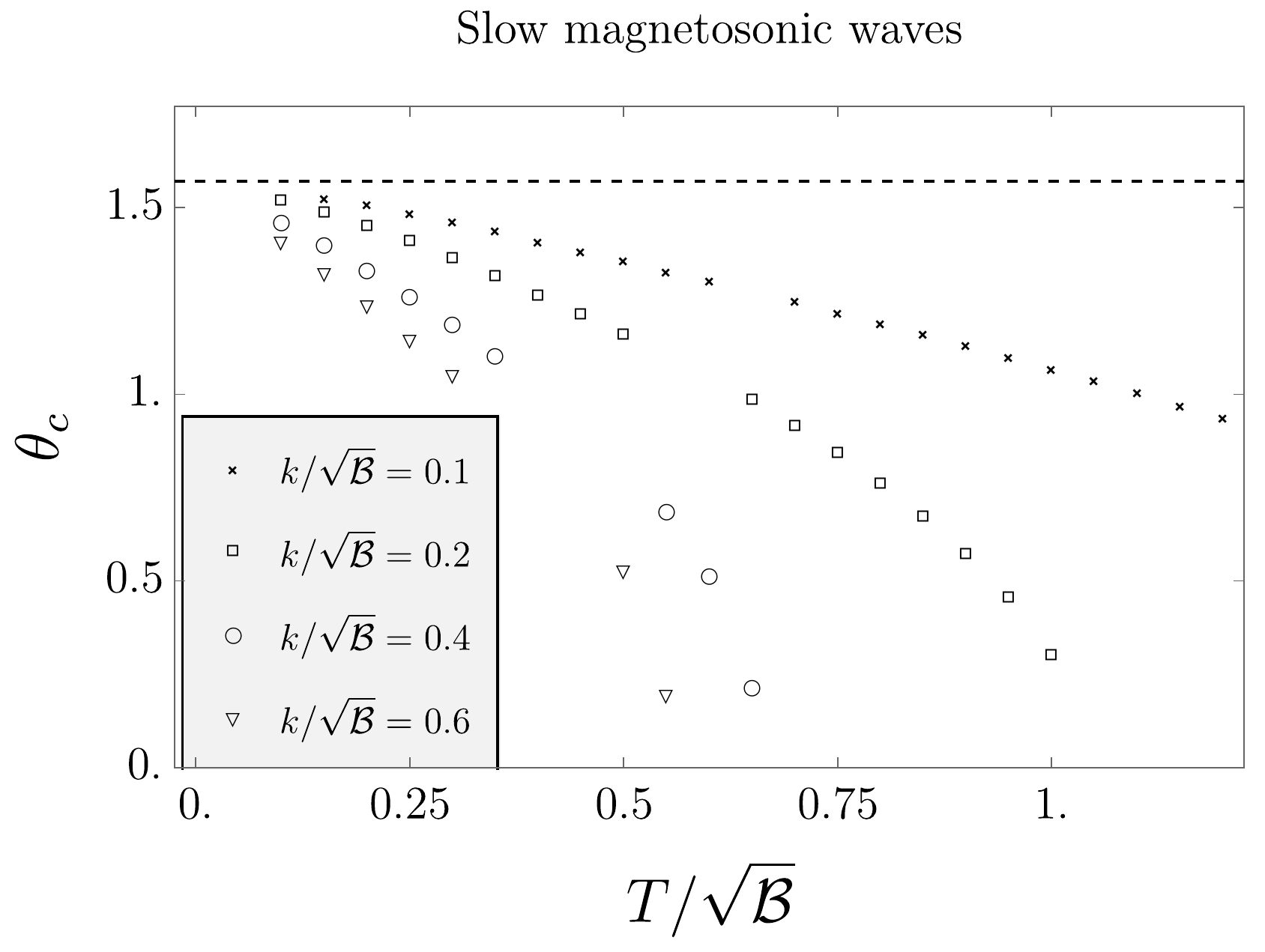}
\caption{The critical angle $\theta_c$ for Alfv\'en waves (left) and slow magnetosonic waves (right), plotted as a function of $T/\sqrt{\CB}$ for $k/\sqrt{\CB} = \{0.1, \, 0.2, \,0.4, \, 0.6 \}$. The dashed line at the top of both sub-figures indicates the value of $\theta_c = \pi/2$.}
\label{fig:CriticalAngleAlfven}
\end{figure}

As can be seen from the structure of the square-root in Eq. \eqref{AlfvenSolFull}, the expansion in small $k$ is only sensible so long as $k^2 \ll \CV_A^2 \cos^2\theta / (\CD_{A,-})^2$. Hence, even for a small finite $k$, this expansion is inapplicable for angles $\theta$ near $\theta = \pi / 2$ where $\cos \theta$ becomes very small. In fact, for 
\begin{equation}\label{eq:criticalThetaAlfven}
\CV_A^2 \cos^2\theta \leq (\CD_{A,-})^2k^2 \,,
\end{equation}
the propagating modes cease to exist altogether and the two modes become purely imaginary (diffusive to $\CO(k^2)$). The transmutation of two propagating Alfv\'{e}n modes into two non-propagating modes occurs when the inequality in \eqref{eq:criticalThetaAlfven} is saturated, i.e. at the critical angle $\theta_c$ when $\text{Re}[\omega] = 0$:
\begin{align}\label{ThetaC}
\frac{ \cos (\theta_c)}{\CD_{A,-}(\theta_c)} =  \frac{k}{\CV_A} .
\end{align}
In other words, the plasma exhibits propagating (sound) modes for $0 \leq \theta < \theta_c$ and non-propagating (diffusive) modes for $\theta_c < \theta \leq \pi / 2$. We plot the dependence of the critical angle $\theta_c$ on $k/\sqrt{\CB}$ and $T/\sqrt{\CB}$ for the Alfv\'{e}n waves in our model in Figure \ref{fig:CriticalAngleAlfven}. What we see is that for small $k / \sqrt{\CB}$ and small $T / \sqrt{\CB}$, the transition to diffusive modes occurs closer to $\theta_c \approx \pi / 2$. For any fixed and finite $T / \sqrt{\CB}$, Eq. \eqref{ThetaC} indeed implies that $\theta_c \to \pi / 2$ as $k\to 0$.

We note that as already pointed out in \cite{Hernandez:2017mch}, the limits of $k\to0$ and $\theta \to \pi /2$ do not commute and we obtain different results depending on which expansion ($k \approx 0$ or $\theta \approx \pi/2$) is performed first.  If one first takes the limit $\theta \to \pi /2$, then Eq. \eqref{AlfvenDispersion} becomes
\begin{equation}\label{AlfvenDispersionOpt2}
-\omega^2 - i  \left( \frac{\mu r_\parallel}{\rho} + \frac{\eta_\perp}{\varepsilon+p} \right) \omega k^2  + \frac{ \mu r_\parallel \eta_\perp }{\rho(\varepsilon+p)} k^4 = 0 \, ,
\end{equation}
which instead of Eq. \eqref{AlfDiffK0First} results in two non-degenerate diffusive modes
\begin{align}
\omega = - i \frac{\eta_\perp}{\varepsilon+p} k^2\,, && \omega = - i \frac{\mu r_\parallel}{\rho} k^2 \, .
\end{align}
The dispersion relation \eqref{AlfvenDispersion2} is therefore only sensible at a finite $T/\sqrt{\CB}$ and infinitesimally small $k / \Lambda_h$.

In the magnetosonic channel, the story is entirely analogous to the one described for the Alfv\'{e}n waves. By expanding around $k\approx 0$ first, we obtain the dispersion relation of \cite{Grozdanov:2016tdf}:  
\begin{align}
\omega = \pm v_M k - i \tau k^2 \, ,
\end{align}
where the speed of magnetosonic wave is given by 
\begin{equation}\label{speedvM}
v_M^2 = \frac{1}{2}\left\{ (\CV_A^2+ \CV_0^2)\cos^2\theta + \CV_S^2\sin^2\theta \pm \sqrt{[(\CV_A^2-\CV_0^2)\cos^2\theta + \CV_S^2\sin^2\theta]^2+ 4\CV^4 \cos^2\theta \sin^2\theta }\right\}.
\end{equation}
The functions $\CV_A$, $\CV_0$, $\CV_S$ and $\CV$ appearing in \eqref{speedvM} are 
\begin{align}
\CV_A^2 &= \frac{\mu\rho}{\varepsilon+p}, & \CV_0^2 &= \frac{s}{T\chi_{11}}, \nn
 \CV_S^2 &= \frac{(s-\rho\chi_{12})(s+\rho\chi_{21})+\rho^2\chi_{11}\chi_{22}}{(\varepsilon+p)\chi_{11}}, & \CV^4 &= \frac{s(s-\rho\chi_{12})(s+\rho\chi_{21})}{T(\varepsilon+p)\chi_{11}^2}.
\end{align}
The susceptibilities are\footnote{Note that these susceptibilities are different from the ones used in \cite{Grozdanov:2016tdf}, where independent thermodynamic quantities were $T$ and $\mu$, not $T$ and $\rho$. For this reason we also use different notation.}
\begin{equation}\label{susceptibilities}
\chi_{11} = \left( \frac{\partial s}{\partial T}\right)_\rho,\quad \chi_{12} = \left( \frac{\partial s}{\partial \rho}\right)_T,\quad \chi_{21} = \left( \frac{\partial \mu}{\partial T}\right)_\rho,\quad \chi_{22} = \left( \frac{\partial \mu}{\partial \rho}\right)_T \,.
\end{equation}
The two types of magnetosonic waves, corresponding to $\pm$ solutions in \eqref{speedvM}, are known as the fast (with $+$) and the slow (with $-$) magnetosonic waves. We refer the reader to Appendix \ref{appendix:magnetosonicSpectrum} for further details regarding the derivation of the magnetosonic modes. Each pair of the propagating slow magnetosonic modes also splits, in analogy with the Alfv\'{e}n waves, into two non-propagating diffusive modes for $\theta \geq \theta_c$. The critical angle $\theta_c$ for magnetosonic modes is also defined as in the Alfv\'{e}n channel: the angle at which $\text{Re}[\omega] = 0$. We plot the numerically-computed dependence of the magnetosonic $\theta_c$ on $k/\sqrt{\CB}$ and $T/\sqrt{\CB}$ in Fig. \ref{fig:CriticalAngleAlfven}. As can be seen from the plot, the critical angles for the two types of waves are independent. However, they show similar qualitative dependence on the parameters that characterise the waves. 

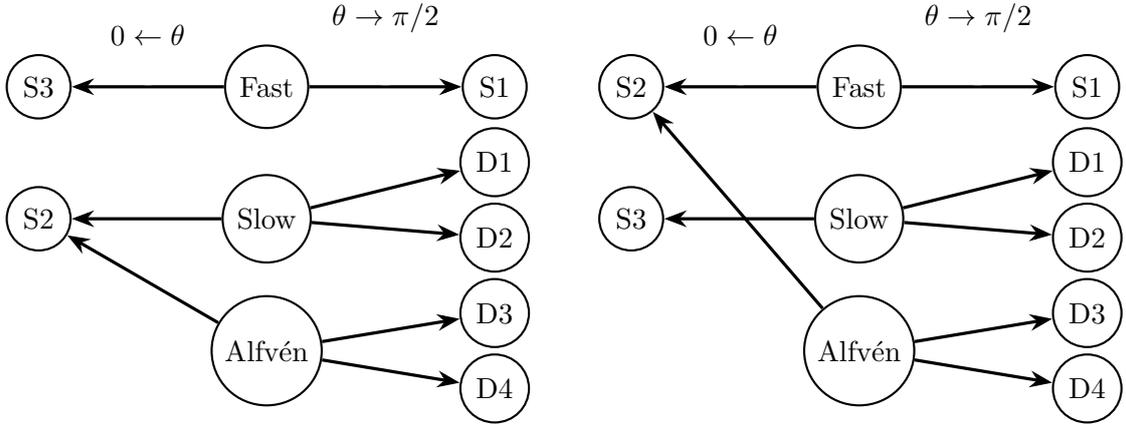
\begin{figure}[tbh]
\centering
\begin{tikzpicture}
\begin{scope}[every node/.style={circle,thick,draw}]
    \node (A) at (0,3.5) {Fast};
    \node (B) at (0,1.75) {Slow};
    \node (C) at (0,0) {Alfv\'en};
    \node (D) at (3,3.5) {S1} ;
    \node (E) at (3,2.5) {D$1$} ;
    \node (F) at (3,1.5) {D$2$} ;
    \node (Ea) at (3,0.5) {D$3$} ;
    \node (Fa) at (3,-0.5) {D$4$} ;
    \node (A-1) at (-3,3.5) {S3};
    \node (B-1) at (-3,1.75) {S2};
\end{scope}

\begin{scope}[>={Stealth[black]},
              every node/.style={fill=white,circle},
              every edge/.style={draw=black,very thick}]
    \path [->] (A) edge node [above] {$\theta \rightarrow \pi/2 $} (D);
    \path [->] (B) edge  (E);
    \path [->] (B) edge  (F);
    \path [->] (C) edge  (Ea);
    \path [->] (C) edge  (Fa);
    \path [->] (A) edge node [above] {$0 \leftarrow \theta $} (A-1);
    \path [->] (B) edge  (B-1);
    \path [->] (C) edge  (B-1);
\end{scope}
\end{tikzpicture}
\qquad
\begin{tikzpicture}
\begin{scope}[every node/.style={circle,thick,draw}]
    \node (A) at (0,3.5) {Fast};
    \node (B) at (0,1.75) {Slow};
    \node (C) at (0,0) {Alfv\'en};
    \node (D) at (3,3.5) {S1} ;
    \node (E) at (3,2.5) {D$1$} ;
    \node (F) at (3,1.5) {D$2$} ;
    \node (Ea) at (3,0.5) {D$3$} ;
    \node (Fa) at (3,-0.5) {D$4$} ;
    \node (A-1) at (-3,3.5) {S2};
    \node (B-1) at (-3,1.75) {S3};
\end{scope}

\begin{scope}[>={Stealth[black]},
              every node/.style={fill=white,circle},
              every edge/.style={draw=black,very thick}]
    \path [->] (A) edge node [above] {$\theta \rightarrow \pi/2 $} (D);
    \path [->] (B) edge  (E);
    \path [->] (B) edge  (F);
    \path [->] (C) edge  (Ea);
    \path [->] (C) edge  (Fa);
    \path [->] (A) edge node [above] {$ 0 \leftarrow \theta$} (A-1);
    \path [->] (B) edge  (B-1);
    \path [->] (C) edge  (A-1);
\end{scope}
\end{tikzpicture}
\vspace{0.5cm}
\caption{Diagrams depicting the $\theta$-dependent pattern of transmutation from sound to diffusive modes for Alfv\'{e}n waves and slow and fast magnetosonic waves. The left and right diagrams correspond to weak- and strong-field regimes. The relevant dispersion relation are stated in Eqs. \eqref{specialSound} and \eqref{specialDiffuse}.}
\label{fig:SummaryHighT}
\end{figure}

We summarise the $\theta$-dependent characteristics of MHD modes in Fig. \ref{fig:SummaryHighT}. We observe the pattern of a transmutation of sound modes into diffusion to be different in the weak- and strong-field regimes. Namely, the two magnetosonic waves interchange their dispersion relations at small $\theta$. Since the complicated expressions for dispersion relations greatly simplify at $\theta = 0$ and $\theta = \pi/2$, we state them below. The sound mode dispersion relations, denoted by S, are
\begin{equation}\label{specialSound}
\begin{aligned}
\text{S1} &:\quad \omega = \pm \CV_S k - \frac{i}{2} \Bigg\{ \frac{\zeta_\perp + \eta_\perp}{\varepsilon+p} \\
&\quad\quad + \frac{r_\perp \left[ (s-\rho\chi_{12})(\mu-T\chi_{21}) - \rho T \chi_{11}\chi_{22}\right]\left[ (s+\rho\chi_{21})(\mu+T\chi_{12}) - \rho T \chi_{11}\chi_{22}\right]}{T^2\chi_{11}\left[(s-\rho\chi_{12})(s+\rho\chi_{21}) + \rho^2\chi_{11}\chi_{22}  \right]}\Bigg\} k^2 \,,\\
\text{S2} &:\quad \omega =\pm \CV_A k - \frac{i}{2}\left( \frac{\eta_\parallel}{\varepsilon+p} +\frac{\mu r_\perp}{\rho}\right) k^2 \, ,\\
\text{S3} &:\quad \omega =\pm \CV_0 k - \frac{i}{2} \frac{\zeta_\parallel}{sT} k^2\,,\\
\end{aligned}
\end{equation}
and the diffusive modes, denoted by D, are
\begin{equation}\label{specialDiffuse}
\begin{aligned}
\text{D1} &:\qquad \omega = -i\frac{\eta_\parallel}{sT}k^2\,,\\
\text{D2} &:\qquad \omega = -\frac{ir_\perp (\varepsilon+p)^2\chi_{22}}{T^2\left[ (s-\rho\chi_{12})(s+\rho\chi_{21})+\rho^2\chi_{11}\chi_{22} \right]}k^2\,,\\
\text{D3} &:\qquad \omega = -i\frac{\eta_\perp}{\varepsilon+p}k^2\,,  \\
\text{D4} &:\qquad \omega = -i\frac{r_\parallel \mu}{\rho}k^2 \,.\\
\end{aligned}
\end{equation}

In the regime of a large $T/\sqrt{\CB}$, the results agree with those of \cite{Hernandez:2017mch}. Furthermore, using the asymptotic form of the thermodynamics quantities and transport coefficients in the $T/\sqrt{\CB}\to \infty$ limit, one can show that these modes reduce to sound and diffusive modes of uncharged relativistic hydrodynamics.

In the strong-field regime, which cannot be described within standard MHD, the speeds of S1 and S3 become large and approach the speed of light in the limit of $T\to0$. It is clear that in the strong-field regime, MHD sound waves can easily violate any causal upper bound on the speed of sound \cite{Cherman:2009tw,Cherman:2009kf,Hohler:2009tv,Hoyos:2016cob}. Furthermore, as discussed above, all diffusion constants vanish and the system becomes controlled by second-order MHD \cite{Grozdanov:2016tdf}, which we do not investigate in this work. All details regarding angle-dependent wave propagation are presented in Section \ref{Thetadep}.

\subsection{Speeds and attenuations of MHD waves}\label{SpeedsAndAtt}

\begin{figure}[tbh]
\center
\includegraphics[width=0.37\textwidth]{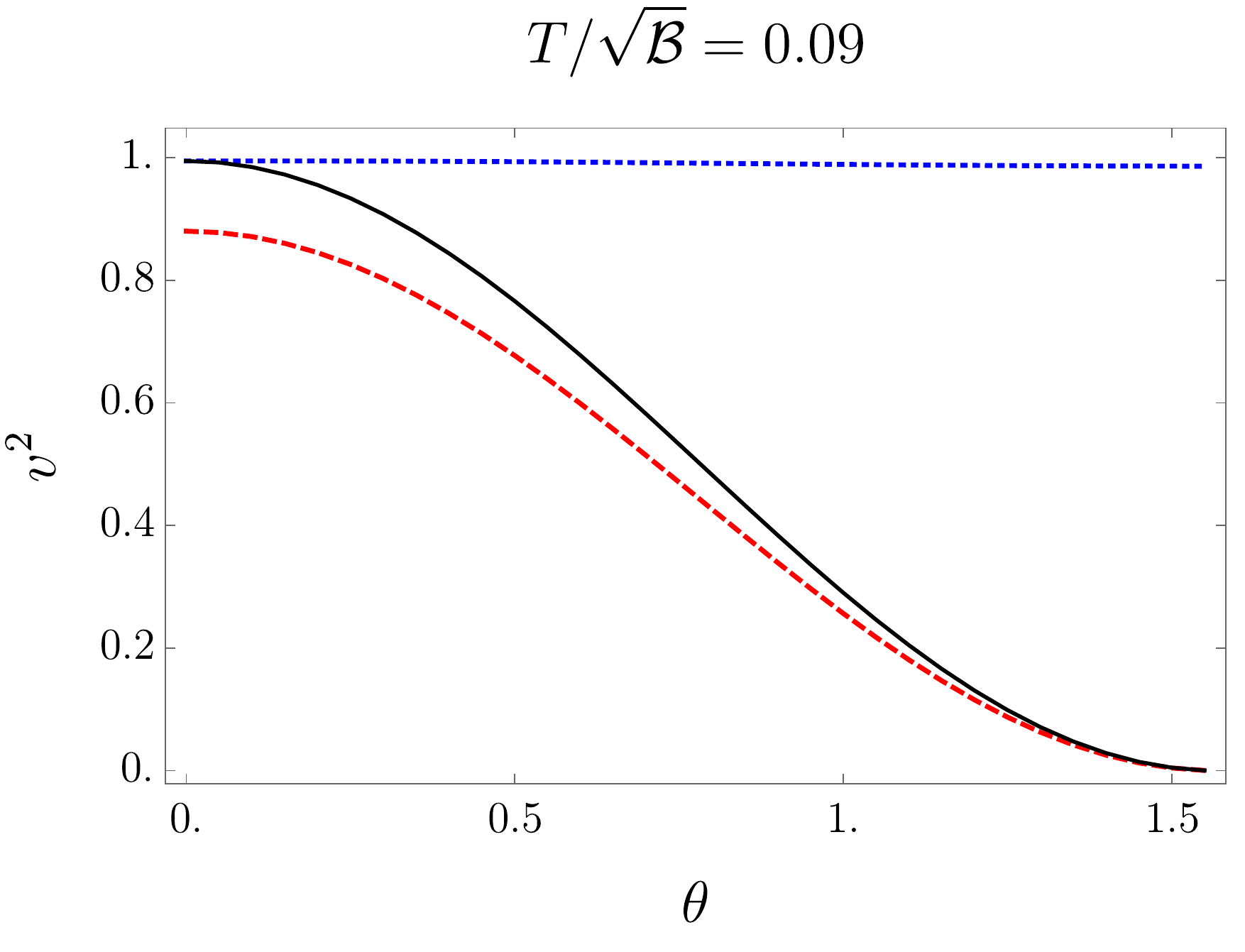}
\hspace{2cm}
\includegraphics[width=0.37\textwidth]{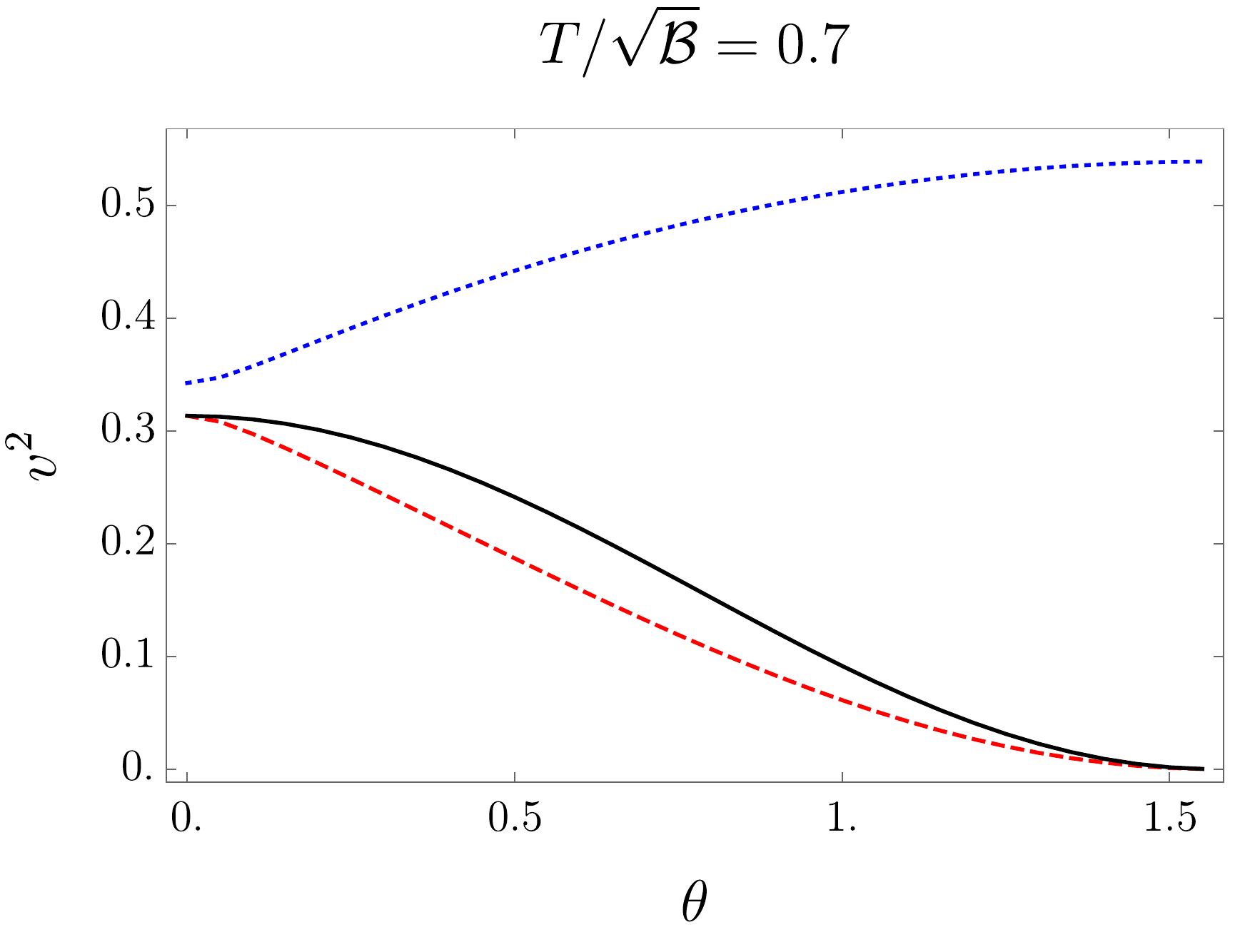}
\includegraphics[width=0.37\textwidth]{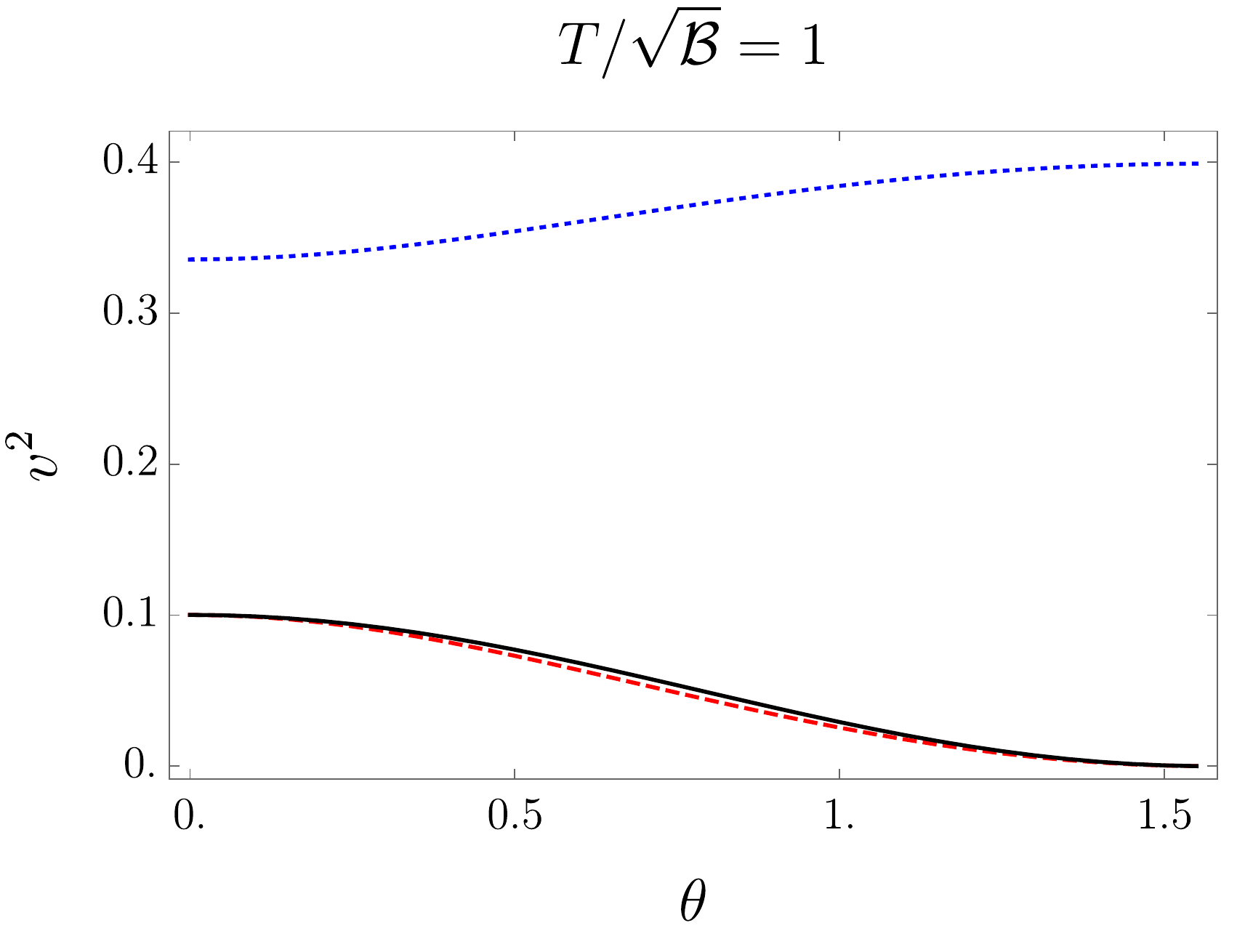}
\caption{Angular dependence of the speeds of Alfv\'{e}n (black, solid), fast (blue, dotted) and slow (red, dashed) magnetosonic waves in the strong-field, the crossover and the weak-field regimes.
}
\label{fig:3Sound}
\end{figure}

\begin{figure}[tbh]
\center
\includegraphics[width=0.32\textwidth]{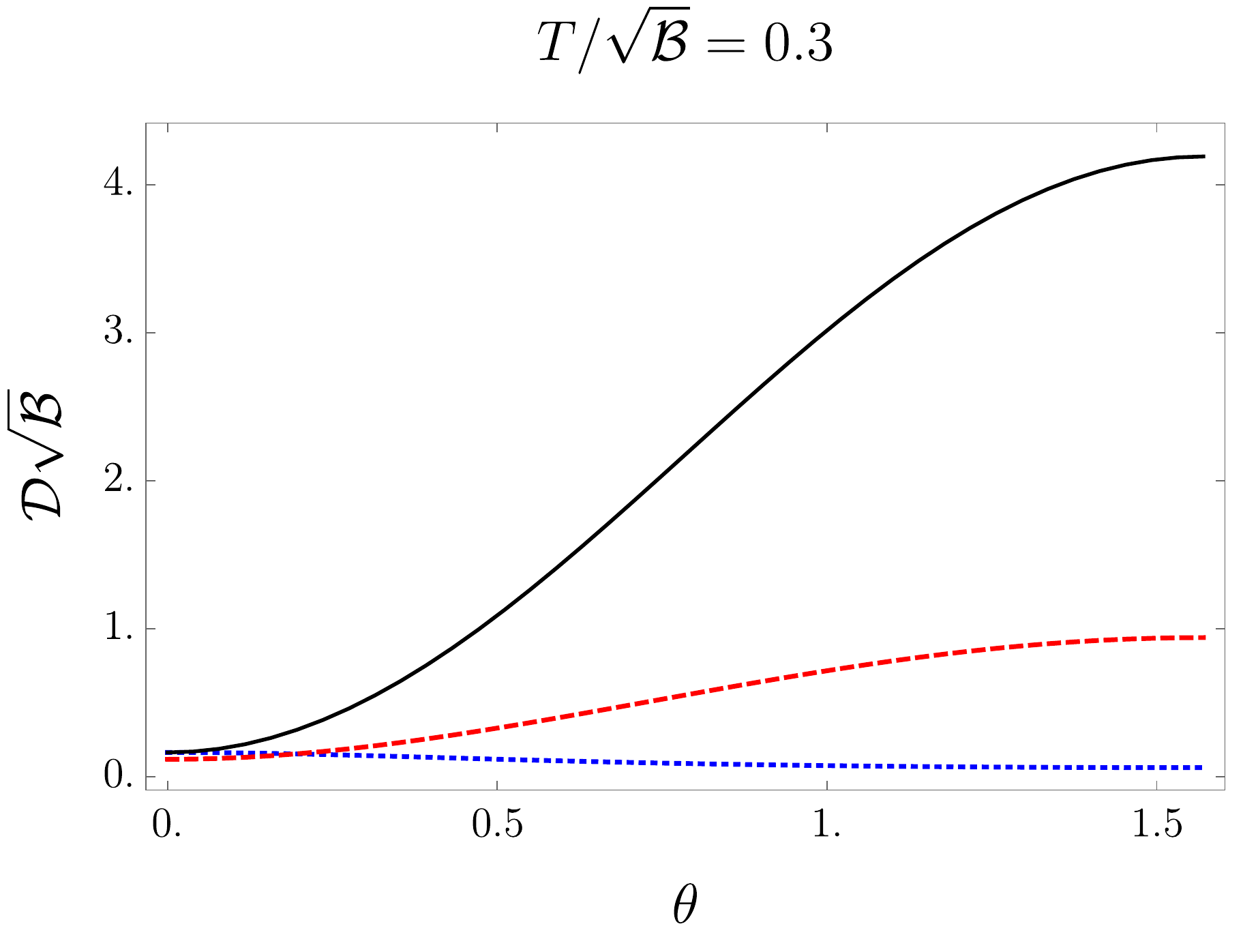}
\includegraphics[width=0.32\textwidth]{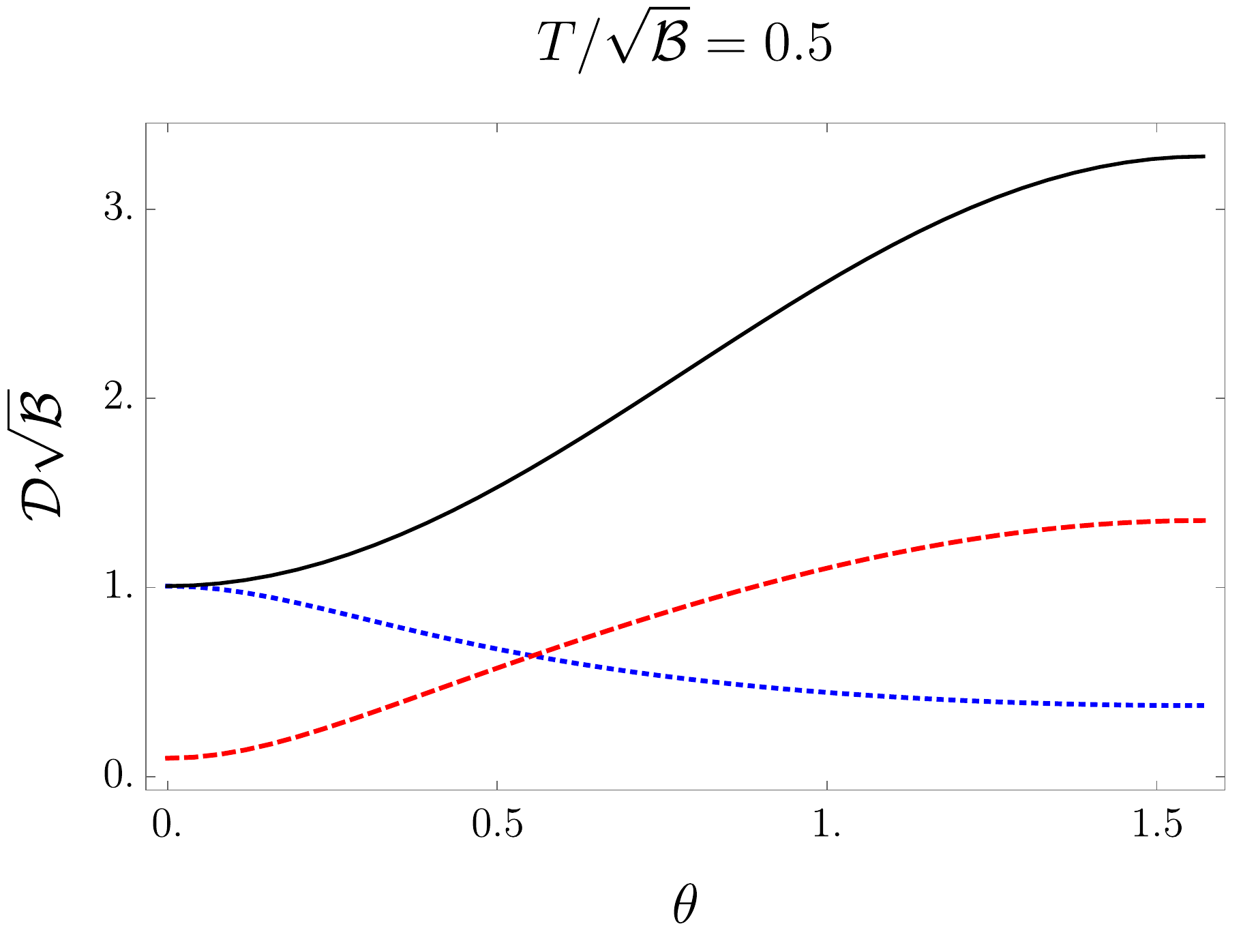}
\includegraphics[width=0.32\textwidth]{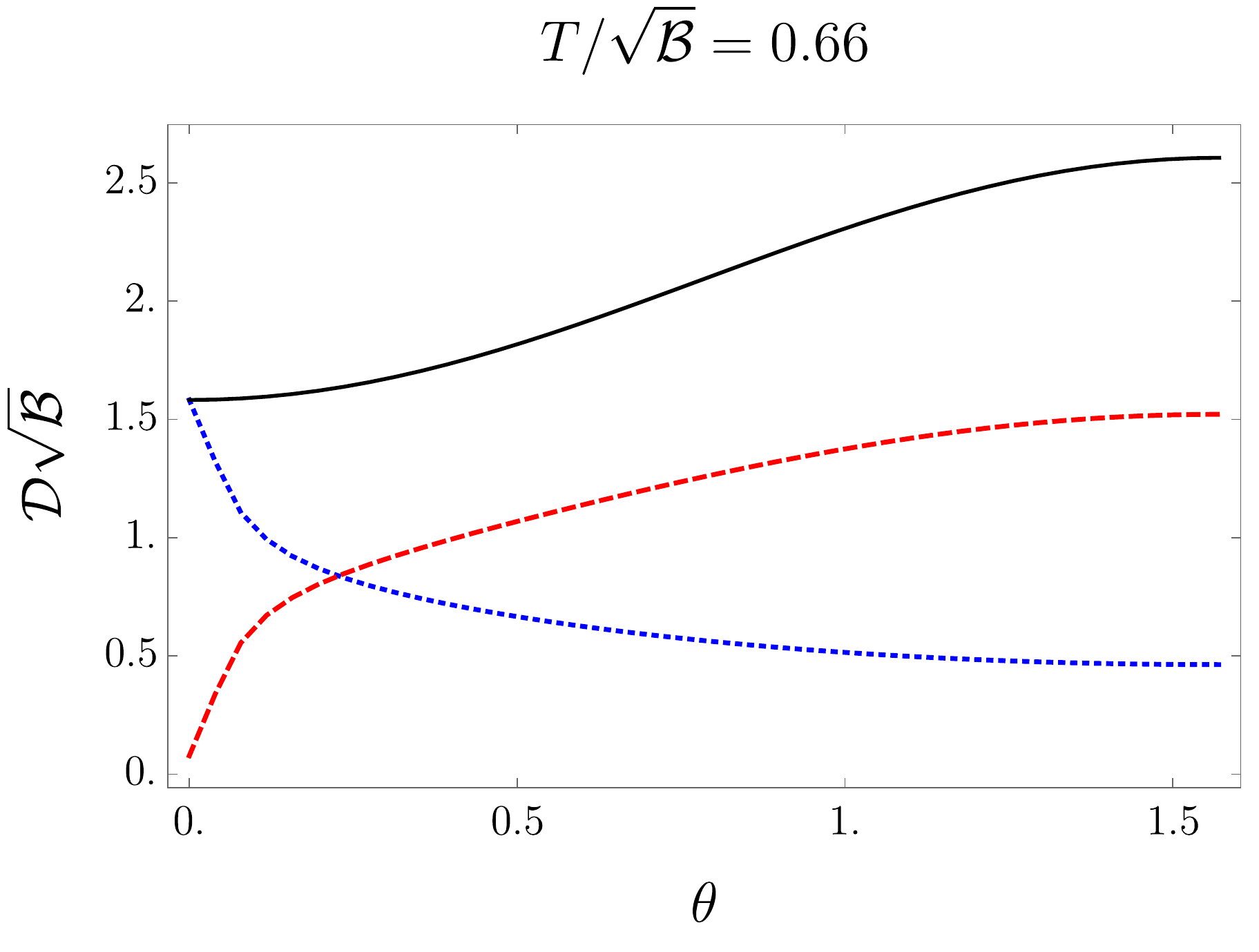}
\includegraphics[width=0.32\textwidth]{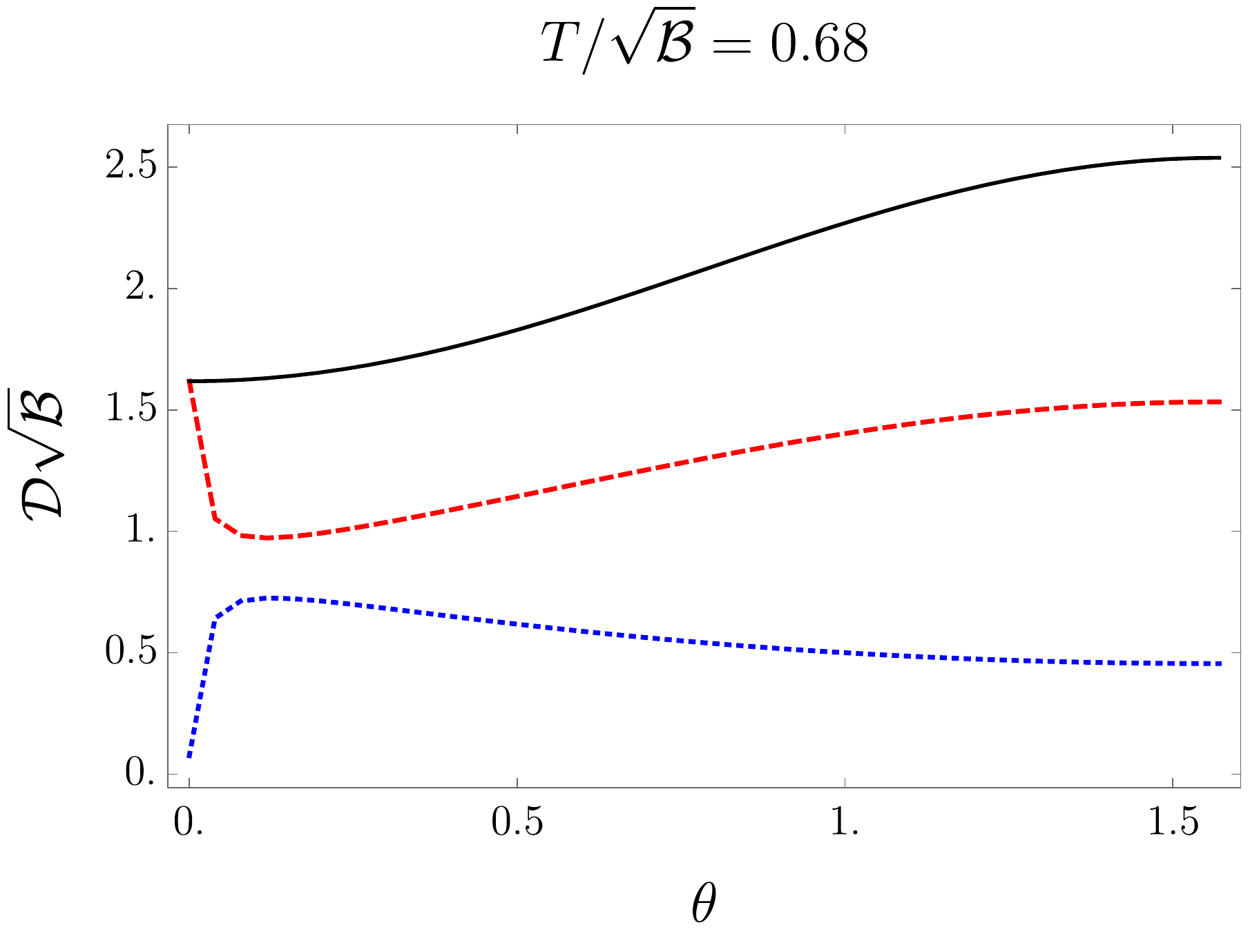}
\includegraphics[width=0.32\textwidth]{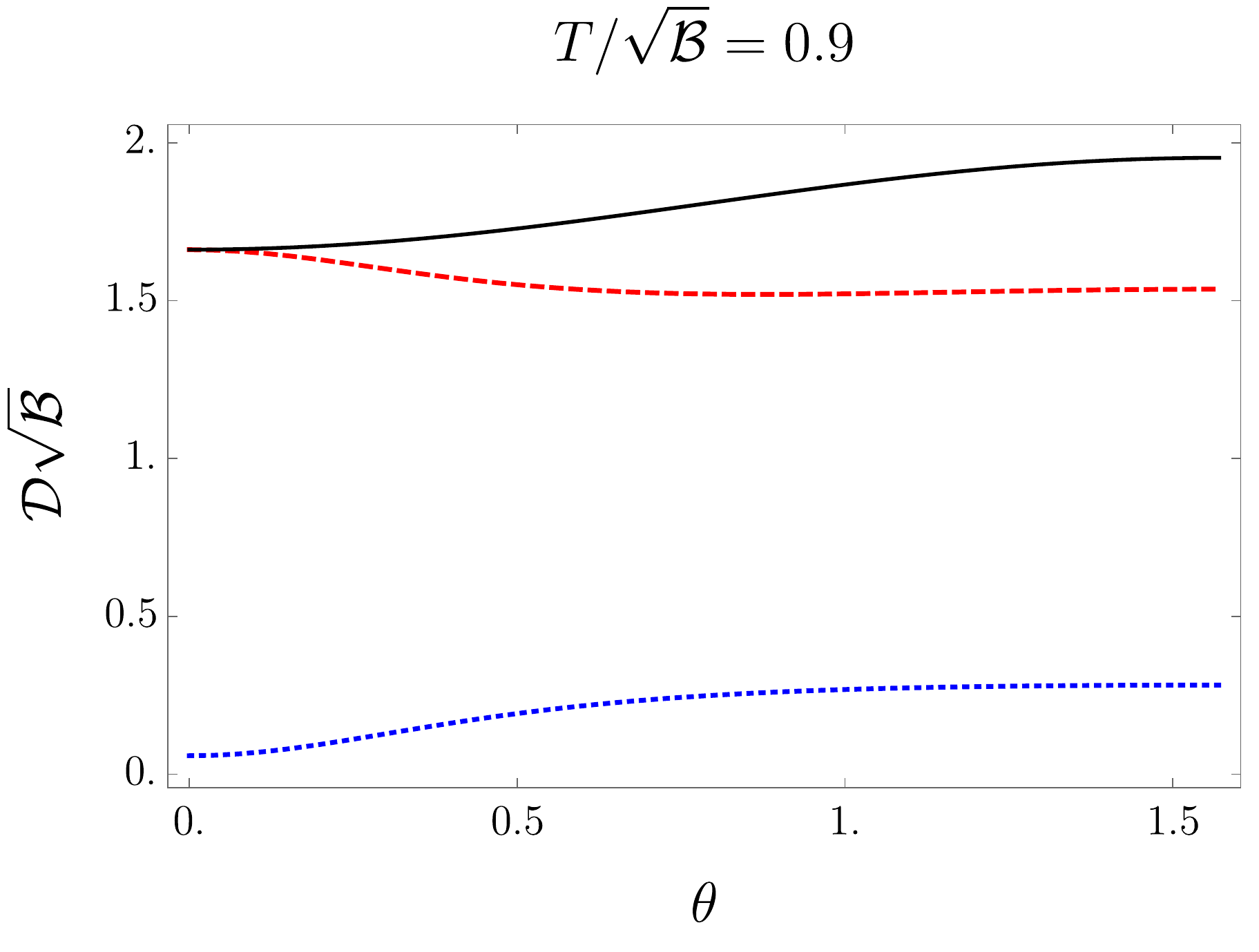}
\includegraphics[width=0.32\textwidth]{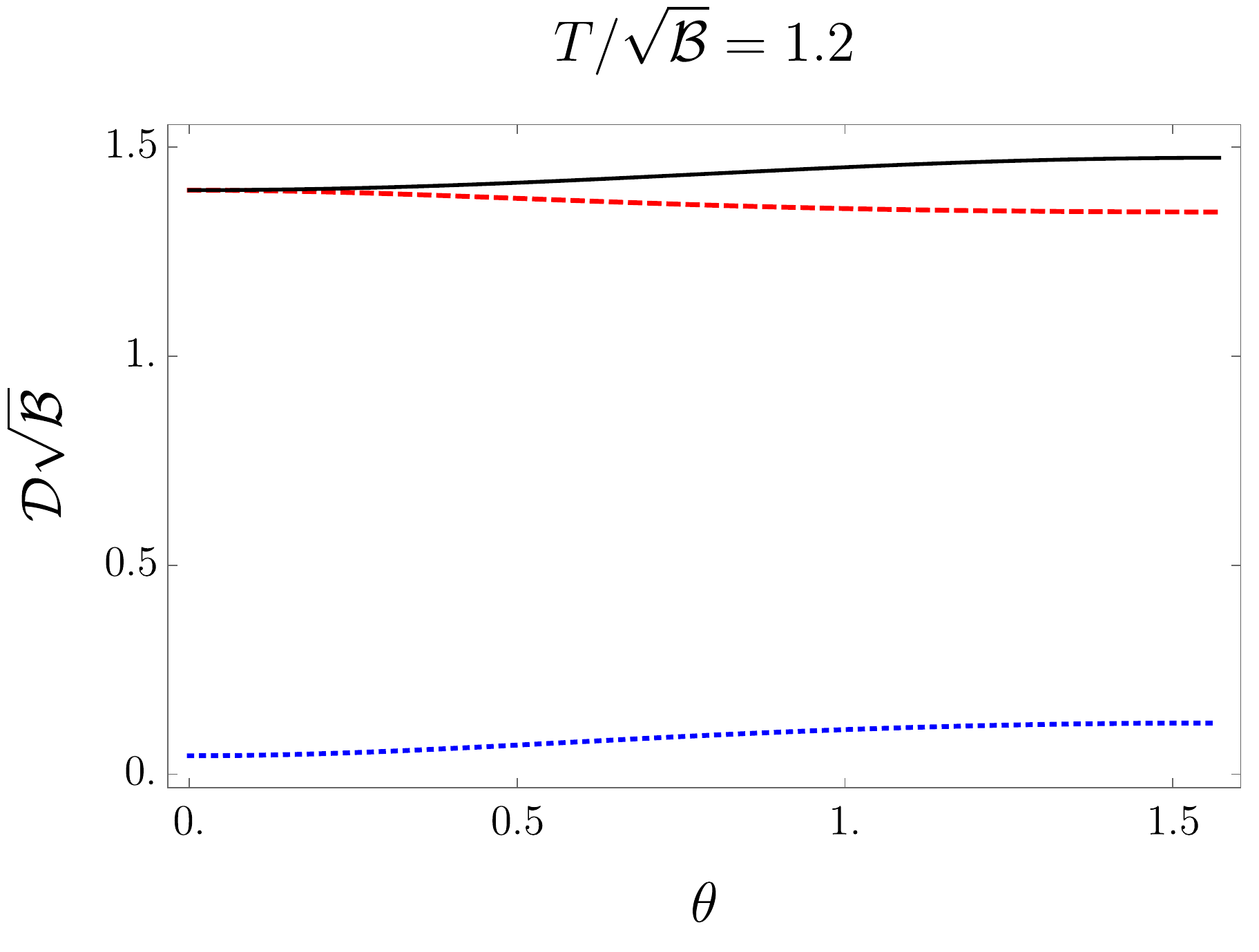}
\caption{Angular dependence of the (dimensionless) attenuation coefficients of Alfv\'{e}n (black, solid), fast (blue, dotted) and slow (red, dashed) magnetosonic waves, $\CD\sqrt{\CB}$, in the strong-field, the crossover and the weak-field regimes.
}
\label{fig:3soundAtten}
\end{figure}

Here, we plot the speeds (phase velocities) and first-order attenuation coefficients of the three types of MHD sound waves: the Alfv\'{e}n and the fast and slow magnetosonic waves for the holographic strongly coupled plasma discussed above. These results assume an infinitesimally small value of momentum $k$, and follow from first expanding the polynomial equation of the type of \eqref{AlfvenDispersion} around $k \approx 0$ and writing each dispersion relation as $\omega = \pm v k - i \CD k^2$. The speeds $v$ (presented in Fig. \ref{fig:3Sound}) and attenuation coefficients $\CD$ (presented in Fig. \ref{fig:3soundAtten}) are then plotted for all $0 \leq \theta \leq \pi / 2$, which, as discussed above, is only physically sensible when $\theta_c \to \pi / 2$, i.e. as $k\to 0$. 

The angular profiles of the speeds and the dissipative attenuation coefficients show distinct behaviour in the strong-, the crossover (cf. Eq. \eqref{Crossover}) and the weak-field regimes. In particular, the speeds of sound enter the weak-field regime, where they reduce to well-known standard MHD results, rapidly after the temperature exceeds $T / \sqrt{\CB} \approx 0.7$. There, Alfv\'{e}n and slow magnetosonic waves travel with very similar speeds for all $\theta$ and their speeds coincide at $\theta = 0 $ and $\theta = \pi/2$. The situation is different in the strong-field regime where the profiles of speeds qualitatively match the strong-field predictions of \cite{Grozdanov:2016tdf}. There, slow magnetosonic and Alfv\'{e}n waves can travel faster at small $\theta$, with speeds comparable to those of fast magnetosonic waves. At $\theta = 0$, the Alfv\'{e}n speed equals that of fast, instead of slow, magnetosonic waves (cf. Fig. \ref{fig:SummaryHighT}). It should also be noted that there exists a value of $T / \sqrt{\CB}$ in the crossover regimes where all three speeds are equal at $\theta = 0$. 

The attenuation coefficients, computed with all seven transport coefficients \cite{Grozdanov:2016tdf,Hernandez:2017mch}, are computed for the first time for a concrete microscopically (holographically) realisable plasma and therefore difficult to compare with other past results. What we observe is that the Alfv\'{e}n waves experience the strongest damping for all values of $T / \sqrt{\CB}$. Beyond that, the qualitative behaviour again displays distinct angle-dependent features in the three regimes, which are apparent from Fig. \ref{fig:3soundAtten}. A noteworthy, but not a surprising fact is that the strength of attenuation appears to be much more strongly dependent on the angle between momentum and magnetic field in the regime of small $T / \sqrt{\CB}$. Furthermore, in the crossover regime, we find that the strengths of fast and slow magnetosonic mode attenuations interchange roles as $T / \sqrt{\CB}$ increases. In plots at $T / \sqrt{\CB} = 0.5$ and $T / \sqrt{\CB} = 0.66$, there exists an angle $\theta$ at which the two attenuation strengths coincide. 

\subsection{MHD modes on a complex frequency plane}\label{Thetadep}

By assuming a finite value of momentum $k$, a full analysis of the spectrum requires us to take into account the transmutation of sound modes into non-propagating diffusive modes. The pattern of this behaviour, as a function of the angle between momentum and the direction of the equilibrium magnetic field $\theta$, was summarised in Fig. \ref{fig:SummaryHighT}. Motivated by holographic quasinormal mode (poles of two-point correlators) analyses, we plot the motion of the MHD modes on the complex frequency plane---here, as a function of $\theta$ and $T/\sqrt{\CB}$. One should consider these plots as a prediction of how the first-order approximation to the hydrodynamic sector of the full quasinormal spectrum computed from the theory \eqref{HoloAction} is expected to behave. 

\begin{figure}[tbh]
\center 
\includegraphics[width=.49\textwidth]{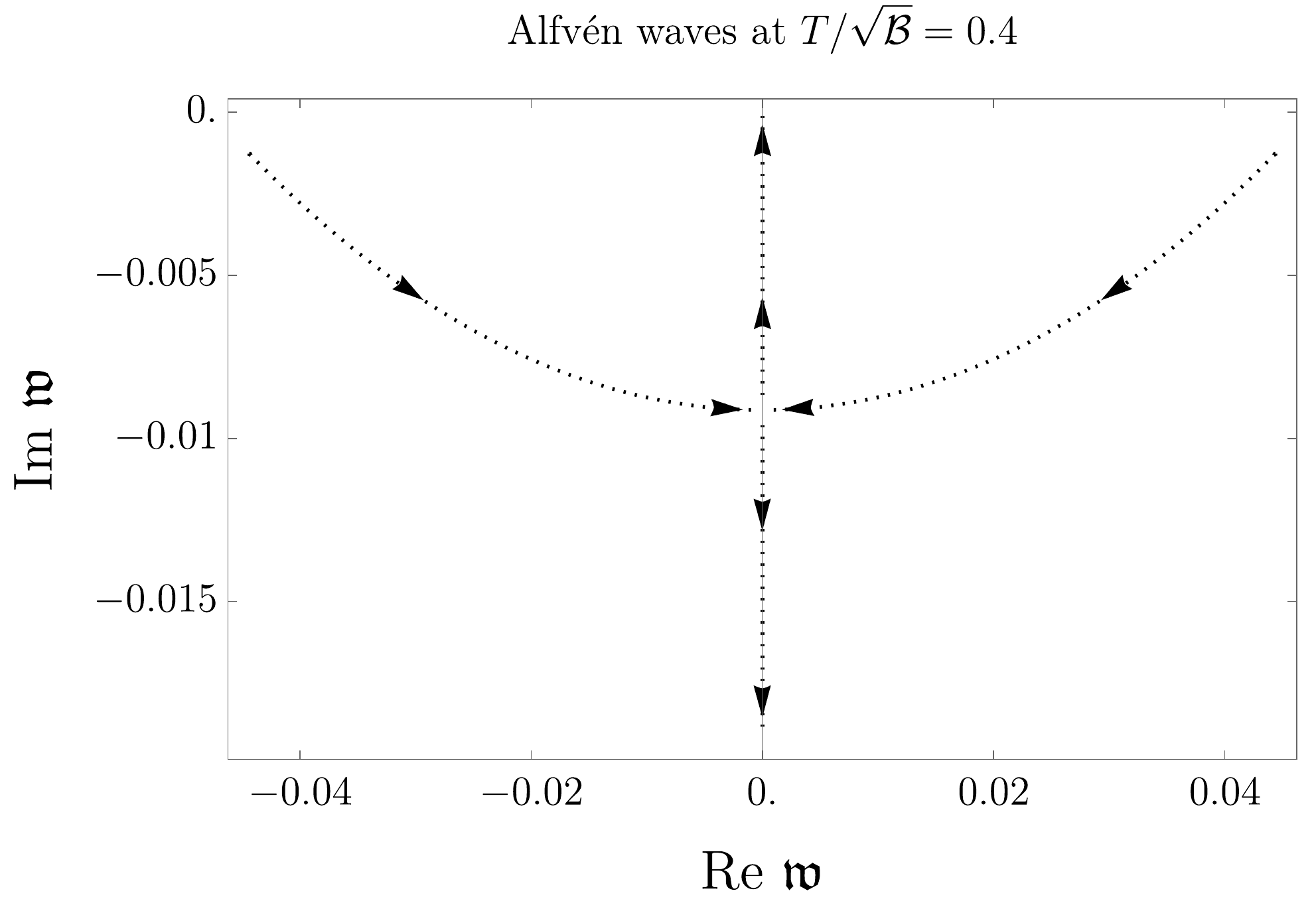}
\includegraphics[width=.49\textwidth]{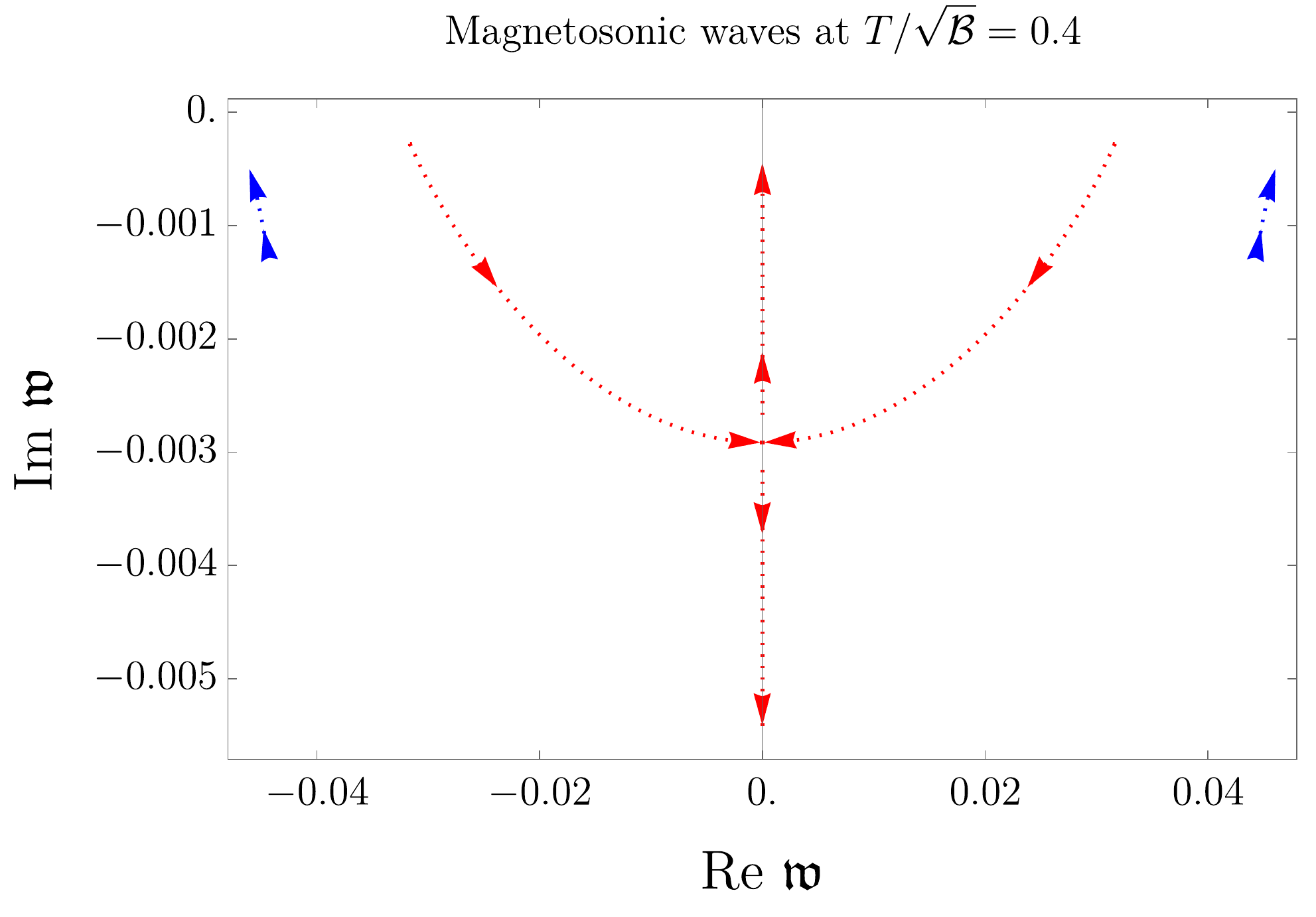}
\includegraphics[width=.491\textwidth]{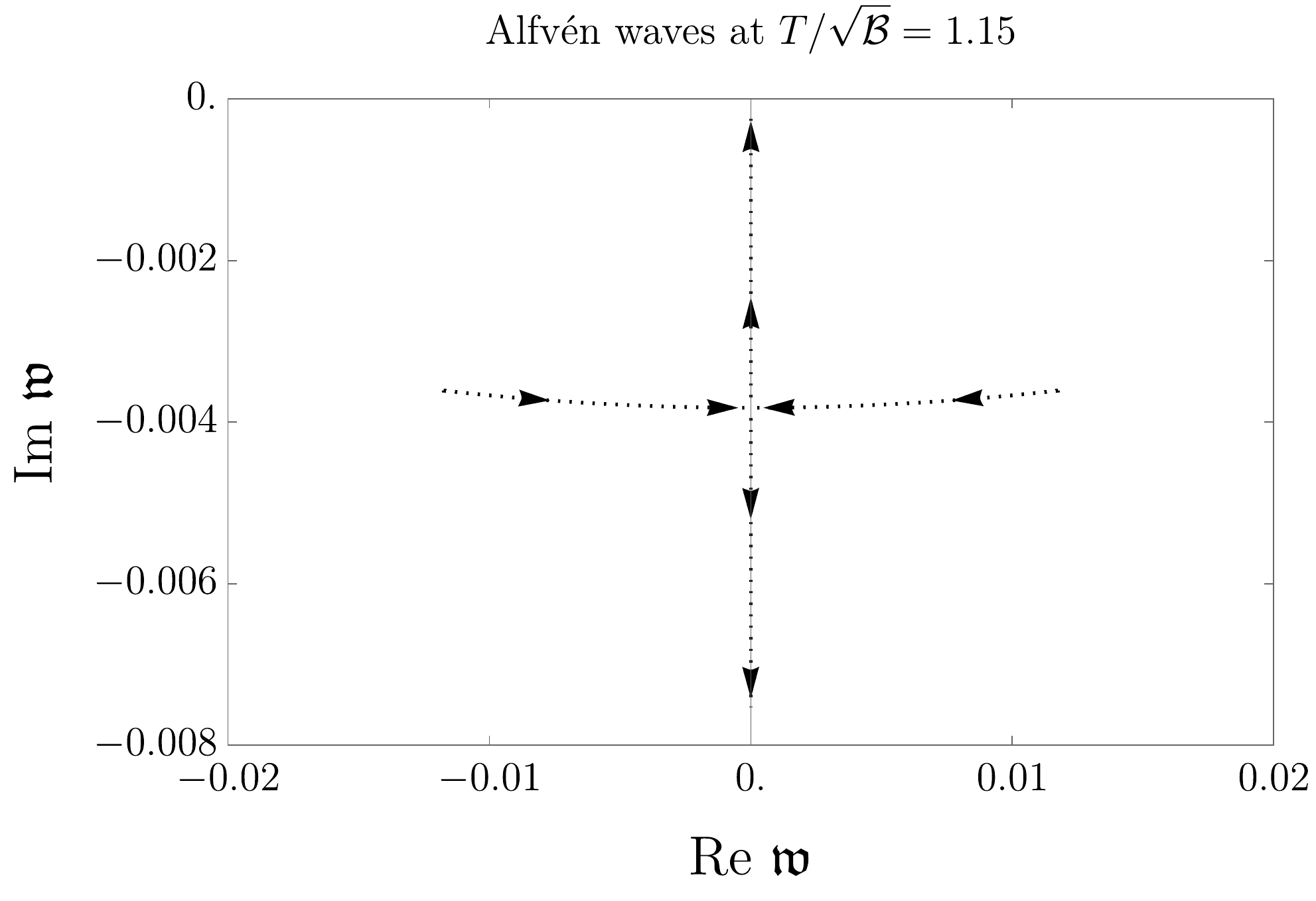}
\includegraphics[width=.49\textwidth]{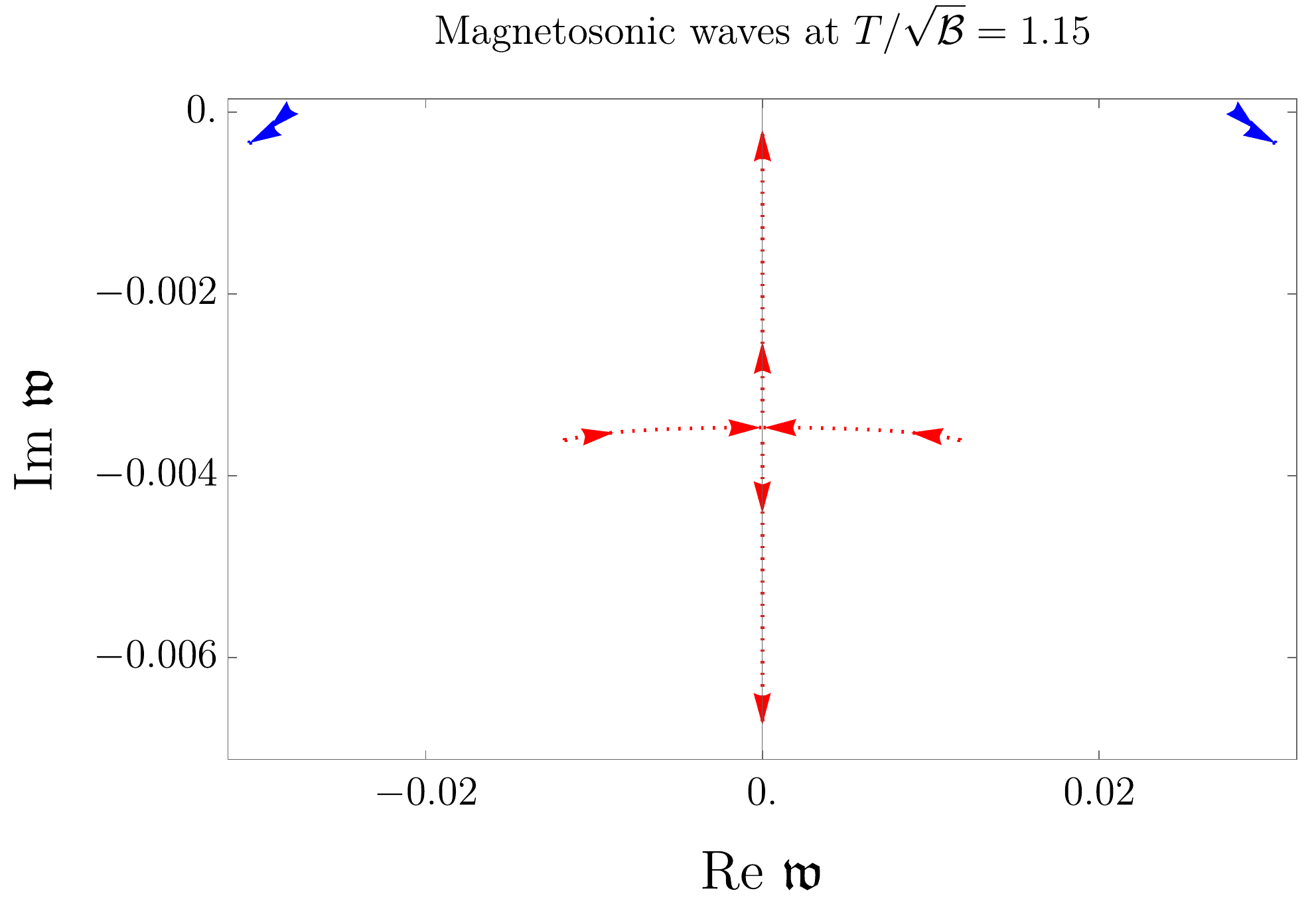}
\caption{Dependence of the complex (dimensionless) frequency $\mathfrak{w}=\omega/\sqrt{\CB}$ on $\theta$, plotted for Alf\'{e}n (black) and fast (blue) and slow (red) magnetosonic waves in the strong- and weak-field regimes with $T/\sqrt{\CB}=0.4$ and  $T / \sqrt{\CB} = 1.15$, respectively. The arrows represent the motion of poles as $\theta$ is tuned from $0$ to $\pi/2$. Momentum is set to $k/\sqrt{\CB}=0.05$.}
\label{fig:complexPlaneVaryThetaExtremeT}
\end{figure}

In Fig. \ref{fig:complexPlaneVaryThetaExtremeT}, we plot the typical $\theta$-dependent trajectories of $\omega(\theta)$ for Alfv\'{e}n and magnetosonic modes in distinctly strong- and weak-field regimes. At all temperatures (except at $T = 0$ where $\CD = 0$), the behaviour is consistent with our previous discussions, including the fact that the transmutation of Alfv\'{e}n and slow magnetosonic waves into diffusive modes occurs at lower $\theta_c$ as $k/\sqrt{\CB}$ increases.

\begin{figure}[tbh]
\center 
\includegraphics[width=.49\textwidth]{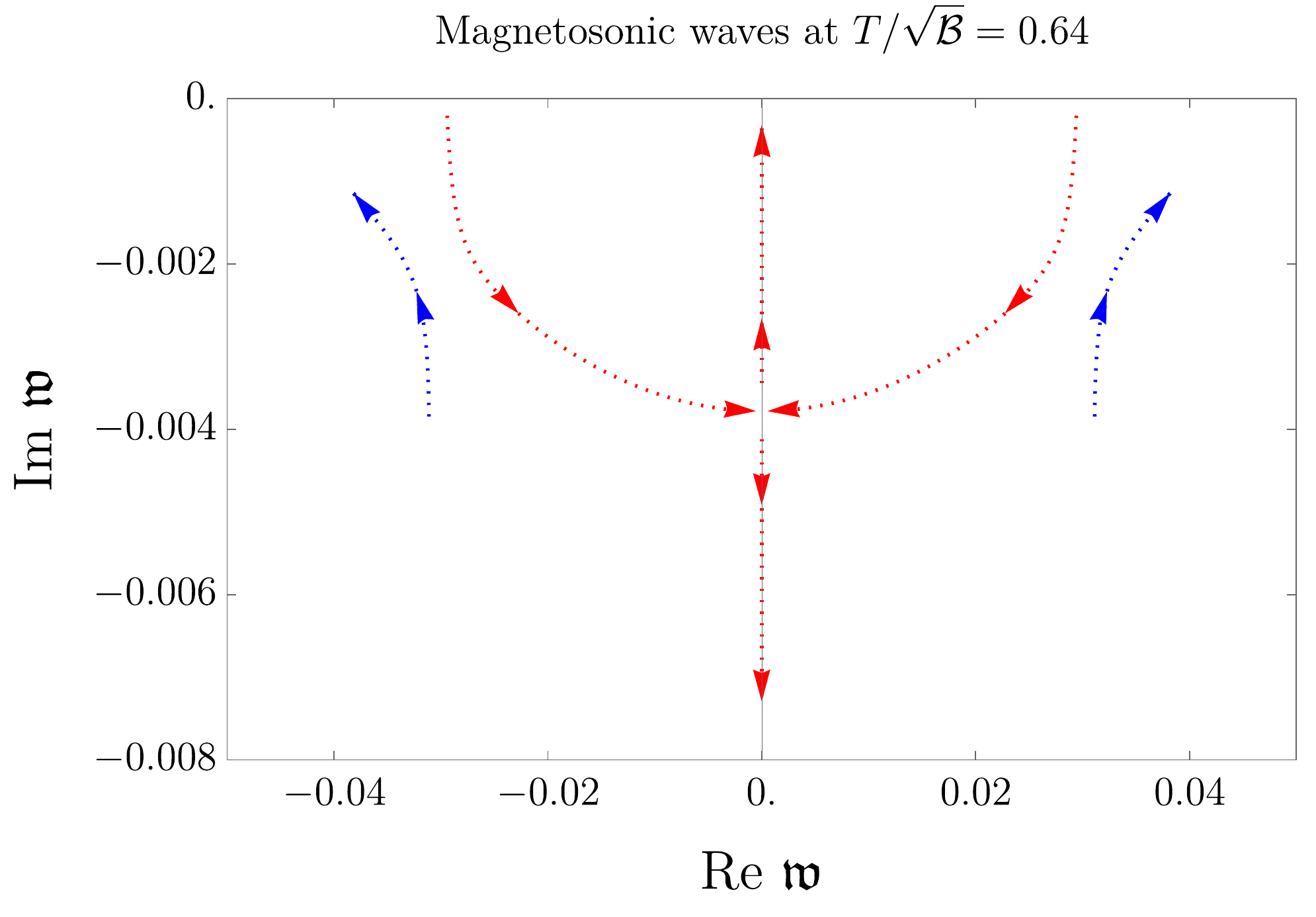}
\includegraphics[width=.49\textwidth]{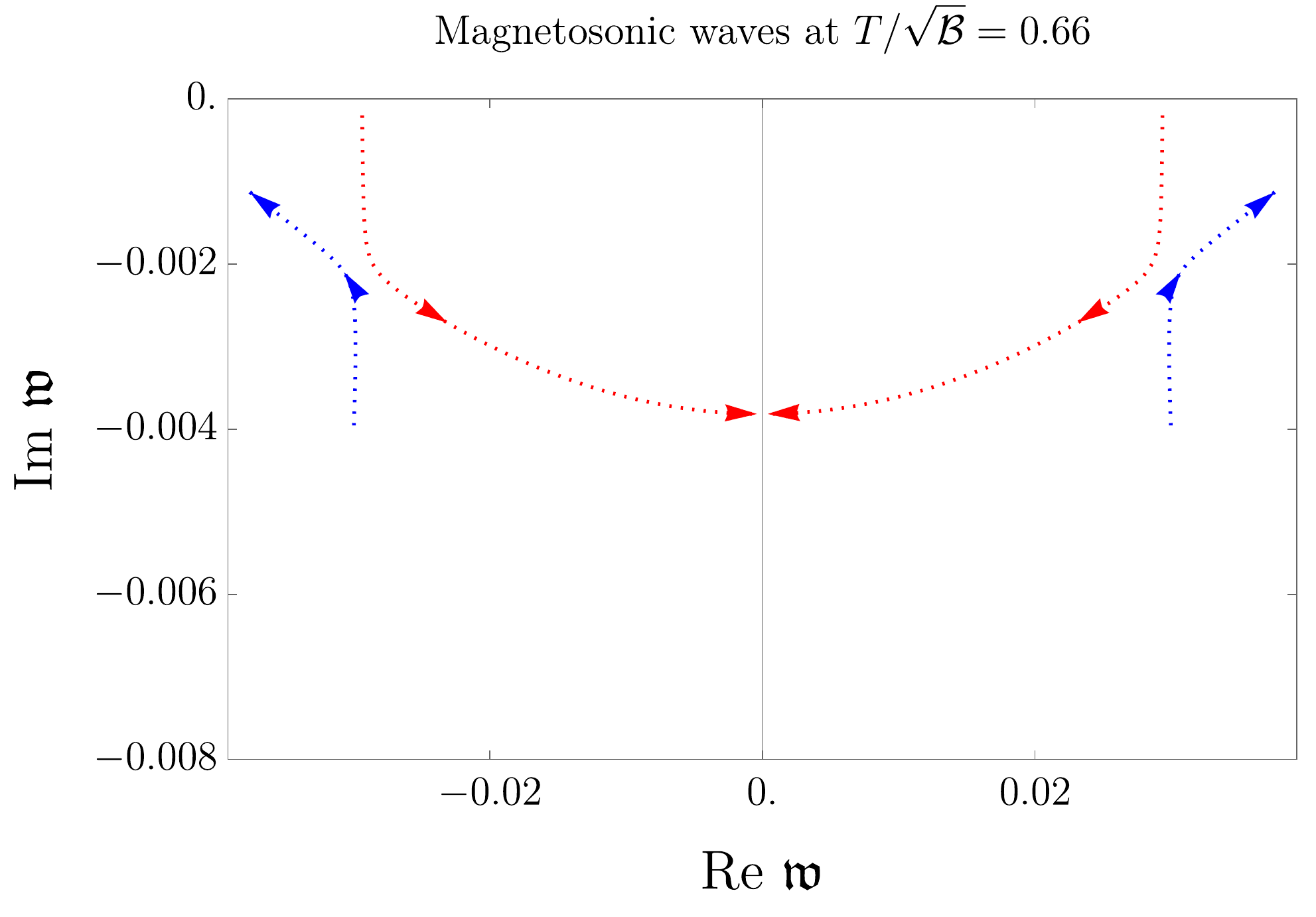}
\includegraphics[width=.49\textwidth]{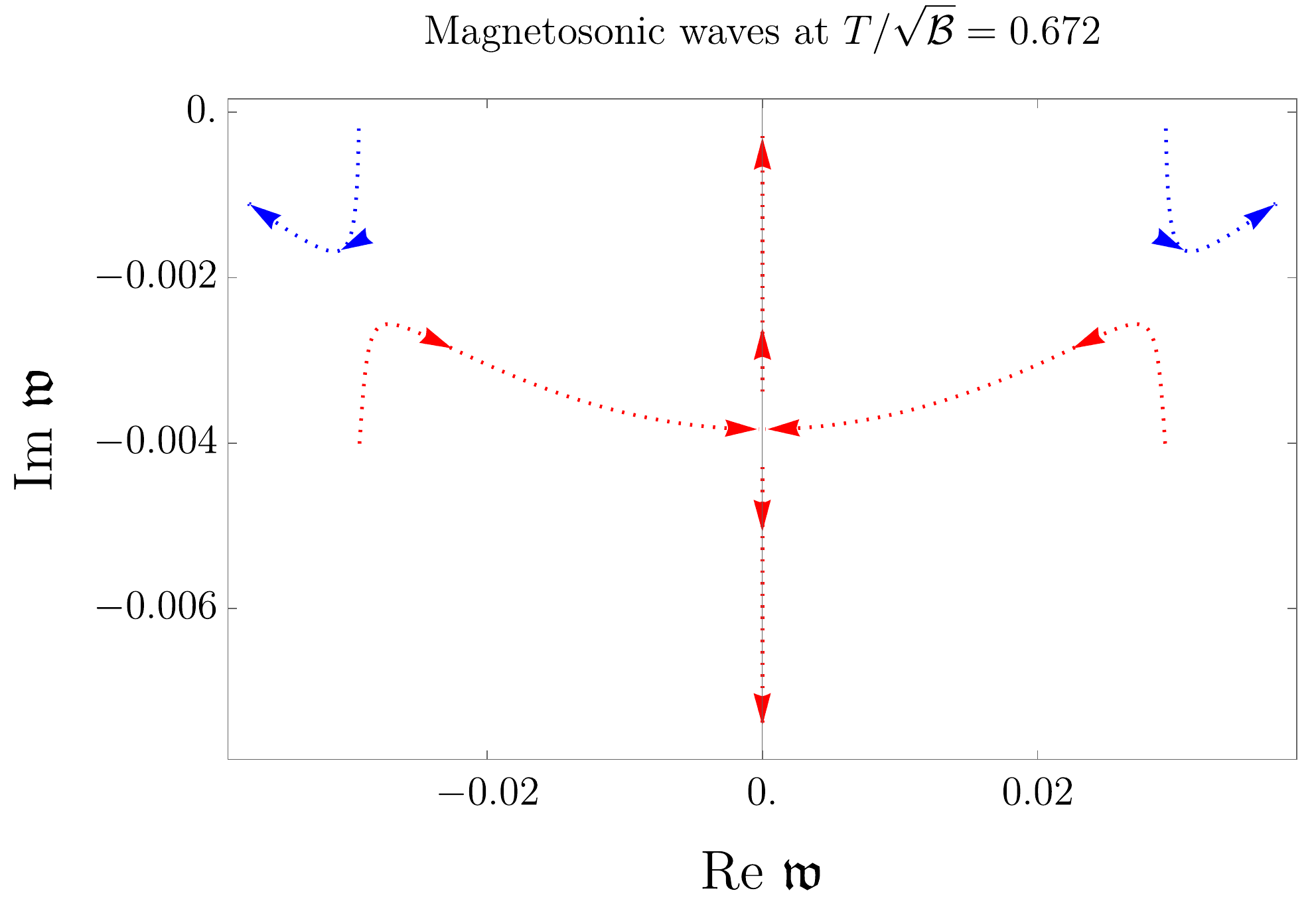}
\includegraphics[width=.49\textwidth]{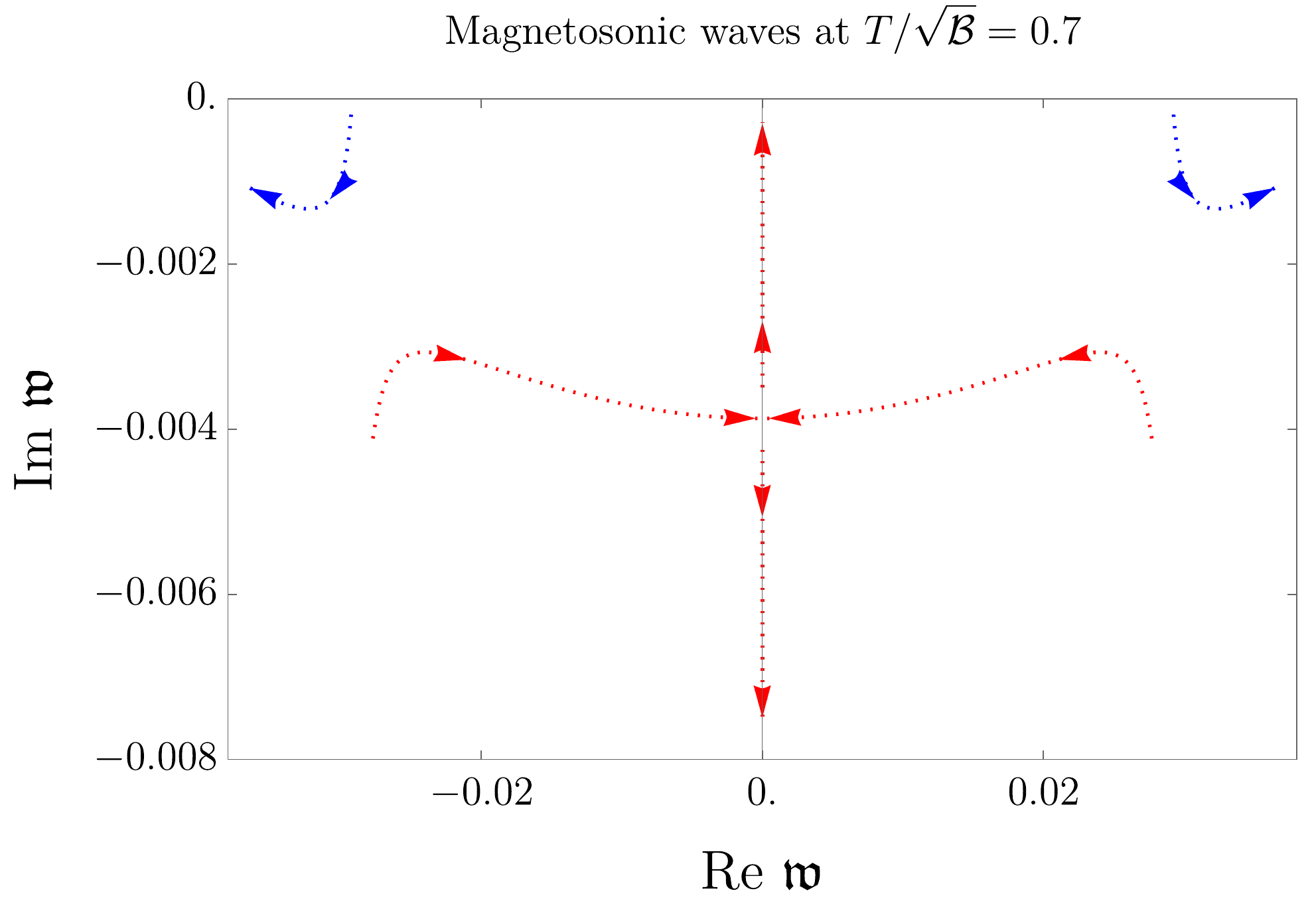}
\caption{Dependence of the complex (dimensionless) frequency $\mathfrak{w}=\omega/\sqrt{\CB}$ of fast (blue) and slow (red) magnetosonic modes on $\theta$ in the crossover regime. The arrows represent the motion of poles as $\theta$ is tuned from $0$ to $\pi/2$. Momentum is set to $k/\sqrt{\CB}=0.05$.}
\label{fig:complexPlaneVaryThetaMidT1234}
\end{figure}

In the crossover temperature regime (around $T/\sqrt{\CB} \approx 0.6$), we can observe in more detail the interplay between fast and slow magnetosonic modes, which was noted in Section \ref{SpeedsAndAtt}. While the speed of fast magnetosonic waves always exceeds that of slow waves, their attenuation strengths exchange roles around $T/\sqrt{\CB} \approx 0.675$, which manifests in a characteristically distinct behaviour for $\theta < \theta_c$, presented in Fig. \ref{fig:complexPlaneVaryThetaMidT1234} (see also Fig. \ref{fig:3soundAtten}). The $\theta$-dependence of Alfv\'en waves remains qualitatively similar to those depicted in Fig. \ref{fig:complexPlaneVaryThetaExtremeT}.

\begin{figure}[tbh]
\center 
\includegraphics[width=.49\textwidth]{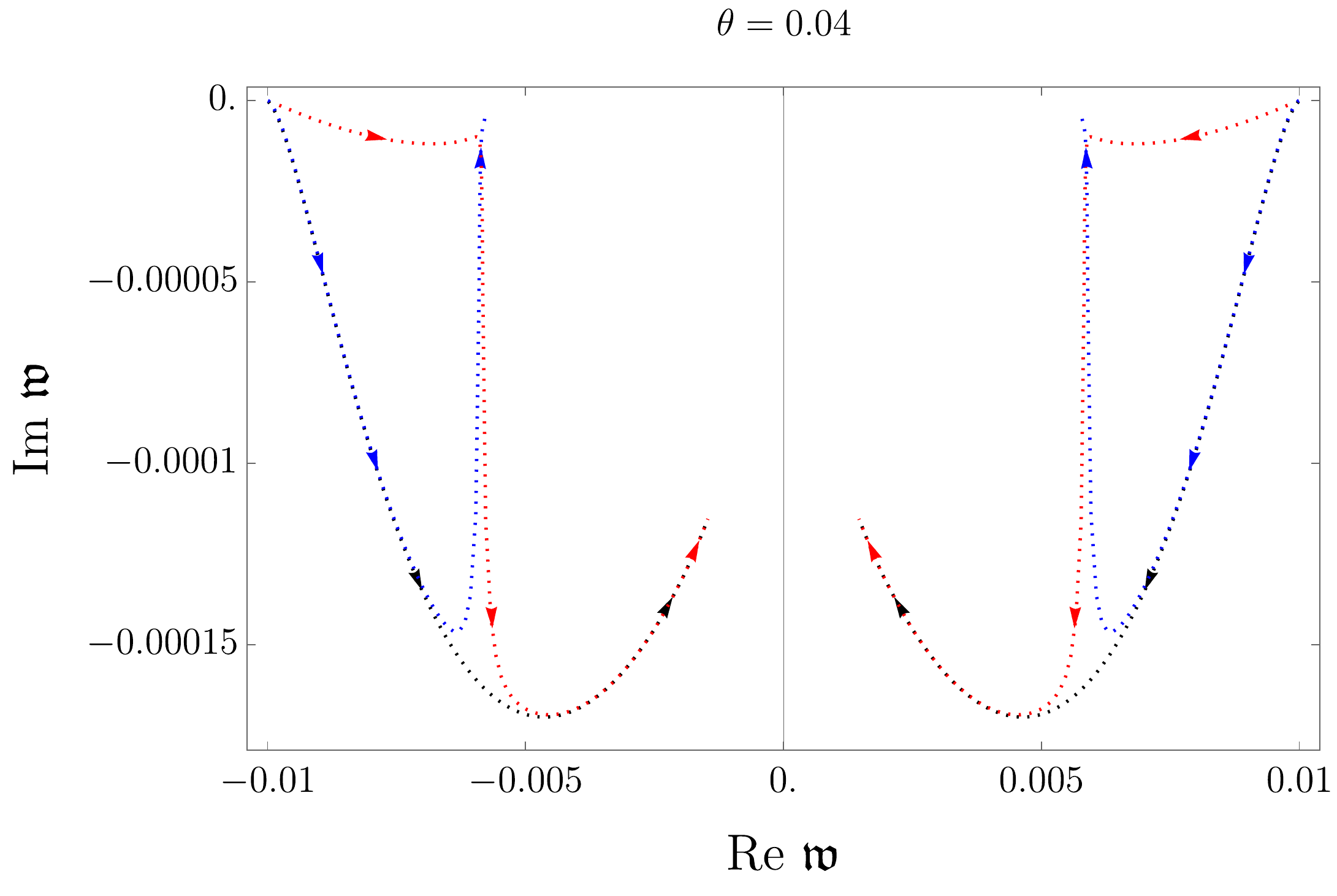}
\includegraphics[width=.49\textwidth]{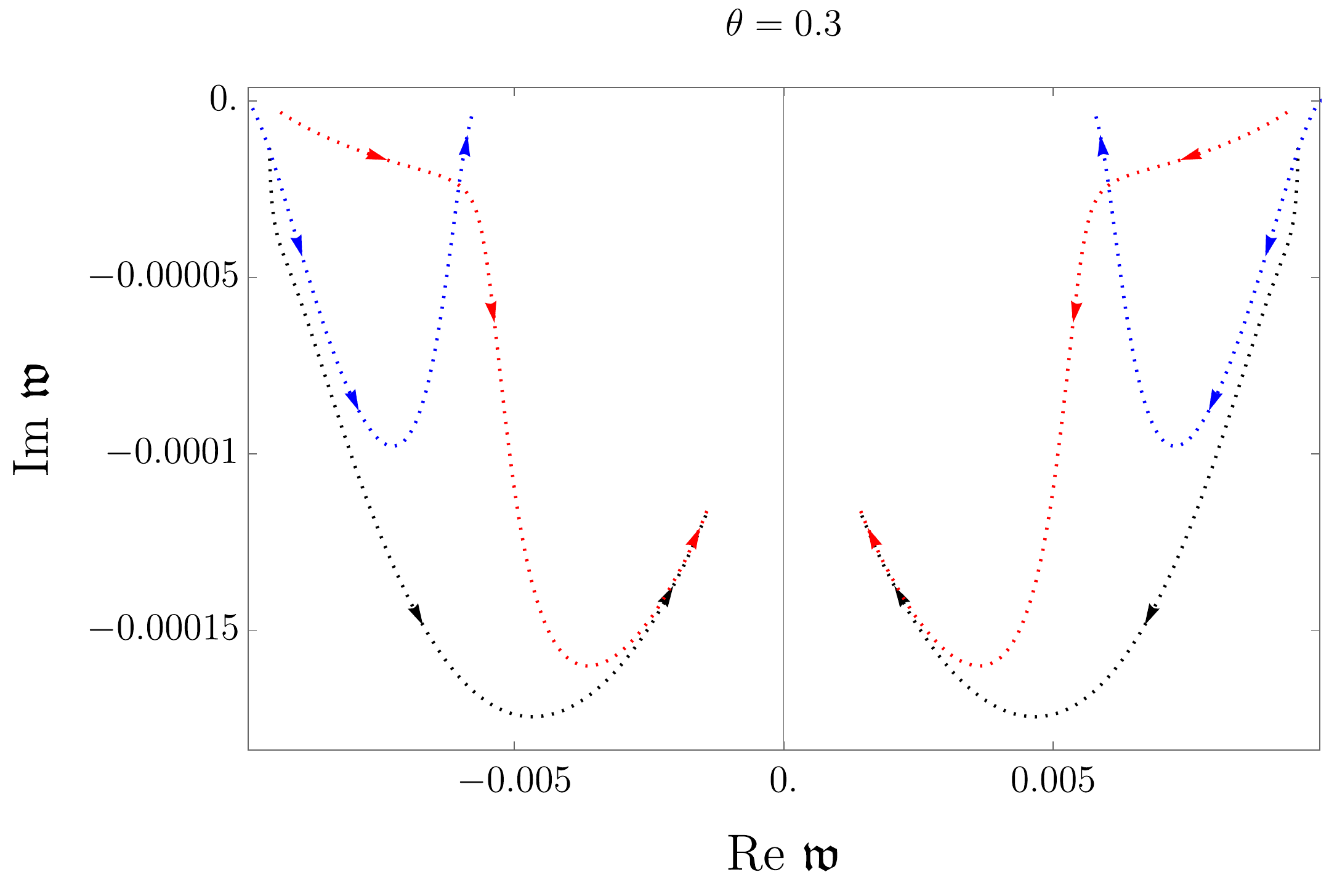}
\includegraphics[width=.49\textwidth]{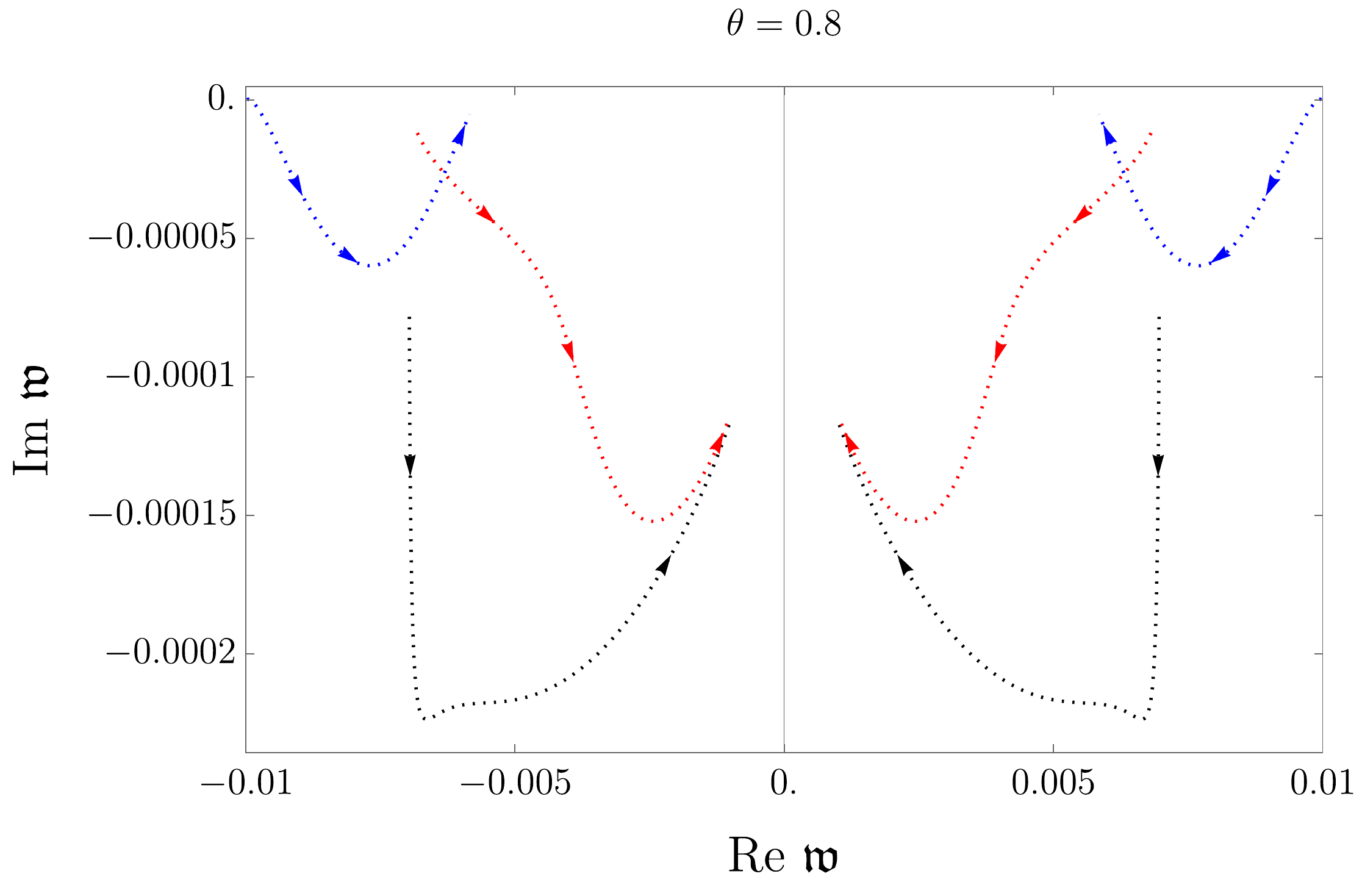}
\includegraphics[width=.49\textwidth]{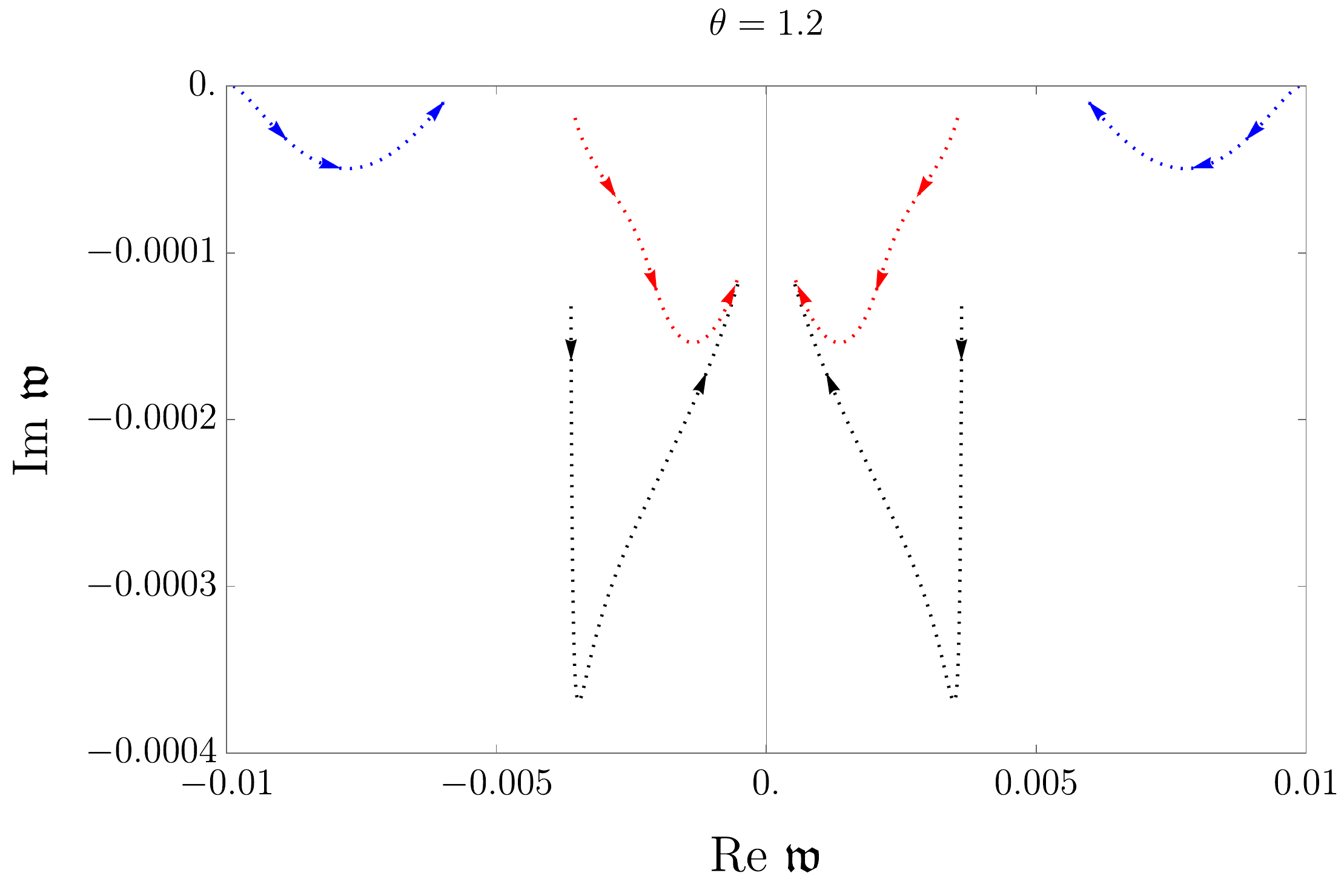}
\caption{Dependence of the complex (dimensionless) frequency $\mathfrak{w}=\omega/\sqrt{\CB}$ on $T/\sqrt{\CB}$, plotted for Alfv\'{e}n (black) and fast (blue) and slow (red) magnetosonic waves for $\theta < \theta_c$. The arrows represent the motion of poles as $T/\sqrt{\CB}$ is tuned from $0$ towards the weak-field regime. Momentum is set to $k/\sqrt{\CB}=0.01$. 
}
\label{fig:complexPlaneVaryTem1}
\end{figure}

For a fixed $\theta < \theta_c$, where $\theta_c$ depends on $k$ and $T/\sqrt{\CB}$, we plot the typical behaviour of $\omega(k)$ as a function of $T/\sqrt{\CB}$ in Fig. \ref{fig:complexPlaneVaryTem1}. At $T=0$, all poles start from the non-dissipative regime (the real $\mathfrak{w}$ axis), with the speed of fast magnetosonic waves given by $v = 1$. As they move towards larger $T/\sqrt{\CB}$, the Alfv\'{e}n and the slow magnetosonic modes again asymptote to each other, eventually transforming into diffusive modes, while the speed of the fast magnetosonic modes gradually converges towards that of neutral conformal sound with $v = 1 / \sqrt{3}$. 

\begin{figure}[tbh]
\center
\includegraphics[width=.49\textwidth]{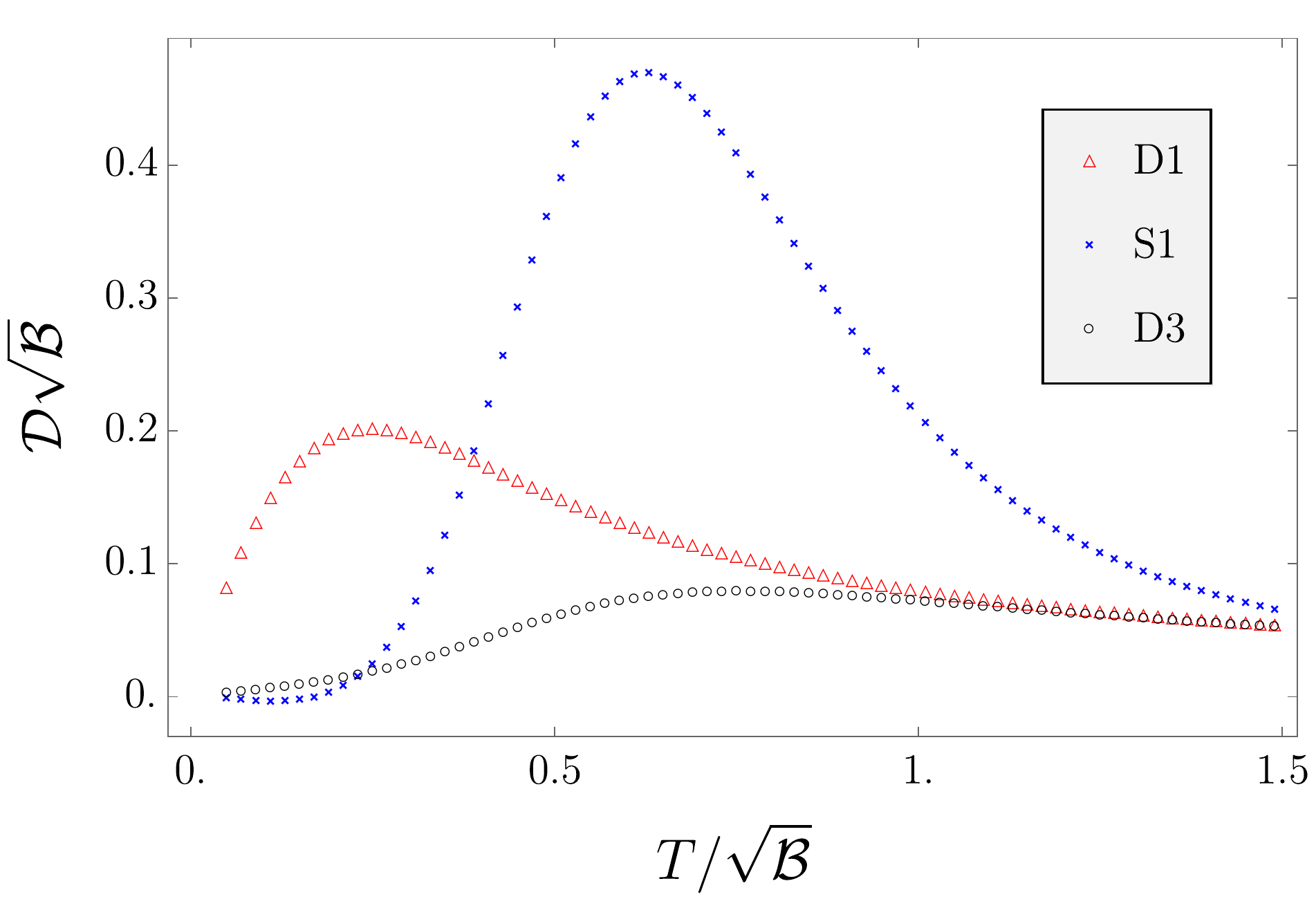}
\includegraphics[width=.476\textwidth]{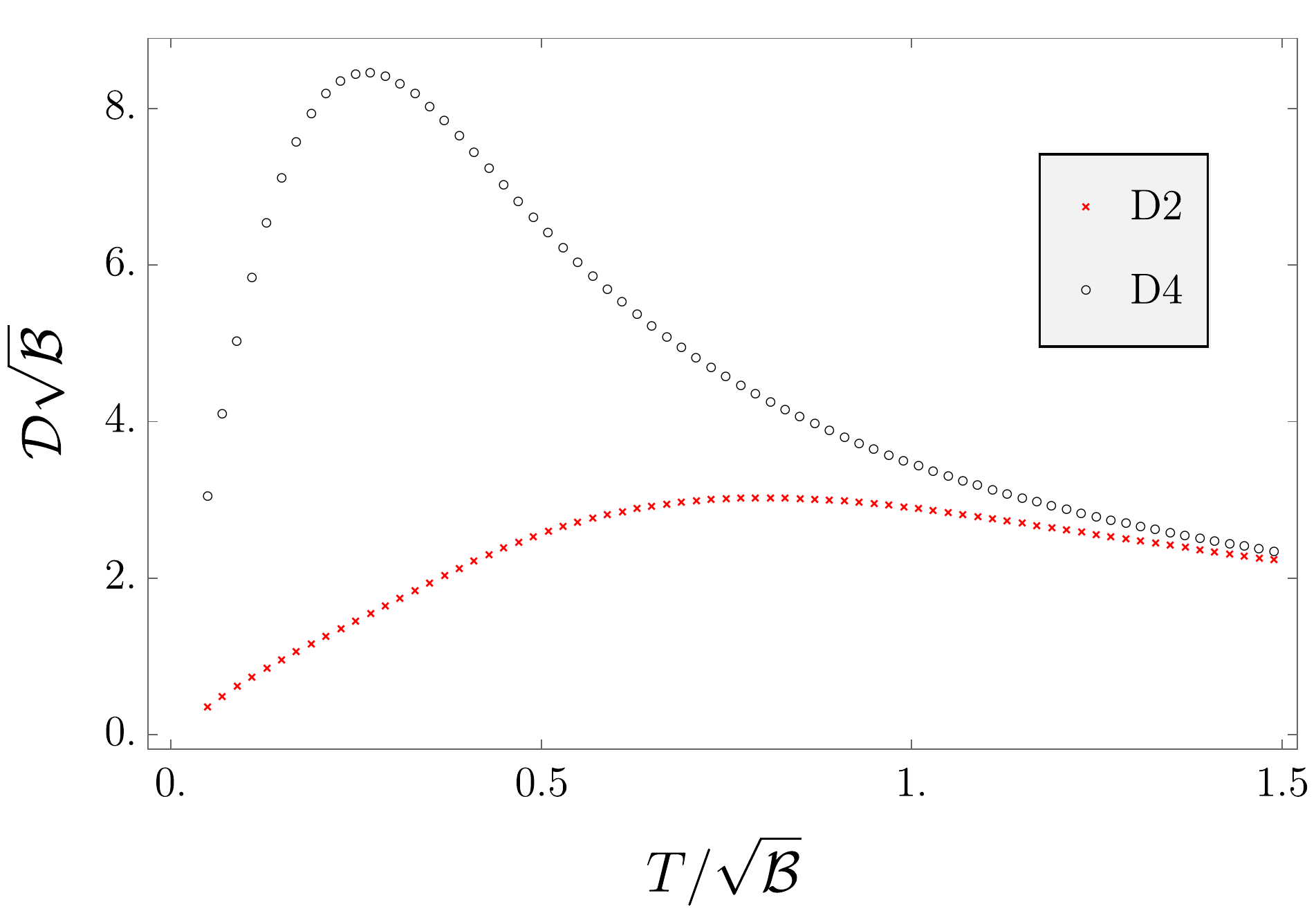}
\caption{Plots of the four diffusion constants (D1, D2, D3, D4) and the sound attenuation (S1) as a function of $T/\sqrt{\CB}$ at $\theta =\pi/2$. Black, red and blue curves depict dissipative coefficients that originate from the Alfv\'{e}n, slow magnetosonic and fast magnetosonic waves, respectively.
}
\label{fig:90degreeModes}
\end{figure}

In the high temperature limit, the ``collision" of the Alfv\'{e}n and, independently, the slow magnetosonic poles on the imaginary axis occurs close to the real axis, which follows from the fact that for both types of waves, 
\begin{align}
 \text{Im}\left[\mathfrak{w}\right] \approx -\frac{1}{2}\left( \frac{\eta}{\varepsilon+p} + \frac{\mu r}{\CB}\right)\sqrt{\CB} \sim -\frac{\sqrt{\CB}}{T} \to 0\,,
\end{align}
as $T / \sqrt{\CB} \to \infty$. The Alfv\'{e}n waves then become the diffusive modes of uncharged conformal hydrodynamics with $\omega = -i\eta k^2/( 2sT )$. As for our final plot, in Fig. \ref{fig:90degreeModes}, we present the dependence of the four diffusion constants and one sound attenuation coefficient on the temperature at $\theta = \pi/2$ (cf. Fig. \ref{fig:SummaryHighT} and Eqs. \eqref{specialSound}--\eqref{specialDiffuse}). The modes D1, D3 and S1 reduce to dispersion relations of uncharged relativistic hydrodynamics. D2 and D4 are new. 

\subsection{Electric charge dependence}\label{sec:alphaDependence}

We end our discussion of MHD dispersion relations by investigating their dependence on the choice of the $U(1)$ coupling constant, or equivalently, the position of the Landau pole, which has so far been set to the ($N_c$-rescaled) $\bar\alpha = 1 / 137$. All dependence on $\bar\alpha$ enters into the expectation value of the stress-energy tensor through the term proportional to $\CH_{\mu\nu}\CH^{\mu\nu} \ln\CC$ (cf. Eq. \eqref{defT}), which contributes no terms linear in $\omega$. For this reason, while the equation of state strongly depends on $\bar\alpha$, the first-order transport coefficients do not. Hence, all speeds of sound and attenuation (and diffusive) coefficients depend on the choice of $\bar\alpha$ through the equation of state and susceptibilities. 

\begin{figure}[tbh]
\center
\includegraphics[width=.49\textwidth]{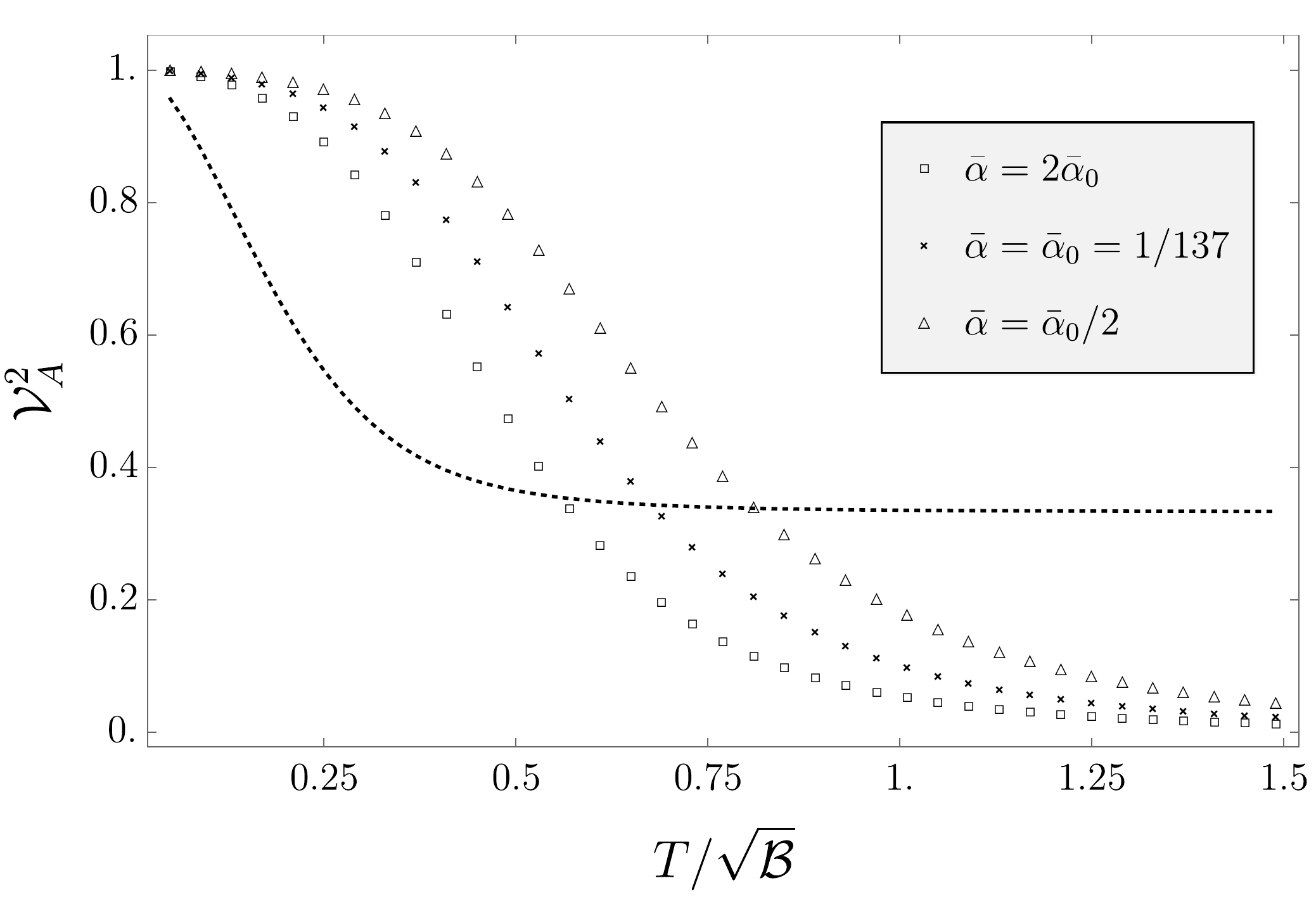}
\includegraphics[width=.49\textwidth]{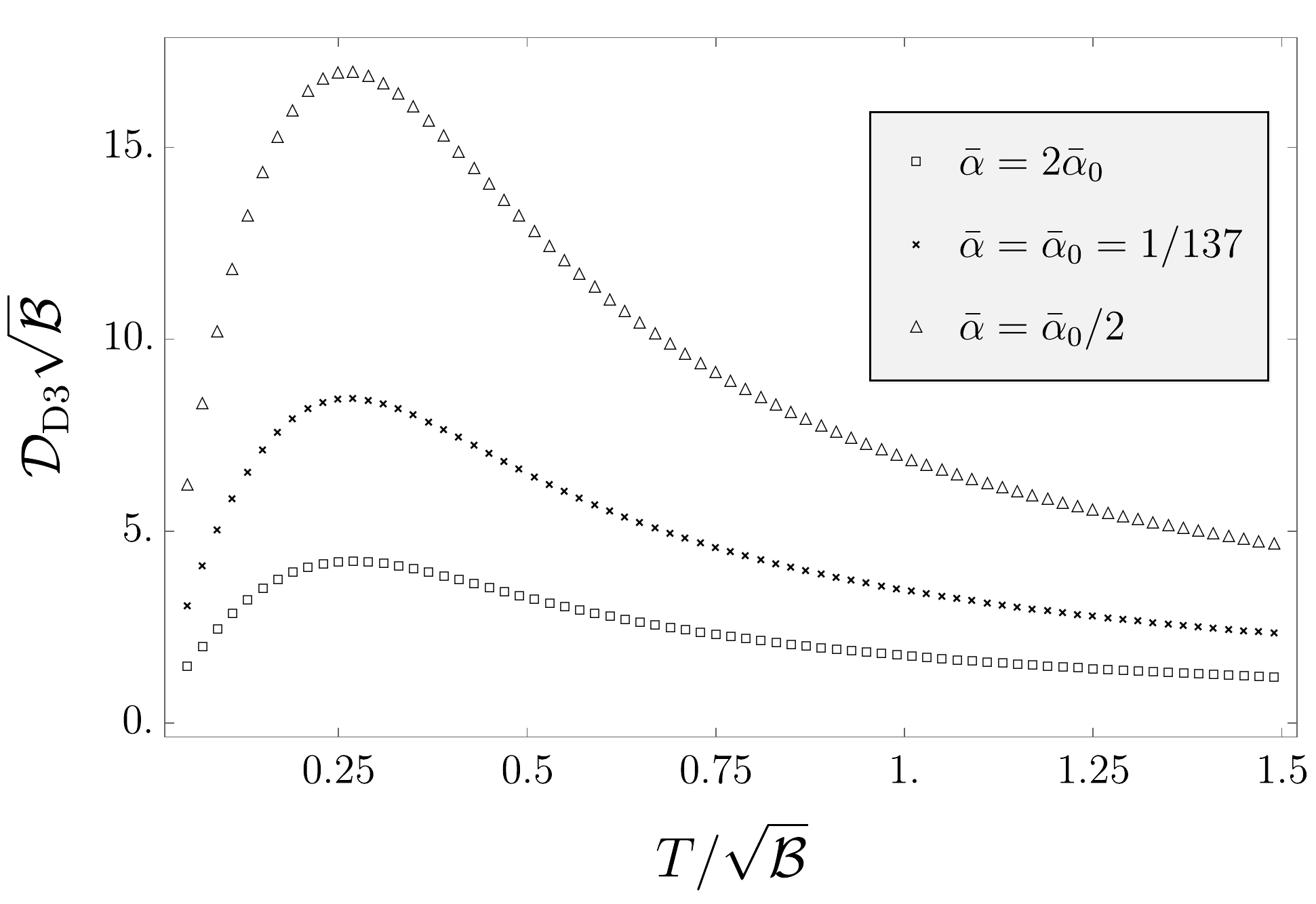}
\caption{The plot of $\CV_A^2$ and the diffusion constant $\CD_{D3}$ at $\bar\alpha = \{ \bar\alpha_0 / 2,\, \bar\alpha_0,\,2\bar\alpha_0 \}$, where $\bar\alpha_0 = 1/137$. The dashed line in the left plot is the $\bar\alpha$-independent speed (squared) of the S3 mode (cf. Fig. \ref{fig:SummaryHighT}), i.e. $\CV_0^2$, which is plotted for comparison.
}
\label{fig:VA-differentAlpha}
\end{figure}

\begin{figure}[tbh]
\center
\includegraphics[width=.55\textwidth]{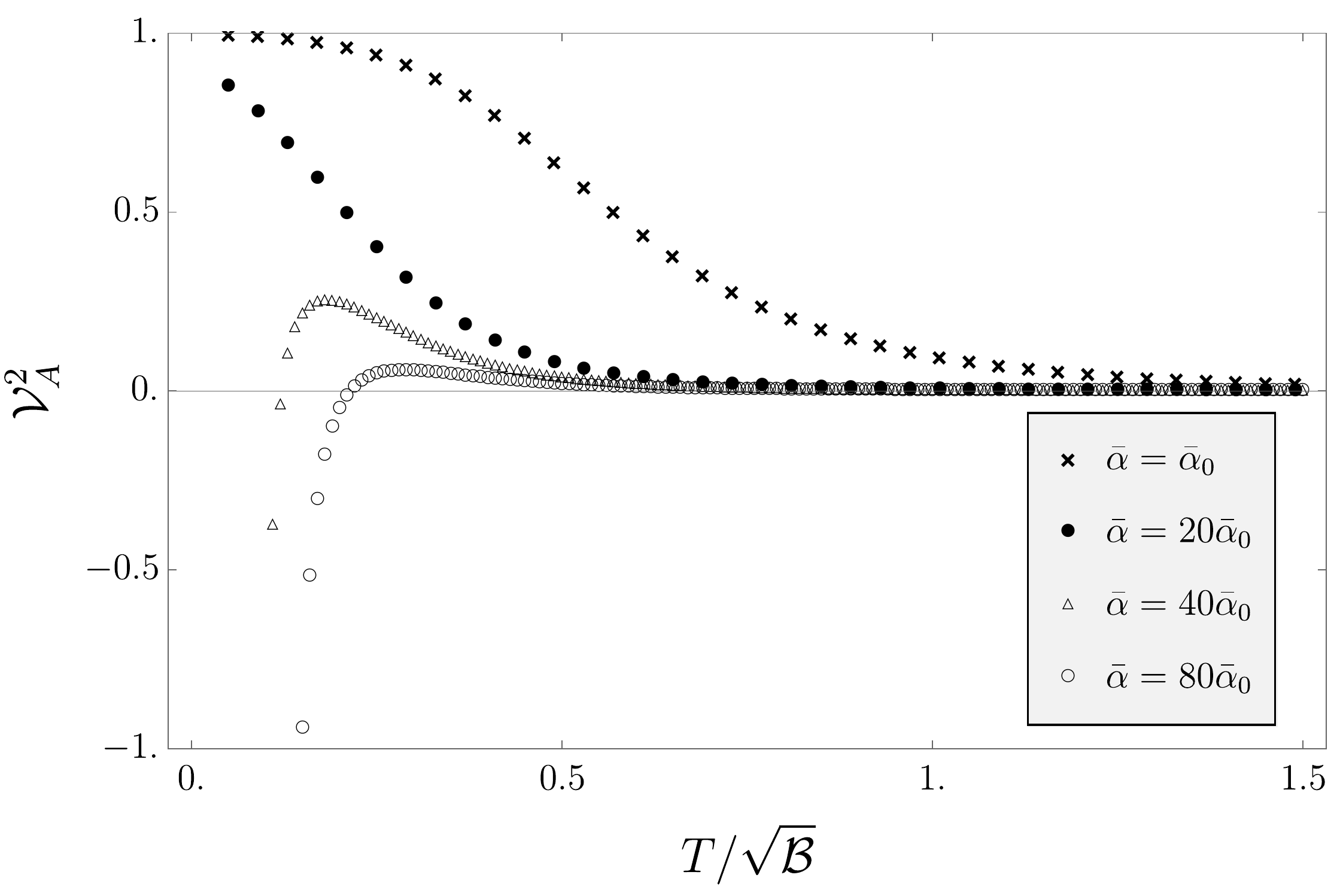}
\caption{The plot of the Alfv\'{e}n $\CV_A^2$ at a varying $\bar\alpha$ ranging from $\bar\alpha = \bar\alpha_0$ to $\bar\alpha = 80 \bar\alpha_0$, where $\bar\alpha_0 = 1/137$. We see that as $\bar\alpha$ increases, the waves develop an instability in the strong-field regime.
}
\label{speedAflvenVaryAlpha}
\end{figure}

What we observe is that the speeds of waves and attenuation coefficients strongly depend on the renormalised electromagnetic coupling, so, unsurprisingly, the strength of electromagnetic interactions plays an important role in the phenomenology of MHD. For concreteness, we only present the detailed behaviour of the Alfv\'{e}n waves (with speed $\CV_A \cos\theta$), which reduce to the neutral hydrodynamic diffusive mode D3 (and D4) at $\theta = \pi/2$. Both $\CV_A$ and the diffusion constant of D3, $\CD_{D3}$, strongly depend on $\bar\alpha$. For a small variation in the values of $\bar\alpha$, we plot the results in Fig. \ref{fig:VA-differentAlpha}.\footnote{We remind the reader that in the boundary Lagrangian, the electromagnetic coupling is scaled out from the covariant derivatives. Thus, only the Maxwell term depends on $e_r$. As we vary $e_r$, we keep the strength of the electromagnetic field fixed.} To show the importance of a sensible choice of the renormalisation condition, we also vary the coupling over a larger range (to $\bar\alpha = 80 / 137$), where we see that the system develops unphysical behaviour with instabilities. As is apparent from Fig. \ref{speedAflvenVaryAlpha}, Alfv\'{e}n waves become unstable at low $T/\sqrt{\CB}$. 

\begin{figure}[tbh]
\center
\includegraphics[width=.49\textwidth]{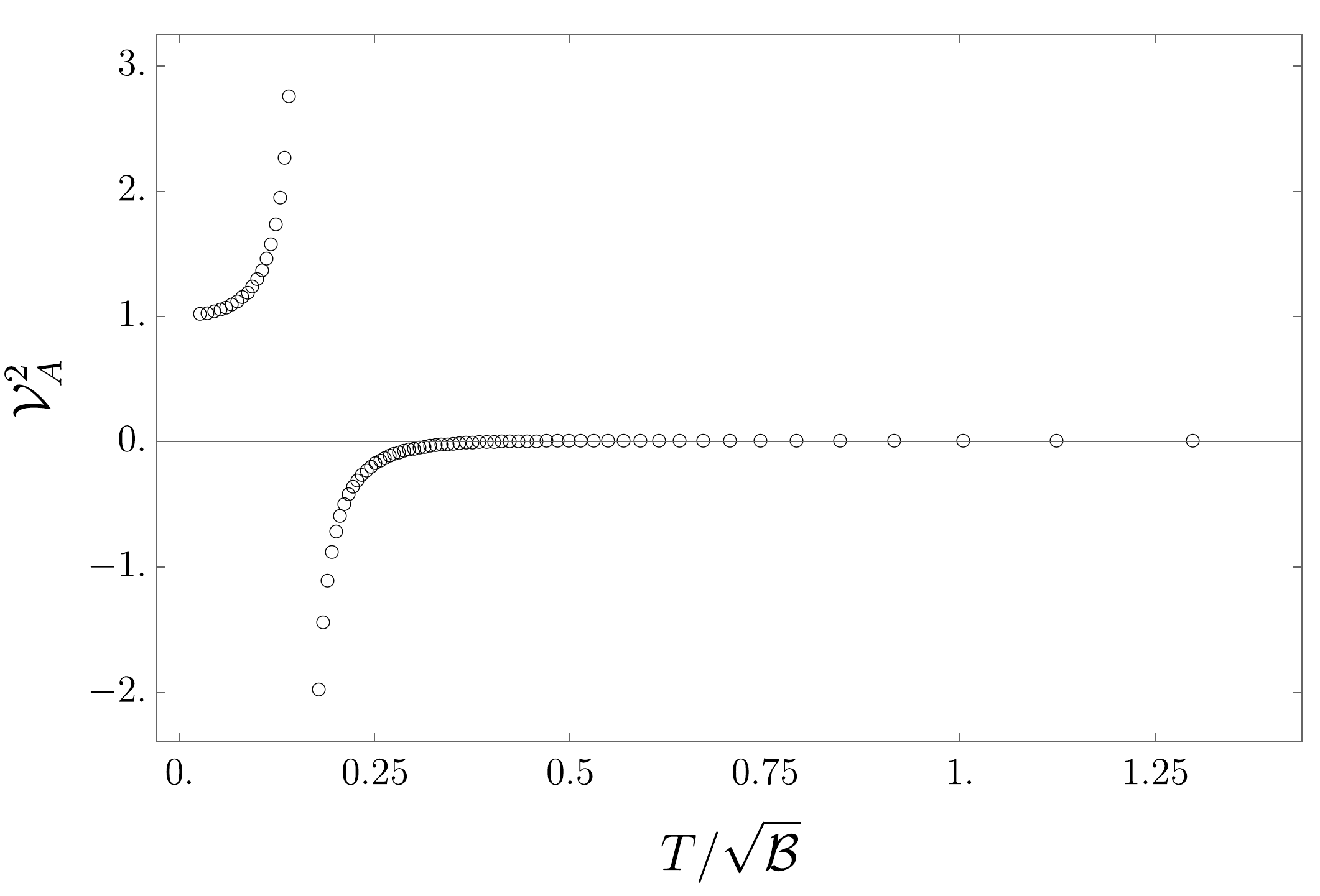}
\includegraphics[width=.49\textwidth]{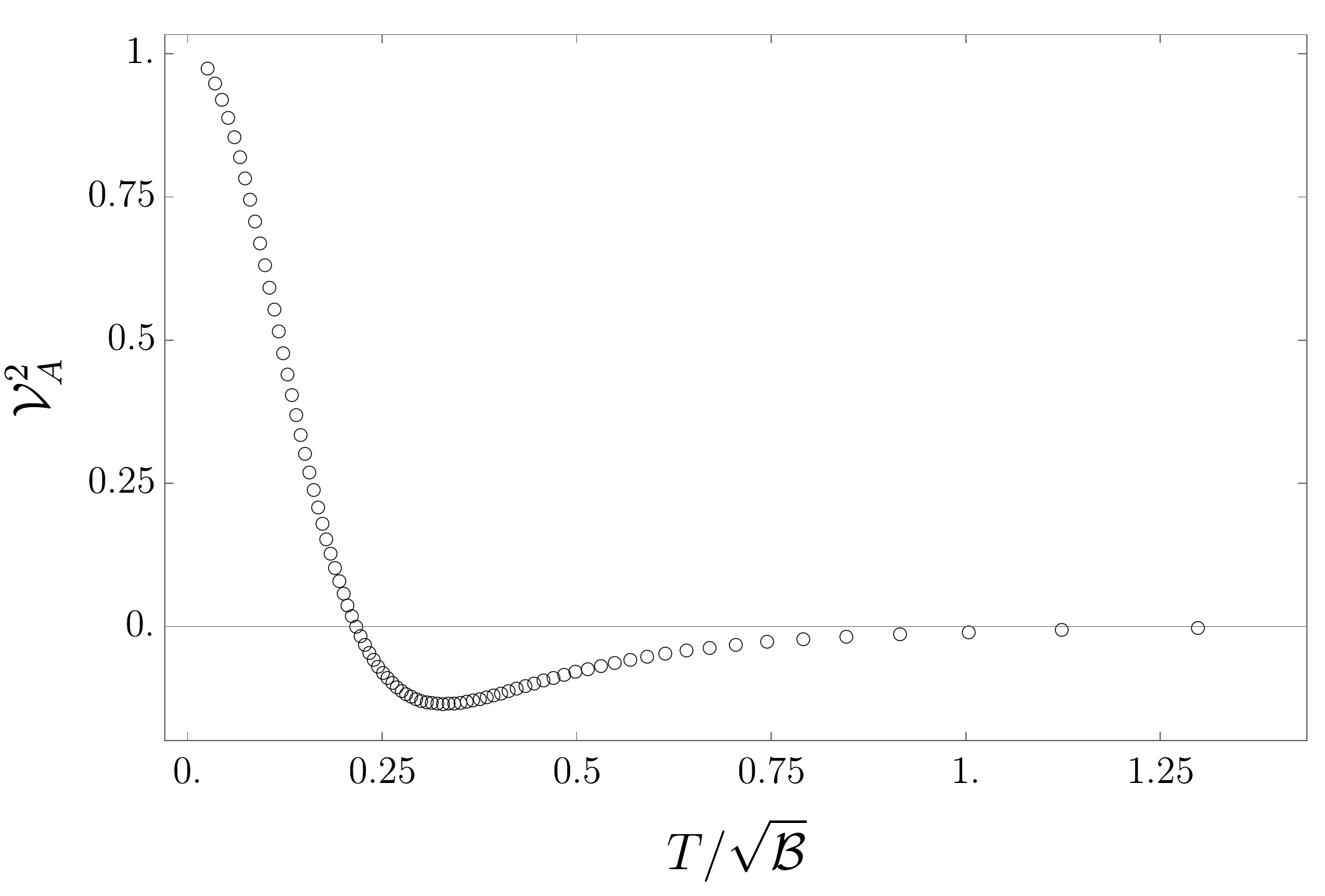}
\caption{The $\theta$-independent factor $\CV_A$ of the Alfv\'en wave speed plotted for the renormalised $e^2_r \to \infty$ (left) from Eq. \eqref{stressChoice1} and for $e^2 \sim \ln \CB$ (right) from Eq. \eqref{stressChoice2}. }
\label{fig:unphysical}
\end{figure}

In all to us known literature, the unavoidable choice of the constant $\CC$, which sets $\bar\alpha$, is made in a different way. $\CC$ is either chosen so that the logarithmic terms vanish altogether, or so that it sets the UV scale to that of the magnetic field, which is convenient when studying strong magnetic fields as e.g. in \cite{Fuini:2015hba,Janiszewski:2015ura}. Here, we wish to point out some of the consequences of setting $\CC$ to either of the two standard options. The first option, which eliminates the logarithmic terms, results in the following thermodynamics quantities: 
\begin{align}\label{stressChoice1}
\varepsilon = \frac{N_c^2}{2\pi^2}\left( -\frac{3}{4}f^b_4 r_h^4 \right) \,,&& p =\frac{N_c^2}{2\pi^2}\left[ \left(-\frac{1}{4}f^b_4 + \frac{v^b_4}{v} \right)r_h^4-\frac{\CB^2}{4}\right],&& \mu\rho =\frac{N_c^2}{2\pi^2} \left( \frac{3v^b_4}{v} r_h^4-\frac{\CB^2}{4} \right).
\end{align}
The second choice results in 
\begin{equation}\label{stressChoice2}
\begin{aligned}
\varepsilon &= \frac{N_c^2}{2\pi^2}\left( -\frac{3}{4}f^b_4r_h^4 + \frac{\CB^2}{4}\ln \CB \right), && p =\frac{N_c^2}{2\pi^2}\left[\left( -\frac{1}{4}f^b_4 + \frac{v^b_4}{v} \right)r_h^4-\frac{\CB^2}{4} + \frac{\CB^2}{4}\ln \CB\right], \\
\mu\rho &= \frac{N_c^2}{2\pi^2}\left( \frac{3v^b_4}{v} r_h^4-\frac{\CB^2}{4} -\frac{\CB^2}{4}\ln\CB\right).
\end{aligned}
\end{equation}
While these two renormalisation conditions are suitable for studying certain physical setups involving static electromagnetic fields, we claim that they lead to unphysical results when the boundary $U(1)$ gauge field is dynamical. By comparing the renormalised stress-energy tensor \eqref{THol1}--\eqref{THol3} to expressions in \eqref{stressChoice1} and \eqref{stressChoice2}, we find that the two choices correspond to the renormalised coupling being $e_r^2 \to \infty$ and $e_r^2 \sim \ln \CB$, respectively. An infinite $U(1)$ coupling is unphysical in a plasma state. The problem with the second choice is that if extrapolated to the weak-field regime, $\ln \CB / M$, where $M$ is some scale, can become negative and $e_r$ imaginary, which is again unphysical. Thus, these choices may lead to instabilities and superluminal propagation, which were absent from our results with $\bar\alpha$ near $1/137$. We plot the Alfv\'{e}n speed parameter $\CV_A$ for the two couplings from \eqref{stressChoice1} and \eqref{stressChoice2} in Fig. \ref{fig:unphysical}. 

\section{Discussion}\label{sec:Discussion}

This work is the first holographic study of states with generalised global (higher-form) symmetries. Moreover, it is the first step in a long road to a better understanding of magnetohydrodynamics in plasmas outside of the regime of validity of standard MHD, be it in the presence of strong magnetic fields or in a strongly interacting (or dense) plasma with a complicated equation of state and transport coefficients---all claimed to be describable within the recent (generalised global) symmetry-based formulation of MHD of Ref. \cite{Grozdanov:2016tdf}. In order to supply a hydrodynamical theory of MHD with the necessary microscopic information of a strongly coupled plasma, we resorted to the simplest, albeit experimentally inaccessible option: holography. Nevertheless, our hope is that in analogy with the myriad of works on holographic conformal hydrodynamics, which have led to important new insights into strongly interacting realistic fluids, holography can also help us understand observable MHD states in the presence of strong fields, high density and of strongly interacting gauge theories, such as QCD.  

With this view, we constructed the simplest theory dual to the operator structure and Ward identities used in MHD of \cite{Grozdanov:2016tdf}, investigated the relevant aspects of the holographic dictionary and used it to compute the equation of state and transport coefficients of the dual plasma state. This information was then used to analyse the dependence of MHD waves---Alfv\'{e}n and magnetosonic waves---on tuneable parameters specifying the state: the strength of the magnetic field, temperature, the angle between momentum of propagation and the equilibrium magnetic field direction, as well as the strength of the $U(1)$ electromagnetic gauge coupling. We believe that the latter feature of our model---dynamical electromagnetism on the boundary---which in the (dual) language of two-form gauge fields in the bulk allows for standard (Dirichet) quantisation, could in its own right be used for holographic studies of $U(1)$-gauged systems, unrelated to MHD. 

Our results have revealed several new qualitative features of MHD waves, particularly in the regime of a strong magnetic field, which is inaccessible to standard MHD methods. Various properties of the equation of state, transport coefficients and dispersion relations found here, may now be compared to those in experimentally realisable plasmas, or at the least, used as a toy model for future studies of MHD. Approximate scalings in the limiting regimes of large and small $T / \sqrt{\CB}$ are collected in Tables \ref{table:EOS} and \ref{table:TC}. Here, we summarise some of the most interesting observations:
\\

$\bullet$ The equation of state and transport coefficients strongly depend on the strength of the magnetic field, i.e. on whether the plasma is in the weak-field, the crossover, or the strong-field regime. 

$\bullet$ In the weak-field regime with $T / \sqrt{\CB} \gg 1$, the system is well-described by standard MHD (see \cite{Hernandez:2017mch} for a full description) with small resistivities (large conductivity regime, which is assumed by ideal Ohm's law) and small effects of anisotropy. As $T/\sqrt{\CB} \to \infty$, the plasma becomes an uncharged, conformal fluid with a single independent transport coefficient, $\eta = s/ 4\pi$. In the strong-field limit of $T / \sqrt{\CB} \to 0$, the plasma limits to a non-dissipative regime with all first-order transport coefficients (along with sound attenuations and diffusion constants) tending to zero. Effects of anisotropy are large. 

$\bullet$ Resistivities have a global maximum in the intermediate $T/\sqrt{\CB}$ regime, which indicates a regime of least conductive plasma. If the assumptions of standard MHD are correct at $T / \sqrt{\CB} \gg 1$ and the symmetry-based predictions of \cite{Grozdanov:2016tdf} are correct at $T / \sqrt{\CB} \ll 1$, such a regime should be generically exhibited by any plasma. 

$\bullet$ Out of the three bulk viscosities, $\zeta_\perp$, $\zeta_\parallel$ and $\zeta_\times$, only one is independent and they saturate the positivity of the entropy production inequality, i.e. they are related by $\zeta_\perp \zeta_\parallel = \zeta_\times^2$. One may speculate on how general this result is and whether it is related to the suppression of entropy production at strong coupling \cite{Grozdanov:2014kva,Haehl:2015pja} or perhaps some form of (holographic) universality at infinite (or strong) coupling.

$\bullet$ Various qualitative features of slow and fast magnetosonic modes are exchanged in the weak- and strong-field regimes (usually at small angles, $\theta$, between momentum and the equilibrium magnetic field direction), such as their asymptotic tendency to the speed of Alfv\'{e}n waves and the strength of sound attenuation.

$\bullet$ For a finite momentum, propagating Alfv\'{e}n and slow magnetosonic modes (sound modes to $\CO(k^2)$) transmute into pairs of non-propagating, diffusive (to $\CO(k^2)$) modes. This occurs at large angles between the direction of momentum propagation and the equilibrium magnetic field, $\theta_c < \theta \leq \pi/2$, where $\theta_c$ is some momentum- and $T/\sqrt{\CB}$-dependent critical angle (cf. Eq. \eqref{ThetaC} for Alfv\'{e}n waves).

$\bullet$ The phenomenology of MHD modes strongly depends on the strength of the electromagnetic coupling (or the position of the Landau pole) and can, for large ranges of the coupling, lead to unstable or superluminal propagation.
\\ 

Beyond the types of waves studied in this work, it would be particularly interesting to better understand the role of finite charge density, as studied in \cite{Hernandez:2017mch}, within the formalism of \cite{Grozdanov:2016tdf}. The important question then is how the phenomenology of such MHD waves, which typically experience gapped propagation and instabilities (e.g. the infamous Weibel instability), becomes altered by strong interactions, strong fields and for more `exotic' field content.   
 
Finally, the holographic setup studied here will need to undergo extensive further tests and analyses in order to unambiguously establish its connection to plasma physics and MHD. In particular, it is essential to study the quasinormal spectrum of the theory to verify that the hydrodynamic modes indeed describe the small-$\omega$ and small-$k$ expansion of the leading infrared poles. Furthermore, it will be interesting to understand the role of higher-frequency spectrum and its interplay with MHD modes. We leave all these and many other interesting questions to the future.

\acknowledgments{The authors would like to thank Debarghya Banerjee, Pavel Kovtun, Alexander Krikun, Chris Rosen, Koenraad Schalm, Andrei Starinets, Giorgio Torrieri, Vincenzo Scopelliti, Phil Szepietowski and Jan Zaanen for stimulating discussions, and Simon Gentle for his comments on the draft of this paper. We are also grateful to Diego Hofman and Nabil Iqbal for numerous discussions on the topic of this work, comments on the manuscript and for sharing the draft of \cite{Hofman:2017Something} prior to publication. S. G. is supported in part by a VICI grant of the Netherlands Organisation for Scientific Research (NWO), and by the Netherlands Organisation for Scientific Research/Ministry of Science and Education (NWO/OCW). The work of N. P. is supported by the DPST scholarship from the Thai government and by Leiden University. }

\appendix

\section{Kubo formulae for first-order transport coefficients}\label{appendix:kubo}

In this appendix, we outline the derivation of the Kubo formulae that have been used to compute the seven first-order transport coefficients of \eqref{MHDstress-energy} (or Eqs. \eqref{visc1}--\eqref{sdef}) in Section \ref{transport-maintext} \cite{Grozdanov:2016tdf,Hernandez:2017mch}. We derive the Kubo formulae by using the variational background field method (see e.g. \cite{Kovtun:2012rj} for a review), which amounts to varying the background metric $g_{\mu\nu}$ and background two-form gauge field $b_{\mu\nu}$, sourcing $T^{\mu\nu}$ and $J^{\mu\nu}$, by writing
\begin{align}
g_{\mu\nu} \to \eta_{\mu\nu} + \int \frac{d\omega}{2\pi} e^{-i\omega t} \delta h_{\mu\nu}(\omega),&& b_{\mu\nu} \to b^{\text{eq}}_{\mu\nu} + \int \frac{d\omega}{2\pi} e^{-i\omega t} \delta b_{\mu\nu}(\omega) \,,
\end{align}
where $\delta h_{\mu\nu}$ and $\delta b_{\mu\nu}$ are small variation, $\eta_{\mu\nu}$ is the flat Minkowski metric and $b^{\text{eq}}_{\mu\nu} = 0$ (no external equilibrium source). These variations of the background fields can be viewed as sources that generate variations of the hydrodynamic variables $T$, $\rho$ (which we use here instead of $\mu$ in \cite{Grozdanov:2016tdf}), $u^\mu$ and $h^\mu$: 
\begin{align}
T(t) \to T + \delta T(t)\,, && \rho \to \rho+\delta \rho(t)\,,&& u^\mu \to u^\mu_\text{eq} +\delta u(t)\,,&& h^\mu \to h^\mu_\text{eq} + \delta h^\mu(t)\,,
\end{align}
where we choose the equilibrium configuration to be $u^\mu_\text{eq} = \delta^\mu_t$ and $h^\mu_\text{eq} =\delta^\mu_z$. The normalisation and orthogonality conditions for the two vectors ($u_\mu u^\mu = -1$, $h_\mu h^\mu = 1$, $u_\mu h^\mu = 0$) imply
\begin{align}
\delta u^t = \frac{1}{2}\delta h_{tt}\,, && \delta h^t = \delta u^z +\delta h_{tz}\,, && \delta h^z = -\frac{1}{2}\delta h_{zz} \,.
\end{align}
After writing $ \delta T$, $\delta \rho$, $\delta u^\mu$ and $\delta h^\mu$ in terms of $\delta h_{\mu\nu}$ and $\delta b_{\mu\nu}$, we can insert these solution into 
\begin{align}
\CT^{\mu\nu} \equiv \sqrt{-g} \, \la T^{\mu\nu}\ra\vert_{g,b}\,,&& \CJ^{\mu\nu} \equiv \sqrt{-g} \, \la J^{\mu\nu}\ra\vert_{g,b}\,, 
\end{align}
which give
\begin{equation}
\begin{aligned}
\text{Im}\,\CT^{xx}+\text{Im}\,\CT^{yy}&= \omega \zeta_\perp (\delta h_{xx}+\delta h_{yy}) +  \omega \zeta_\times^{(1)}\delta h_{zz}+\CO(\omega^2,\delta h^2,\delta b^2)\, ,\\
\text{Im}\, \CT^{zz} &= \frac{1}{2}\omega \zeta_\parallel \delta h_{zz} + \frac{1}{2}\omega\zeta_\times^{(2)} \left( \delta h_{xx} + \delta h_{yy} \right)+\CO(\omega^2,\delta h^2,\delta b^2),\\
\text{Im}\, \CT^{xy} &= \omega \eta_\perp \delta h_{xy}+\CO(\omega^2,\delta h^2,\delta b^2),\\
\text{Im}\, \CT^{xz} &= \omega \eta_\parallel \delta h_{xz}+\CO(\omega^2,\delta h^2,\delta b^2),\\
\text{Im}\, \CJ^{xy} &= 2\omega r_\parallel \delta b_{xy}+\CO(\omega^2,\delta h^2,\delta b^2),\\
\text{Im}\, \CJ^{xz} &= 2\omega r_\perp \delta b_{xz}+\CO(\omega^2,\delta h^2,\delta b^2) \,,\\
\end{aligned}
\label{practicalKubo}
\end{equation}
where we have not imposed the Onsager relation equating $\zeta^{(1)}_\times$ with $\zeta^{(2)}_\times$ \cite{Grozdanov:2016tdf,Hernandez:2017mch}. By using the linear response formulae relating the variations of one-point functions to retarded two-point Green's functions,
\begin{equation}
\begin{aligned}
\delta \CT^{\mu\nu}(\omega,\mathbf{k}) &= -\frac{1}{2} G^{\mu\nu,\lambda\sigma}_{TT}(\omega,\mathbf{k}) \delta h_{\lambda\sigma}(\omega,\mathbf{k})-\frac{1}{2} G^{\mu\nu,\lambda\sigma}_{TJ}(\omega,\mathbf{k}) \delta b_{\lambda\sigma}(\omega,\mathbf{k}) +\CO(\delta h^2,\delta b^2),\\
\delta \CJ^{\mu\nu}(\omega,\mathbf{k}) &= -\frac{1}{2} G^{\mu\nu,\lambda\sigma}_{JT}(\omega,\mathbf{k}) \delta h_{\lambda\sigma}(\omega,\mathbf{k})- G^{\mu\nu,\lambda\sigma}_{JJ}(\omega,\mathbf{k}) \delta b_{\lambda\sigma}(\omega,\mathbf{k}) +\CO(\delta h^2,\delta b^2),
\end{aligned}
\end{equation}
it is then easy to extract the relevant Kubo formulae for the seven transport coefficients \cite{Grozdanov:2016tdf,Hernandez:2017mch}, which we used in this work:
\begin{align}
& \eta_{\parallel} = \lim_{\omega \to 0} \frac{G_{TT}^{xz,xz}(\omega,0) }{-i\omega}\, , & \eta_{\perp} = \lim_{\omega \to 0} \frac{G_{TT}^{xy,xy}(\omega,0)}{-i\omega} \,,   \label{KuboShVis} \\
& \zeta_{\parallel} = \lim_{\omega \to 0} \frac{G_{TT}^{zz,zz}(\omega,0) }{- i\omega} \,, & \zeta_{\perp} + \eta_{\perp} = \lim_{\omega \to 0} \frac{G_{TT}^{xx,xx}(\omega,0)}{-i\omega} \, , 
\end{align}
as well as 
\begin{align}
\zeta_{\times} = \lim_{\omega \to 0} \frac{G_{TT}^{zz,xx}(\omega,0)}{-i\omega} = \lim_{\omega \to 0}\frac{G_{TT}^{xx,zz}(\omega,0)}{-i\omega} \, . \label{KuboBulkVisX}
\end{align}
and 
\begin{align}
r_{\parallel} = \lim_{\omega \to 0} \frac{G_{JJ}^{xy,xy}(\omega,0) }{-i\omega}\, , && r_{\perp} = \lim_{\omega \to 0} \frac{G_{JJ}^{xz,xz}(\omega,0)}{-i\omega} \,. \label{resKubo}
\end{align}

\section{Further details regarding the derivation of the transport coefficients}\label{appendix:transport}

Here, we show the details of the derivation of horizon formulae for all remaining transport coefficient: $\eta_{\perp}$, $\eta_{\parallel}$, $\zeta_\perp$, $\zeta_\parallel$, $\zeta_\times$ and $r_\parallel$. The computations are analogous to the calculation of $r_\perp$ in Section \ref{transport-maintext}.

\begin{itemize}
\item[(i)] \textit{Shear viscosity $\eta_\perp$}
\end{itemize}

The only relevant bulk fluctuation for $\eta_\perp$ is $\delta G_{xy}$ with the equation of motion 
\begin{equation}
\delta {G^y_{~x}}'' + \left( \frac{3}{2u} + \frac{F'}{F} + 2\CV' + \CW' \right) \delta {G^y_{~x}}' + \frac{\omega^2}{4r_h^2u^3F^2}\delta G^y_{~x} = 0\,.
\end{equation}
The solution to leading order in the frequency $\omega$ can be found analytically and its near-boundary expansion gives 
\begin{equation}
\delta G^y_{~x} = \delta h_{xy} \left( 1+\frac{i\omega u^2}{ 4r_h v\sqrt{w}} + \CO(u^3)\right),
\end{equation}
where $\delta h_{xy}$ sets the Dirichlet boundary condition and is the boundary theory source. If we plug this solution into to the stress-energy tensor, we find that 
\begin{equation}
\begin{aligned}
\la \delta T^{xy}\ra &= \frac{N_c^2}{2\pi^2}\left( \frac{r_h^4 e^{2\CV}\sqrt{uF}}{2v} \delta {G^y_x}' \right) + \ldots \\
&= \frac{N_c^2}{2\pi^2}\left( \frac{i\omega r_h^3}{4v\sqrt{w}} \right) \delta h_{xy} +\ldots \, .
\end{aligned}
\end{equation}
Using Eq. \eqref{practicalKubo}, we find that
\begin{align}
\eta_\perp = \frac{N_c^2}{2\pi^2} \left( \frac{r_h^3}{4v\sqrt{w}}\right) = \frac{1}{4\pi} s \,,
\end{align}
as stated in Eq. \eqref{horizonFormulae}.

\begin{itemize}
\item[(ii)] \textit{Shear viscosity $\eta_\parallel$}
\end{itemize}

Similarly to the computation of $r_\perp$, the $xu$-component of the two-form gauge field fluctuation equation can be used to reduce the two coupled second-order differential equations coupling $\delta G_{xz}$ and $\delta B_{tx}$ to a single equation:
\begin{equation}
{\delta G^z_{~x}}'' + \left( \frac{3}{2u} + \frac{F'}{F} + 3\CW'\right) {\delta G^z_{~x}}' + \frac{\omega^2}{4r_h^2u^3F^2} \delta G^z_{~x} = 0\, .
\end{equation}
The solution to linear order in $\omega$ can again be found analytically and in the near-boundary region yields
\begin{equation}
\delta G^z_{~x} = \delta h^z_{~x} \left( 1+ \frac{i\omega}{4r_h w^{3/2}} u^2 + \CO(u^3)\right).
\end{equation}
The relevant component of the stress-energy tensor is then
\begin{equation}
\la T^{xz}\ra = \frac{N_c^2}{2\pi^2} \left( \frac{i\omega r_h^3}{4w^{3/2}}\right)\delta h_{xz} + \ldots\,,
\end{equation}
which gives
\begin{align}
\eta_\parallel = \frac{N_c^2}{2\pi^2} \left( \frac{r_h^3}{4w^{3/2}} \right) = \frac{1}{4\pi} \frac{v}{w} s \,,
\end{align}
as stated in Eq. \eqref{horizonFormulae}.

\begin{itemize}
\item[(iii)] \textit{Resistivity $r_\parallel$}
\end{itemize}

The only equation of motion in this channel is 
\begin{equation}
\delta B_{xy}'' + \left( \frac{3}{u} + \frac{F'}{F}-2\CV' + \CW'\right)\delta B_{xy}' + \frac{\omega^2}{4r_h^2u^3F^2}\delta B_{xy} = 0\,,
\end{equation}
which leads to the near-boundary solution
\begin{equation}
\delta B_{xy} = \delta B^{(0)}_{xy} \left( 1+ \frac{i\omega v}{2r_h \sqrt{w}} \ln\, u + \CO(u)\right) .
\end{equation}
The two-form current can then be written as 
\begin{equation}
\la \delta J^{xy}\ra = \frac{2\pi^2}{N_c^2} \left( \frac{2i\omega v}{r_h\sqrt{w}} \right) \delta b_{xy} + \CO(\omega^2) \,,
\end{equation}
which yields 
\begin{align}
r_\parallel =  \frac{2\pi^2}{N_c^2} \left(  \frac{v}{r_h\sqrt{w}} \right) ,
\end{align}
as stated in Eq. \eqref{horizonFormulae}.

\begin{itemize}
\item[(iii)] \textit{Bulk viscosities $\zeta_\perp$, $\zeta_\parallel$ and $\zeta_\times$}
\end{itemize}

By counting the number of the relevant degrees of freedom, it turns out that there is only one dynamical mode in this decoupled system coming from $4\,\times\,$(2$^{\text{nd}}$-order ODE's for $\delta g_{tt}, \delta g_{aa}, \delta g_{zz},\delta b_{tz}$) $-$ $3\,\times\,$(1$^\text{st}$-order ODE's for $\delta g_{tu}, \delta g_{uu}, \delta b_{zu}$). To find the dynamical mode, we start by solving the algebraic equations for $\delta g_{tu}$, $\delta g_{uu}$ and $\delta b_{zu}$ from the $tu$ and $uu$ components of Einstein's equations combined with the $zu$ component of Maxwell's equations. Plugging these solutions into the four second-order equations involving $\delta g_{tt}$, $\delta g_{aa}$, $\delta g_{zz}$ and $\delta b_{tz}$, we find that the remaining two non-trivial equations involve only $\delta g_{aa}$ and $\delta g_{zz}$. The single resulting equation of motion can then be expressed in terms of the gauge-invariant variable $Z_s(u)$ defined as 
\begin{equation}
Z_s(u) = \delta G^a_{~a} -\frac{2\CV'}{\CW'} \delta G^z_{~z} \,,
\end{equation}
where $\delta g_{aa} = \delta g_{xx} + \delta g_{yy}$. The equation of motion for $Z_s$ can be written 
\begin{equation}
Z_s''(u) + C_1(\omega,u) Z_s'(u) + C_2(\omega,u) Z_s(u) = 0 \,,
\end{equation}
where 
\begin{equation}
\begin{aligned}
C_1  &= \frac{3}{2u} + \frac{F'}{F} + \frac{2\CW''}{W'} + 2\CV'+\CW' - 2 \left(\frac{2\CV''+ \CW''}{2\CV' + \CW'}\right), \\
C_2 &= -\frac{b^2 e^{-4\CV}}{3u^3F\CW'}\left( \frac{F'}{F} + 4\CW'\right) +
\frac{\omega^2}{4 r_h^2u^3 F^2}-\frac{2F'^2}{3F^2\CW'}(\CV'-\CW') +
\frac{4\CV'F' (\CV'-\CW')^2}{3F \CW'(2\CV'+\CW')}\\
&\quad + \frac{8 \CV'^2(\CV'+2\CW')(\CV'-\CW')}{2\CW' (2\CV'+\CW')}\,.
\end{aligned}
\end{equation}

Now, suppose that the time-independent solution for $Z_s$ is $\mathfrak{Z}^{(-)}$, so that $\mathfrak{Z}^{(-)}(u\to 0) =Z^{(0)} \equiv \delta h_{aa} -2 \delta h_{zz}$ (note that $\CV'/\CW' \to 1$ and $ u\to 0$). The second solution, denoted as $\mathfrak{Z}^{(+)}$, contains the time-dependent information and can be found from the Wronskian
\begin{align}
\mathfrak{Z}^{(+)}(u) = \mathfrak{Z}^{(-)}(u) \int_u^1 du'\, \frac{
W_R (u')}{ \left(\mathfrak{Z}^{(-)}(u')\right)^{2}} \, ,&&  W_R = \left( \frac{2\CV'+\CW'}{\CW'}\right)^2\frac{e^{2\CV+\CW}}{u^{3/2}F} \,.
\end{align}
We then find that the near-boundary and the near-horizon expansions for $\mathfrak{Z}^{(+)}$ are
\begin{equation}
\mathfrak{Z}^{(+)} = \begin{dcases}
\frac{9}{2v\sqrt{w}}\left[ \mathfrak{Z}^{(-)}(0)\right]^{-1}u^2 + \CO(u^3),& \text{near}\;\;
u\to 0 \,,\\
-9 r_h \left( \frac{6+B^2}{6-B^2} \right)^2\left[2\pi T\mathfrak{Z}^{(-)}(1) \right]^{-1}\ln (1-u) +\CO(1-u),  &\text{near}\;\;
u\to 1\,.
\end{dcases}
\label{eqn:asymptZ}
\end{equation}
The full solution is a linear combination, $Z_s(u) = \mathfrak{Z}^{(-)} + \alpha \mathfrak{Z}^{(+)}$, and the ingoing boundary condition sets the frequency-dependent function $\alpha(\omega)$ to be  
\begin{equation}
\alpha(\omega) = \frac{i\omega}{2r_h} \left(\frac{(6+B^2)}{3(6-B^2)}\right)^2\left[\mathfrak{Z}^{(-)}(1) \right]^2 ,
\end{equation}
which allows us to write the solution for $Z_s$ near the boundary as 
\begin{equation}
Z_s = Z^{(0)} \left( 1+ \frac{i\omega}{4 r_h v\sqrt{w}}\left( \frac{6+B^2}{6-B^2}\right)^2\left[ \frac{\mathfrak{Z}^{(-)}(1)}{\mathfrak{Z}^{(-)}(0)} \right]^2 u^2  \right) + \ldots \, .
\end{equation}

This expression can then be used to compute the bulk viscosities, for which we follow the approach by \cite{Edalati:2010pn} and their analysis of the Green's function in the sound channel. In summary, we first find the expression for $\la \delta T^{xx}+\delta T^{yy}\ra$ and $\la \delta T^{zz}\ra$ in terms of $\delta h_{tt}$, $\delta h_{aa}$, $\delta h_{zz}$ and $\delta b_{tz}$, and then relate the near-boundary data of the bulk modes $\delta G_{tt}$, $\delta G_{aa}$, $\delta G_{zz}$ and $\delta B_{tz}$ to those of $Z_s$. Then, we impose the radial gauge, $\delta G_{u\mu}=0$ and $\delta B_{u\mu}=0$, and solve the equations of motion near the boundary. The first-order equations of motion give the following relations:
\begin{equation}\label{radialGaugeConstrains}
\begin{aligned}
h_{aa}^{(2)} + h_{tt}^{(2)} + h_{zz}^{(2)} + \frac{B^2h_{aa}^{(0)}}{36 v^2} &= 0\, ,\\
2 \left( h_{aa}^{(2)} + h_{zz}^{(2)}\right) +\frac{v_4^b}{v} \left(h^{(0)}_{aa}-2h^{(0)}_{zz}\right) - f^b_4
\left(h_{aa}^{(0)}+ h_{zz}^{(0)}\right) &=0\,.
\end{aligned}
\end{equation}
The coefficients $h_{\mu\nu}^{(n)}$ are defined through the near-boundary expansion of the metric fluctuation. By using the second-order dynamical equation and the radial gauge, the solutions are 
\begin{equation}\label{radialGaugeNbExpansion}
\begin{aligned}
\delta G^a_{~a} &= h^{(0}_{aa} + h_{aa}^{(2)} u^2 + \frac{h_{aa}^{(0)} B^2}{10 v^2}u^2 \ln
u + \CO(\omega^2,u^3),\\
\delta G^{t}_{~t} &= h^{(0)}_{tt} + h_{tt}^{(2)} u^2 - \frac{h_{aa}^{(0)}B^2}{20
  v^2}u^2\ln u +\CO(\omega^2,u^3),\\
\delta G^z_{~z} &= h_{zz}^{(0)} + h_{zz}^{(2)}u^2 - \frac{h_{aa}^{(0)}B^2}{20
  v^2}u^2\ln u +\CO(\omega^2,u^3),
\end{aligned}
\end{equation}
where $h_{\mu\nu}^{(0)} \equiv \delta h_{\mu\nu}$ is the metric perturbation used throughout the paper. By combining Eqs. \eqref{radialGaugeConstrains} and \eqref{radialGaugeNbExpansion}, and using the definition of the gauge-invariant mode $Z_s$, we find that 
\begin{align}
Z_s = Z^{(0)} + \frac{Z^{(0)} \omega^2}{6 r_h^2} + Z^{(2)}u^2 +
\frac{h_{aa}^{(0)}B^2}{5v^2} u^2 \ln u +\CO(\omega^4) \,,
\end{align}
where 
\begin{align}
Z^{(0)} = h_{aa}^{(0)} - 2 h_{zz}^{(0)},&& Z^{(2)} = -3
h_{zz}^{(2)}-\frac{v^b_4}{v}Z^{(0)}+ f^b_4 \left(h_{aa}^{(0)}+h_{zz}^{(0)}\right) \,.
\end{align}
It is most convenient to extract the transport coefficients from $\la \delta T^{zz}\ra $: 
\begin{equation*}
\begin{aligned}
\la \delta T^{zz}\ra  &= -\frac{N_c^2}{2\pi^2}
\frac{r_h^4e^{2\CW}}{w}\left(\frac{1}{2}\sqrt{uF}\left({\delta G^a_a}' + {\delta G^t_t}' \right)+ \left( \frac{3}{2u} + \frac{\sqrt{u}F'}{2F} +2 \sqrt{uF}\CV' \right)\delta G^z_z\right)+ \ldots \\
&=\frac{N_c^2}{2\pi^2}r_h^4h_{zz}^{(2)}  + \ldots \\
&= -\frac{N_c^2}{2\pi^2}\left( \frac{i\omega r_h^3}{12 v\sqrt{w}} \left( \frac{6+B^2}{6-B^2}\right)^2
  \left[\mathfrak{Z}^{(-)}(1)/\mathfrak{Z}^{(-)}(0) \right]^2 \right)\left(\delta
  h_{aa}-2h_{zz} \right) + \ldots\,.
\end{aligned}
\end{equation*}
Using the Kubo formula \eqref{practicalKubo}, we find that 
\begin{equation}
\begin{aligned}
\zeta_\parallel &=\frac{N_c^2}{2\pi^2}\left( \frac{r_h^3}{3 v\sqrt{w}} \left( \frac{6+B^2}{6-B^2}\right)^2
  \left[\mathfrak{Z}^{(-)}(1)/\mathfrak{Z}^{(-)}(0) \right]^2 \right)\\ &= \frac{s}{4\pi} \left( \frac{4}{3} \left( \frac{6+B^2}{6-B^2}\right)^2
  \left[\mathfrak{Z}^{(-)}(1)/\mathfrak{Z}^{(-)}(0) \right]^2 \right),
\end{aligned}
\end{equation}
and $\zeta_\times^{(2)} = - \zeta_\parallel / 2$. 

Similarly, we can extract $\zeta_\perp$ and $\zeta^{(1)}_\times$ from 
\begin{equation*}
\begin{aligned}
\la \delta T^{xx}\ra + \la \delta T^{yy}\ra  =& - \frac{N_c^2}{2\pi^2} 
\frac{r_h^4e^{2\CV}}{v}      \left[ \frac{1}{2}\sqrt{uF}\left({\delta G^a_a}' + {\delta G^t_t}' +{\delta G^z_z}' \right) \right.\\
& \left. + \left( \frac{3}{2u} + \frac{\sqrt{u}F'}{2F} + \sqrt{uF}(\CV' +\CW')\right)\delta G^z_z\right] + \ldots \\
=&\, \frac{N_c^2}{2\pi^2}\left( \frac{i\omega r_h^3}{12 v\sqrt{w}} \left( \frac{6+B^2}{6-B^2}\right)^2
  \left[\mathfrak{Z}^{(-)}(1)/\mathfrak{Z}^{(-)}(0) \right]^2 \right)\left(\delta
  h_{aa}-2h_{zz} \right) + \ldots\,,
\end{aligned}
\end{equation*}
which gives $\zeta_\perp = \zeta_\parallel  / 4= - \zeta_\times^{(1)} / 2 $. Hence, we find that $\zeta_\times^{(1)} = \zeta_\times^{(2)}$, which is the manifestation of the Onsager relation imposed in \cite{Grozdanov:2016tdf,Hernandez:2017mch}. This completes the derivation of expressions stated in Eq. \eqref{horizonFormulae}.

As a simple check of our results, one can show that in the zero magnetic field limit, 
\begin{equation}\label{Check1}
\zeta_\parallel = -\lim_{\omega\to 0}\partial_\omega \text{Im}G^{zz,zz}_{TT}(\omega,0)  =-\frac{4}{3}\lim_{\omega\to 0}\partial_\omega \text{Im}G^{xy,xy}_{TT}(\omega,0) = \frac{4}{3} \eta \,,
\end{equation}
which is consistent with the fact that as $\CB \to 0$, our plasma should become described by conformal hydrodynamics. By using standard relations between two-point functions in a neutral CFT fluid, \eqref{Check1} is equivalent to the statement that bulk viscosity vanishes in conformal relativistic hydrodynamics (see e.g. \cite{Kovtun:2012rj}).\footnote{For a neutral relativistic fluid, one can show that $\text{Im} \la\delta T^{xx}\ra+\text{Im}\la\delta T^{yy}\ra = \omega \left( \frac{\eta}{3}+\zeta \right)\delta h_{aa} + \omega \left(  \zeta- \frac{2}{3}\eta \right)\delta h_{zz} + \ldots\,$, and that $\text{Im}\la \delta T^{zz}\ra = \frac{1}{2}\omega \left(\zeta - \frac{2}{3}\eta  \right)\delta h_{aa} + \omega \left( \frac{2}{3}\eta+\frac{1}{2}\zeta \right) \delta h_{zz} +\ldots\,$. In a conformal fluid with $\zeta = 0$, one therefore finds that $\lim_{\omega\to 0}\frac{1}{2}\partial_\omega \text{Im}G^{aa,aa}_{TT}(\omega,0) =\lim_{\omega \to 0} \frac{1}{4} \partial_\omega \text{Im}G^{zz,zz}_{TT}(\omega,0)= -\eta/3$. The relation $ \zeta_\perp = \zeta_\parallel/4$ arises from equations $\lim_{\omega\to0}\partial_\omega \text{Im}G^{aa,aa}_{TT}(\omega,0) = -2\zeta_\perp$ and $\lim_{\omega\to0}\partial_\omega \text{Im}G^{zz,zz}_{TT}(\omega,0) = -\zeta_\parallel$ (see Eq. \eqref{practicalKubo}).} For another check, one can write the relation $\zeta_\parallel = 4 \,\zeta_\perp$ in the language of two-point functions and obtain the relation $\lim_{\omega \to 0}\left[\partial_\omega G^{aa,aa}_{TT}(\omega,0) - \frac{1}{2} \partial_\omega G^{zz,zz}_{TT}(\omega,0)\right] = 0$, which is also satisfied by conformal relativistic hydrodynamics. Interestingly, this relation holds for all strengths of the magnetic field in the model studied in this work. 

\section{Dispersion relations of magnetosonic waves}\label{appendix:magnetosonicSpectrum}

In the magnetosonic channel, the polynomial equation in $\omega$ and $k$, which needs to be solved in order for us to find the dispersion relations $\omega(k)$ is a quartic equations in $\omega$, which can be written in the following form:
\begin{equation}\label{DetMS}
\text{Det}\left[-i\omega \mathds{1}+\mathds{M}\right] = 0 \,,
\end{equation}
with $\mathds{1}$ the $4\times4$ identity matrix and the non-zero components $M_{ij}$ of the matrix $\mathds{M}$ given by
\begin{equation}
\begin{aligned}
M_{11} &= r_\perp k^2 \sin^2\theta \mathcal{A}_{11}\,, && M_{12} = -r_\perp k^2 \CA_{12}\,, &&
M_{13} = ik \sin\theta \CA_{13} \,, && M_{14}=i k \frac{s\cos\theta}{\chi_{11}} \,,\\
M_{21} &= -r_\perp k^2\sin^2\theta \CA_{21}\, , && M_{22} = r_\perp k^2\CA_{22} \,, && M_{23} = ik \rho\sin\theta \,,\\
M_{31} &= ik\sin\theta \CA_{31}\,, && M_{32} = ik \CA_{32}\,, &&M_{33} = \CA_{33} k^2, && \CM_{34} = \eta_\perp k^2 \CA_{34} \,,\\
M_{41} &= i\frac{k}{T}\cos\theta\,, && M_{43} = \eta_\perp k^2 \CA_{43}\,, &&M_{44} =k^2\CA_{44} \,.
\end{aligned}
\end{equation}
The coefficients $\CA_{ij}$ are
\begin{equation}\label{eq:coCA}
\begin{aligned}
\CA_{11} &= \frac{1}{2T^2\chi_{11}}(\mu+T\chi_{12})(\mu-T\mu_{21}), &&
\CA_{12} = \frac{1}{2T\rho\chi_{11}}(\mu+T\chi_{12})(\mu\cos^2\theta + \rho\chi_{22}\sin^2\theta),\\
\CA_{13} &= \frac{s-\rho\chi_{12}}{\chi_{11}} \,, &&
\CA_{21} = \frac{\mu-T\chi_{21}}{2T}\,,\\
\CA_{22} &= \frac{1}{2\rho}\left( \mu\cos^2\theta + \rho\chi_{22}\sin^2\theta\right),&&
\CA_{31} = \frac{s+\rho\chi_{21}}{\varepsilon+p}\,,\\
\CA_{32} &= \frac{2\rho}{(\varepsilon+p)\sin\theta}\CA_{22} \,,&&
\CA_{33} =\left( \frac{\eta_\parallel \cos^2\theta +(\eta_\perp+\zeta_\perp)\sin^2\theta}{\varepsilon+p} \right),\\
\CA_{34} &=\frac{\cos\theta\sin\theta}{\varepsilon+p}\,,&&
\CA_{43} = \frac{\varepsilon+p}{sT} \CA_{34}\,,\\
\CA_{44} &= \frac{2\zeta_\parallel\cos^2\theta + \eta_\parallel\sin^2\theta}{sT} \,.
\end{aligned}
\end{equation}
By computing the determinant in \eqref{DetMS}, the resulting quartic equation is
\begin{equation}\label{MSQuartic}
\omega^4+ c_3  \omega^3 + c_2  \omega^2 + c_1 \omega + c_0 = 0 \,,
\end{equation}
where $c_i$ are functions of thermodynamics quantities, transport coefficients, $k$ and $\theta$. The expressions for $c_i$ in terms of $\CA_{ij}$ in \eqref{eq:coCA} are
\begin{equation}
\begin{aligned}
&c_3 = ik^2 \left(\CA_{33}+\CA_{44} + A_{22}r_\perp + A_{11}r_\perp \sin^2\theta  \right),\\
&c_2= -\frac{k^2}{T\chi_{11}} \left( s\cos^2\theta  + T\chi_{11}\sin\theta(\CA_{32}\rho + \CA_{13}\CA_{31}\sin\theta) \right)-k^4 \Big[ \CA_{22}\CA_{44}r_\perp \\ 
&\,\, + \CA_{33}(\CA_{44}+\CA_{22}r_\perp) - \CA_{34}\eta_\perp^2 +r_\perp\left( \CA_{11}(\CA_{33}+\CA_{44}) + r_\perp\sin^2\theta(\CA_{11}\CA_{22}-\CA_{12}\CA_{21})\right) \Big],\\
&c_1 =- i\frac{k^4}{T\chi_{11}}\Bigg\{ s(r_\perp\CA_{22}+\CA_{33})\cos^2\theta - \eta_\perp\cos\theta\sin\theta(sT \CA_{31}+\chi_{11}\CA_{13}\CA_{34}) \\
&\,\, +\chi_{11}T\sin\theta \Big[ \rho\CA_{32}\CA_{44}  + \CA_{31}\sin\theta\left( \CA_{13}\CA_{44}+r_\perp\CA_{13}\CA_{22} +r_\perp \rho \CA_{12} \right) \\
&\,\, +r_\perp\CA_{32}\sin^2\theta (\CA_{13}\CA_{21}+\rho\CA_{11}) \Big]\Bigg\} -i r_\perp k^6\Bigg\{ \CA_{22}(\CA_{33}\CA_{44}-\CA_{34}\eta_\perp^2) \\
&\,\, +\sin^2\theta  \Big[  -  r_\perp\CA_{12}\CA_{21}(\CA_{33}+\CA_{44})  \\
&\,\, + \CA_{11}\sin^2\theta\left( \CA_{33}\CA_{44}+r_\perp\CA_{22}\CA_{33}+r_\perp\CA_{22}\CA_{44}-\eta_\perp^2\CA_{34} \right)\Big]\Bigg\},\\
&c_0 = \left( \frac{s\rho \CA_{23}\cos^2\theta\sin^2\theta}{T\chi_{11}}\right)k^4 + \frac{r_\perp k^6}{T\chi_{11}}\Bigg\{s \CA_{22}\CA_{33} \cos^2\theta + \chi_{11}\CA_{32}\CA_{44}(\CA_{13}\CA_{21} + \rho\CA_{11})\sin^3\theta\\
&\,\, + \chi_{11}\CA_{13}\CA_{31}\CA_{44}(\CA_{13}\CA_{22}+\rho\CA_{12})+\eta_\perp\cos\theta\sin\theta\Big[ sT \CA_{22}\CA_{31} +\chi_{11}\CA_{13}\CA_{22}\CA_{34}\\
&\,\, + \chi_{11}\rho \CA_{12}\CA_{34} + sT \CA_{21}\CA_{32}\sin\theta \Big]\Bigg\} + r_\perp^2(\CA_{12}\CA_{21}-\CA_{11}\CA_{22})(\CA_{33}\CA_{34}\eta_\perp^2)k^8\sin^2\theta \, .
\end{aligned}
\end{equation}
In principle, Eq. \eqref{MSQuartic} gives a closed-form solution for the four $\omega(k)$. In practice, the explicit solutions are extremely lengthy so it is often more convenient to find the roots of \eqref{MSQuartic} numerically (our equations of state and transport coefficients are in any case given numerically), or by using various expansions, e.g. small $k / T$ or small $k / \sqrt{\CB}$.

\bibliographystyle{JHEP}
\bibliography{biblio.bib}

\providecommand{\href}[2]{#2}\begingroup\raggedright\begin{thebibliography}{10}

\bibitem{bellan2008fundamentals}
P.~M. Bellan, \emph{Fundamentals of plasma physics}. Cambridge University
  Press, 2008.

\bibitem{freidberg2014}
J.~P. Freidberg, \emph{Ideal MHD:}. Cambridge University Press, Cambridge, 006,
  2014,
  \href{https://doi.org/10.1017/CBO9780511795046}{10.1017/CBO9780511795046}.

\bibitem{goedbloed2004principles}
J.~Goedbloed and S.~Poedts, \emph{Principles of Magnetohydrodynamics: With
  Applications to Laboratory and Astrophysical Plasmas}. Cambridge University
  Press, 2004.

\bibitem{goedbloed2010advanced}
J.~Goedbloed, R.~Keppens and S.~Poedts, \emph{Advanced Magnetohydrodynamics:
  With Applications to Laboratory and Astrophysical Plasmas}. Cambridge
  University Press, 2010.

\bibitem{Dubovsky:2011sj}
S.~Dubovsky, L.~Hui, A.~Nicolis and D.~T. Son, \emph{{Effective field theory
  for hydrodynamics: thermodynamics, and the derivative expansion}},
  \href{https://doi.org/10.1103/PhysRevD.85.085029}{\emph{Phys. Rev.}
  {\bfseries D85} (2012) 085029}
  [\href{https://arxiv.org/abs/1107.0731}{{\ttfamily 1107.0731}}].

\bibitem{Endlich:2012vt}
S.~Endlich, A.~Nicolis, R.~A. Porto and J.~Wang, \emph{{Dissipation in the
  effective field theory for hydrodynamics: First order effects}},
  \href{https://doi.org/10.1103/PhysRevD.88.105001}{\emph{Phys. Rev.}
  {\bfseries D88} (2013) 105001}
  [\href{https://arxiv.org/abs/1211.6461}{{\ttfamily 1211.6461}}].

\bibitem{Grozdanov:2013dba}
S.~Grozdanov and J.~Polonyi, \emph{{Viscosity and dissipative hydrodynamics
  from effective field theory}},
  \href{https://doi.org/10.1103/PhysRevD.91.105031}{\emph{Phys. Rev.}
  {\bfseries D91} (2015) 105031}
  [\href{https://arxiv.org/abs/1305.3670}{{\ttfamily 1305.3670}}].

\bibitem{Nicolis:2013lma}
A.~Nicolis, R.~Penco and R.~A. Rosen, \emph{{Relativistic Fluids, Superfluids,
  Solids and Supersolids from a Coset Construction}},
  \href{https://doi.org/10.1103/PhysRevD.89.045002}{\emph{Phys. Rev.}
  {\bfseries D89} (2014) 045002}
  [\href{https://arxiv.org/abs/1307.0517}{{\ttfamily 1307.0517}}].

\bibitem{Kovtun:2014hpa}
P.~Kovtun, G.~D. Moore and P.~Romatschke, \emph{{Towards an effective action
  for relativistic dissipative hydrodynamics}},
  \href{https://doi.org/10.1007/JHEP07(2014)123}{\emph{JHEP} {\bfseries 07}
  (2014) 123} [\href{https://arxiv.org/abs/1405.3967}{{\ttfamily 1405.3967}}].

\bibitem{Harder:2015nxa}
M.~Harder, P.~Kovtun and A.~Ritz, \emph{{On thermal fluctuations and the
  generating functional in relativistic hydrodynamics}},
  \href{https://doi.org/10.1007/JHEP07(2015)025}{\emph{JHEP} {\bfseries 07}
  (2015) 025} [\href{https://arxiv.org/abs/1502.03076}{{\ttfamily
  1502.03076}}].

\bibitem{Grozdanov:2015nea}
S.~Grozdanov and J.~Polonyi, \emph{{Dynamics of the electric current in an
  ideal electron gas: A sound mode inside the quasiparticles}},
  \href{https://doi.org/10.1103/PhysRevD.92.065009}{\emph{Phys. Rev.}
  {\bfseries D92} (2015) 065009}
  [\href{https://arxiv.org/abs/1501.06620}{{\ttfamily 1501.06620}}].

\bibitem{Crossley:2015evo}
M.~Crossley, P.~Glorioso and H.~Liu, \emph{{Effective field theory of
  dissipative fluids}},  \href{https://arxiv.org/abs/1511.03646}{{\ttfamily
  1511.03646}}.

\bibitem{Glorioso:2017fpd}
P.~Glorioso, M.~Crossley and H.~Liu, \emph{{Effective field theory for
  dissipative fluids (II): classical limit, dynamical KMS symmetry and entropy
  current}},  \href{https://arxiv.org/abs/1701.07817}{{\ttfamily 1701.07817}}.

\bibitem{Haehl:2015foa}
F.~M. Haehl, R.~Loganayagam and M.~Rangamani, \emph{{The Fluid Manifesto:
  Emergent symmetries, hydrodynamics, and black holes}},
  \href{https://doi.org/10.1007/JHEP01(2016)184}{\emph{JHEP} {\bfseries 01}
  (2016) 184} [\href{https://arxiv.org/abs/1510.02494}{{\ttfamily
  1510.02494}}].

\bibitem{Haehl:2015uoc}
F.~M. Haehl, R.~Loganayagam and M.~Rangamani, \emph{{Topological sigma models
  \& dissipative hydrodynamics}},
  \href{https://doi.org/10.1007/JHEP04(2016)039}{\emph{JHEP} {\bfseries 04}
  (2016) 039} [\href{https://arxiv.org/abs/1511.07809}{{\ttfamily
  1511.07809}}].

\bibitem{Torrieri:2016dko}
D.~Montenegro and G.~Torrieri, \emph{{Lagrangian formulation of relativistic
  Israel-Stewart hydrodynamics}},
  \href{https://doi.org/10.1103/PhysRevD.94.065042}{\emph{Phys. Rev.}
  {\bfseries D94} (2016) 065042}
  [\href{https://arxiv.org/abs/1604.05291}{{\ttfamily 1604.05291}}].

\bibitem{Glorioso:2016gsa}
P.~Glorioso and H.~Liu, \emph{{The second law of thermodynamics from symmetry
  and unitarity}},  \href{https://arxiv.org/abs/1612.07705}{{\ttfamily
  1612.07705}}.

\bibitem{Gao:2017bqf}
P.~Gao and H.~Liu, \emph{{Emergent Supersymmetry in Local Equilibrium
  Systems}},  \href{https://arxiv.org/abs/1701.07445}{{\ttfamily 1701.07445}}.

\bibitem{Jensen:2017kzi}
K.~Jensen, N.~Pinzani-Fokeeva and A.~Yarom, \emph{{Dissipative hydrodynamics in
  superspace}},  \href{https://arxiv.org/abs/1701.07436}{{\ttfamily
  1701.07436}}.

\bibitem{Baier:2007ix}
R.~Baier, P.~Romatschke, D.~T. Son, A.~O. Starinets and M.~A. Stephanov,
  \emph{{Relativistic viscous hydrodynamics, conformal invariance, and
  holography}},
  \href{https://doi.org/10.1088/1126-6708/2008/04/100}{\emph{JHEP} {\bfseries
  04} (2008) 100} [\href{https://arxiv.org/abs/0712.2451}{{\ttfamily
  0712.2451}}].

\bibitem{Bhattacharyya:2008jc}
S.~Bhattacharyya, V.~E. Hubeny, S.~Minwalla and M.~Rangamani, \emph{{Nonlinear
  Fluid Dynamics from Gravity}},
  \href{https://doi.org/10.1088/1126-6708/2008/02/045}{\emph{JHEP} {\bfseries
  02} (2008) 045} [\href{https://arxiv.org/abs/0712.2456}{{\ttfamily
  0712.2456}}].

\bibitem{Romatschke:2009kr}
P.~Romatschke, \emph{{Relativistic Viscous Fluid Dynamics and Non-Equilibrium
  Entropy}}, \href{https://doi.org/10.1088/0264-9381/27/2/025006}{\emph{Class.
  Quant. Grav.} {\bfseries 27} (2010) 025006}
  [\href{https://arxiv.org/abs/0906.4787}{{\ttfamily 0906.4787}}].

\bibitem{Grozdanov:2015kqa}
S.~Grozdanov and N.~Kaplis, \emph{{Constructing higher-order hydrodynamics: The
  third order}}, \href{https://doi.org/10.1103/PhysRevD.93.066012}{\emph{Phys.
  Rev.} {\bfseries D93} (2016) 066012}
  [\href{https://arxiv.org/abs/1507.02461}{{\ttfamily 1507.02461}}].

\bibitem{Grozdanov:2016tdf}
S.~Grozdanov, D.~M. Hofman and N.~Iqbal, \emph{{Generalized global symmetries
  and dissipative magnetohydrodynamics}},
  \href{https://doi.org/10.1103/PhysRevD.95.096003}{\emph{Phys. Rev.}
  {\bfseries D95} (2017) 096003}
  [\href{https://arxiv.org/abs/1610.07392}{{\ttfamily 1610.07392}}].

\bibitem{Kovtun:2012rj}
P.~Kovtun, \emph{{Lectures on hydrodynamic fluctuations in relativistic
  theories}}, \href{https://doi.org/10.1088/1751-8113/45/47/473001}{\emph{J.
  Phys.} {\bfseries A45} (2012) 473001}
  [\href{https://arxiv.org/abs/1205.5040}{{\ttfamily 1205.5040}}].

\bibitem{Schubring:2014iwa}
D.~Schubring, \emph{{Dissipative String Fluids}},
  \href{https://doi.org/10.1103/PhysRevD.91.043518}{\emph{Phys. Rev.}
  {\bfseries D91} (2015) 043518}
  [\href{https://arxiv.org/abs/1412.3135}{{\ttfamily 1412.3135}}].

\bibitem{Hernandez:2017mch}
J.~Hernandez and P.~Kovtun, \emph{{Relativistic magnetohydrodynamics}},
  \href{https://arxiv.org/abs/1703.08757}{{\ttfamily 1703.08757}}.

\bibitem{Huang:2011dc}
X.-G. Huang, A.~Sedrakian and D.~H. Rischke, \emph{{Kubo formulae for
  relativistic fluids in strong magnetic fields}},
  \href{https://doi.org/10.1016/j.aop.2011.08.001}{\emph{Annals Phys.}
  {\bfseries 326} (2011) 3075}
  [\href{https://arxiv.org/abs/1108.0602}{{\ttfamily 1108.0602}}].

\bibitem{Critelli:2014kra}
R.~Critelli, S.~I. Finazzo, M.~Zaniboni and J.~Noronha, \emph{{Anisotropic
  shear viscosity of a strongly coupled non-Abelian plasma from magnetic
  branes}}, \href{https://doi.org/10.1103/PhysRevD.90.066006}{\emph{Phys. Rev.}
  {\bfseries D90} (2014) 066006}
  [\href{https://arxiv.org/abs/1406.6019}{{\ttfamily 1406.6019}}].

\bibitem{Finazzo:2016mhm}
S.~I. Finazzo, R.~Critelli, R.~Rougemont and J.~Noronha, \emph{{Momentum
  transport in strongly coupled anisotropic plasmas in the presence of strong
  magnetic fields}},
  \href{https://doi.org/10.1103/PhysRevD.94.054020}{\emph{Phys. Rev.}
  {\bfseries D94} (2016) 054020}
  [\href{https://arxiv.org/abs/1605.06061}{{\ttfamily 1605.06061}}].

\bibitem{Kovtun:2016lfw}
P.~Kovtun, \emph{{Thermodynamics of polarized relativistic matter}},
  \href{https://doi.org/10.1007/JHEP07(2016)028}{\emph{JHEP} {\bfseries 07}
  (2016) 028} [\href{https://arxiv.org/abs/1606.01226}{{\ttfamily
  1606.01226}}].

\bibitem{Montenegro:2017rbu}
D.~Montenegro, L.~Tinti and G.~Torrieri, \emph{{The ideal relativistic fluid
  limit for a medium with polarization}},
  \href{https://arxiv.org/abs/1701.08263}{{\ttfamily 1701.08263}}.

\bibitem{Gaiotto:2014kfa}
D.~Gaiotto, A.~Kapustin, N.~Seiberg and B.~Willett, \emph{{Generalized Global
  Symmetries}}, \href{https://doi.org/10.1007/JHEP02(2015)172}{\emph{JHEP}
  {\bfseries 02} (2015) 172} [\href{https://arxiv.org/abs/1412.5148}{{\ttfamily
  1412.5148}}].

\bibitem{Policastro:2001yc}
G.~Policastro, D.~T. Son and A.~O. Starinets, \emph{{The Shear viscosity of
  strongly coupled N=4 supersymmetric Yang-Mills plasma}},
  \href{https://doi.org/10.1103/PhysRevLett.87.081601}{\emph{Phys. Rev. Lett.}
  {\bfseries 87} (2001) 081601}
  [\href{https://arxiv.org/abs/hep-th/0104066}{{\ttfamily hep-th/0104066}}].

\bibitem{Policastro:2002se}
G.~Policastro, D.~T. Son and A.~O. Starinets, \emph{{From AdS / CFT
  correspondence to hydrodynamics}},
  \href{https://doi.org/10.1088/1126-6708/2002/09/043}{\emph{JHEP} {\bfseries
  09} (2002) 043} [\href{https://arxiv.org/abs/hep-th/0205052}{{\ttfamily
  hep-th/0205052}}].

\bibitem{Policastro:2002tn}
G.~Policastro, D.~T. Son and A.~O. Starinets, \emph{{From AdS / CFT
  correspondence to hydrodynamics. 2. Sound waves}},
  \href{https://doi.org/10.1088/1126-6708/2002/12/054}{\emph{JHEP} {\bfseries
  12} (2002) 054} [\href{https://arxiv.org/abs/hep-th/0210220}{{\ttfamily
  hep-th/0210220}}].

\bibitem{Hofman:2017Something}
D.~M. Hofman and N.~Iqbal, \emph{{Generalized global symmetries and
  holography}},  \href{https://arxiv.org/abs/1707.08577}{{\ttfamily
  1707.08577}}.

\bibitem{Fuini:2015hba}
J.~F. Fuini and L.~G. Yaffe, \emph{{Far-from-equilibrium dynamics of a strongly
  coupled non-Abelian plasma with non-zero charge density or external magnetic
  field}}, \href{https://doi.org/10.1007/JHEP07(2015)116}{\emph{JHEP}
  {\bfseries 07} (2015) 116}
  [\href{https://arxiv.org/abs/1503.07148}{{\ttfamily 1503.07148}}].

\bibitem{Weinberg:1995mt}
S.~Weinberg, \emph{{The Quantum Theory of Fields. Vol. 1: Foundations}}.
  Cambridge University Press, 2005.

\bibitem{Weinberg:1996kr}
S.~Weinberg, \emph{{The Quantum Theory of Fields. Vol. 2: Modern
  Applications}}. Cambridge University Press, 2013.

\bibitem{Peskin:1995ev}
M.~E. Peskin and D.~V. Schroeder, \emph{{An Introduction to quantum field
  theory}}. 1995.

\bibitem{Yamada:2006rx}
D.~Yamada and L.~G. Yaffe, \emph{{Phase diagram of N=4 super-Yang-Mills theory
  with R-symmetry chemical potentials}},
  \href{https://doi.org/10.1088/1126-6708/2006/09/027}{\emph{JHEP} {\bfseries
  09} (2006) 027} [\href{https://arxiv.org/abs/hep-th/0602074}{{\ttfamily
  hep-th/0602074}}].

\bibitem{Cherman:2013rla}
A.~Cherman, S.~Grozdanov and E.~Hardy, \emph{{Searching for Fermi Surfaces in
  Super-QED}}, \href{https://doi.org/10.1007/JHEP06(2014)046}{\emph{JHEP}
  {\bfseries 06} (2014) 046} [\href{https://arxiv.org/abs/1308.0335}{{\ttfamily
  1308.0335}}].

\bibitem{Freedman:1998tz}
D.~Z. Freedman, S.~D. Mathur, A.~Matusis and L.~Rastelli, \emph{{Correlation
  functions in the CFT(d) / AdS(d+1) correspondence}},
  \href{https://doi.org/10.1016/S0550-3213(99)00053-X}{\emph{Nucl. Phys.}
  {\bfseries B546} (1999) 96}
  [\href{https://arxiv.org/abs/hep-th/9804058}{{\ttfamily hep-th/9804058}}].

\bibitem{Anselmi:1997ys}
D.~Anselmi, J.~Erlich, D.~Z. Freedman and A.~A. Johansen, \emph{{Positivity
  constraints on anomalies in supersymmetric gauge theories}},
  \href{https://doi.org/10.1103/PhysRevD.57.7570}{\emph{Phys. Rev.} {\bfseries
  D57} (1998) 7570} [\href{https://arxiv.org/abs/hep-th/9711035}{{\ttfamily
  hep-th/9711035}}].

\bibitem{D'Hoker:2009mmn}
E.~D'Hoker and P.~Kraus, \emph{{Magnetic Brane Solutions in AdS}},
  \href{https://doi.org/10.1088/1126-6708/2009/10/088}{\emph{JHEP} {\bfseries
  10} (2009) 088} [\href{https://arxiv.org/abs/0908.3875}{{\ttfamily
  0908.3875}}].

\bibitem{Janiszewski:2015ura}
S.~Janiszewski and M.~Kaminski, \emph{{Quasinormal modes of magnetic and
  electric black branes versus far from equilibrium anisotropic fluids}},
  \href{https://doi.org/10.1103/PhysRevD.93.025006}{\emph{Phys. Rev.}
  {\bfseries D93} (2016) 025006}
  [\href{https://arxiv.org/abs/1508.06993}{{\ttfamily 1508.06993}}].

\bibitem{Ammon:2017ded}
M.~Ammon, M.~Kaminski, R.~Koirala, J.~Leiber and J.~Wu, \emph{{Quasinormal
  modes of charged magnetic black branes \& chiral magnetic transport}},
  \href{https://doi.org/10.1007/JHEP04(2017)067}{\emph{JHEP} {\bfseries 04}
  (2017) 067} [\href{https://arxiv.org/abs/1701.05565}{{\ttfamily
  1701.05565}}].

\bibitem{Witten:2001ua}
E.~Witten, \emph{{Multitrace operators, boundary conditions, and AdS / CFT
  correspondence}},  \href{https://arxiv.org/abs/hep-th/0112258}{{\ttfamily
  hep-th/0112258}}.

\bibitem{Pomoni:2008de}
E.~Pomoni and L.~Rastelli, \emph{{Large N Field Theory and AdS Tachyons}},
  \href{https://doi.org/10.1088/1126-6708/2009/04/020}{\emph{JHEP} {\bfseries
  04} (2009) 020} [\href{https://arxiv.org/abs/0805.2261}{{\ttfamily
  0805.2261}}].

\bibitem{Heemskerk:2010hk}
I.~Heemskerk and J.~Polchinski, \emph{{Holographic and Wilsonian
  Renormalization Groups}},
  \href{https://doi.org/10.1007/JHEP06(2011)031}{\emph{JHEP} {\bfseries 06}
  (2011) 031} [\href{https://arxiv.org/abs/1010.1264}{{\ttfamily 1010.1264}}].

\bibitem{Faulkner:2010jy}
T.~Faulkner, H.~Liu and M.~Rangamani, \emph{{Integrating out geometry:
  Holographic Wilsonian RG and the membrane paradigm}},
  \href{https://doi.org/10.1007/JHEP08(2011)051}{\emph{JHEP} {\bfseries 08}
  (2011) 051} [\href{https://arxiv.org/abs/1010.4036}{{\ttfamily 1010.4036}}].

\bibitem{Grozdanov:2011aa}
S.~Grozdanov, \emph{{Wilsonian Renormalisation and the Exact Cut-Off Scale from
  Holographic Duality}},
  \href{https://doi.org/10.1007/JHEP06(2012)079}{\emph{JHEP} {\bfseries 06}
  (2012) 079} [\href{https://arxiv.org/abs/1112.3356}{{\ttfamily 1112.3356}}].

\bibitem{Witten:2003ya}
E.~Witten, \emph{{SL(2,Z) action on three-dimensional conformal field theories
  with Abelian symmetry}},
  \href{https://arxiv.org/abs/hep-th/0307041}{{\ttfamily hep-th/0307041}}.

\bibitem{Marolf:2006nd}
D.~Marolf and S.~F. Ross, \emph{{Boundary Conditions and New Dualities: Vector
  Fields in AdS/CFT}},
  \href{https://doi.org/10.1088/1126-6708/2006/11/085}{\emph{JHEP} {\bfseries
  0611} (2006) 085} [\href{https://arxiv.org/abs/hep-th/0606113}{{\ttfamily
  hep-th/0606113}}].

\bibitem{Jokela:2013hta}
N.~Jokela, G.~Lifschytz and M.~Lippert, \emph{{Holographic anyonic
  superfluidity}}, \href{https://doi.org/10.1007/JHEP10(2013)014}{\emph{JHEP}
  {\bfseries 10} (2013) 014} [\href{https://arxiv.org/abs/1307.6336}{{\ttfamily
  1307.6336}}].

\bibitem{Faulkner:2012gt}
T.~Faulkner and N.~Iqbal, \emph{{Friedel oscillations and horizon charge in 1D
  holographic liquids}},
  \href{https://doi.org/10.1007/JHEP07(2013)060}{\emph{JHEP} {\bfseries 07}
  (2013) 060} [\href{https://arxiv.org/abs/1207.4208}{{\ttfamily 1207.4208}}].

\bibitem{Gubser:1999vj}
S.~S. Gubser, \emph{{AdS / CFT and gravity}},
  \href{https://doi.org/10.1103/PhysRevD.63.084017}{\emph{Phys. Rev.}
  {\bfseries D63} (2001) 084017}
  [\href{https://arxiv.org/abs/hep-th/9912001}{{\ttfamily hep-th/9912001}}].

\bibitem{D'Hoker:2009bc}
E.~D'Hoker and P.~Kraus, \emph{{Charged Magnetic Brane Solutions in AdS (5) and
  the fate of the third law of thermodynamics}},
  \href{https://doi.org/10.1007/JHEP03(2010)095}{\emph{JHEP} {\bfseries 03}
  (2010) 095} [\href{https://arxiv.org/abs/0911.4518}{{\ttfamily 0911.4518}}].

\bibitem{Iqbal:2008by}
N.~Iqbal and H.~Liu, \emph{{Universality of the hydrodynamic limit in AdS/CFT
  and the membrane paradigm}},
  \href{https://doi.org/10.1103/PhysRevD.79.025023}{\emph{Phys. Rev.}
  {\bfseries D79} (2009) 025023}
  [\href{https://arxiv.org/abs/0809.3808}{{\ttfamily 0809.3808}}].

\bibitem{Gubser:2008sz}
S.~S. Gubser, S.~S. Pufu and F.~D. Rocha, \emph{{Bulk viscosity of strongly
  coupled plasmas with holographic duals}},
  \href{https://doi.org/10.1088/1126-6708/2008/08/085}{\emph{JHEP} {\bfseries
  08} (2008) 085} [\href{https://arxiv.org/abs/0806.0407}{{\ttfamily
  0806.0407}}].

\bibitem{Saremi:2011ab}
O.~Saremi and D.~T. Son, \emph{{Hall viscosity from gauge/gravity duality}},
  \href{https://doi.org/10.1007/JHEP04(2012)091}{\emph{JHEP} {\bfseries 04}
  (2012) 091} [\href{https://arxiv.org/abs/1103.4851}{{\ttfamily 1103.4851}}].

\bibitem{Donos:2014cya}
A.~Donos and J.~P. Gauntlett, \emph{{Thermoelectric DC conductivities from
  black hole horizons}},
  \href{https://doi.org/10.1007/JHEP11(2014)081}{\emph{JHEP} {\bfseries 11}
  (2014) 081} [\href{https://arxiv.org/abs/1406.4742}{{\ttfamily 1406.4742}}].

\bibitem{Banks:2015wha}
E.~Banks, A.~Donos and J.~P. Gauntlett, \emph{{Thermoelectric DC conductivities
  and Stokes flows on black hole horizons}},
  \href{https://doi.org/10.1007/JHEP10(2015)103}{\emph{JHEP} {\bfseries 10}
  (2015) 103} [\href{https://arxiv.org/abs/1507.00234}{{\ttfamily
  1507.00234}}].

\bibitem{Davison:2015taa}
R.~A. Davison, B.~Gout\'eraux and S.~A. Hartnoll, \emph{{Incoherent transport
  in clean quantum critical metals}},
  \href{https://doi.org/10.1007/JHEP10(2015)112}{\emph{JHEP} {\bfseries 10}
  (2015) 112} [\href{https://arxiv.org/abs/1507.07137}{{\ttfamily
  1507.07137}}].

\bibitem{Gursoy:2014boa}
U.~Gursoy and J.~Tarrio, \emph{{Horizon universality and anomalous
  conductivities}}, \href{https://doi.org/10.1007/JHEP10(2015)058}{\emph{JHEP}
  {\bfseries 10} (2015) 058} [\href{https://arxiv.org/abs/1410.1306}{{\ttfamily
  1410.1306}}].

\bibitem{Grozdanov:2016ala}
S.~Grozdanov and N.~Poovuttikul, \emph{{Universality of anomalous
  conductivities in theories with higher-derivative holographic duals}},
  \href{https://doi.org/10.1007/JHEP09(2016)046}{\emph{JHEP} {\bfseries 09}
  (2016) 046} [\href{https://arxiv.org/abs/1603.08770}{{\ttfamily
  1603.08770}}].

\bibitem{deHaro:2000vlm}
S.~de~Haro, S.~N. Solodukhin and K.~Skenderis, \emph{{Holographic
  reconstruction of space-time and renormalization in the AdS / CFT
  correspondence}}, \href{https://doi.org/10.1007/s002200100381}{\emph{Commun.
  Math. Phys.} {\bfseries 217} (2001) 595}
  [\href{https://arxiv.org/abs/hep-th/0002230}{{\ttfamily hep-th/0002230}}].

\bibitem{Taylor:2000xw}
M.~Taylor, \emph{{More on counterterms in the gravitational action and
  anomalies}},  \href{https://arxiv.org/abs/hep-th/0002125}{{\ttfamily
  hep-th/0002125}}.

\bibitem{xAct}
J.~M. Mart\'in-Garc\'ia, ``{xAct: Efficient Tensor Computer Algebra},
  {\url{http://www.xact.es/}},.''

\bibitem{Denef:2009yy}
F.~Denef, S.~A. Hartnoll and S.~Sachdev, \emph{{Quantum oscillations and black
  hole ringing}}, \href{https://doi.org/10.1103/PhysRevD.80.126016}{\emph{Phys.
  Rev.} {\bfseries D80} (2009) 126016}
  [\href{https://arxiv.org/abs/0908.1788}{{\ttfamily 0908.1788}}].

\bibitem{Denef:2009kn}
F.~Denef, S.~A. Hartnoll and S.~Sachdev, \emph{{Black hole determinants and
  quasinormal modes}},
  \href{https://doi.org/10.1088/0264-9381/27/12/125001}{\emph{Class. Quant.
  Grav.} {\bfseries 27} (2010) 125001}
  [\href{https://arxiv.org/abs/0908.2657}{{\ttfamily 0908.2657}}].

\bibitem{CaronHuot:2009iq}
S.~Caron-Huot and O.~Saremi, \emph{{Hydrodynamic Long-Time tails From Anti de
  Sitter Space}}, \href{https://doi.org/10.1007/JHEP11(2010)013}{\emph{JHEP}
  {\bfseries 11} (2010) 013} [\href{https://arxiv.org/abs/0909.4525}{{\ttfamily
  0909.4525}}].

\bibitem{Arnold:2016dbb}
P.~Arnold, P.~Szepietowski and D.~Vaman, \emph{{Computing black hole partition
  functions from quasinormal modes}},
  \href{https://doi.org/10.1007/JHEP07(2016)032}{\emph{JHEP} {\bfseries 07}
  (2016) 032} [\href{https://arxiv.org/abs/1603.08994}{{\ttfamily
  1603.08994}}].

\bibitem{Castro:2017mfj}
A.~Castro, C.~Keeler and P.~Szepietowski, \emph{{Tweaking one-loop determinants
  in AdS$_3$}},  \href{https://arxiv.org/abs/1707.06245}{{\ttfamily
  1707.06245}}.

\bibitem{Stricker:2013lma}
S.~A. Stricker, \emph{{Holographic thermalization in N=4 Super Yang-Mills
  theory at finite coupling}},
  \href{https://doi.org/10.1140/epjc/s10052-014-2727-4}{\emph{Eur. Phys. J.}
  {\bfseries C74} (2014) 2727}
  [\href{https://arxiv.org/abs/1307.2736}{{\ttfamily 1307.2736}}].

\bibitem{Waeber:2015oka}
S.~Waeber, A.~Schaefer, A.~Vuorinen and L.~G. Yaffe, \emph{{Finite coupling
  corrections to holographic predictions for hot QCD}},
  \href{https://doi.org/10.1007/JHEP11(2015)087}{\emph{JHEP} {\bfseries 11}
  (2015) 087} [\href{https://arxiv.org/abs/1509.02983}{{\ttfamily
  1509.02983}}].

\bibitem{Grozdanov:2016vgg}
S.~Grozdanov, N.~Kaplis and A.~O. Starinets, \emph{{From strong to weak
  coupling in holographic models of thermalization}},
  \href{https://doi.org/10.1007/JHEP07(2016)151}{\emph{JHEP} {\bfseries 07}
  (2016) 151} [\href{https://arxiv.org/abs/1605.02173}{{\ttfamily
  1605.02173}}].

\bibitem{Grozdanov:2016zjj}
S.~Grozdanov and W.~van~der Schee, \emph{{Coupling constant corrections in
  holographic heavy ion collisions}},
  \href{https://arxiv.org/abs/1610.08976}{{\ttfamily 1610.08976}}.

\bibitem{Grozdanov:2016fkt}
S.~Grozdanov and A.~O. Starinets, \emph{{Second-order transport, quasinormal
  modes and zero-viscosity limit in the Gauss-Bonnet holographic fluid}},
  \href{https://doi.org/10.1007/JHEP03(2017)166}{\emph{JHEP} {\bfseries 03}
  (2017) 166} [\href{https://arxiv.org/abs/1611.07053}{{\ttfamily
  1611.07053}}].

\bibitem{Son:2002sd}
D.~T. Son and A.~O. Starinets, \emph{{Minkowski space correlators in AdS / CFT
  correspondence: Recipe and applications}},
  \href{https://doi.org/10.1088/1126-6708/2002/09/042}{\emph{JHEP} {\bfseries
  09} (2002) 042} [\href{https://arxiv.org/abs/hep-th/0205051}{{\ttfamily
  hep-th/0205051}}].

\bibitem{Herzog:2002pc}
C.~P. Herzog and D.~T. Son, \emph{{Schwinger-Keldysh propagators from AdS/CFT
  correspondence}},
  \href{https://doi.org/10.1088/1126-6708/2003/03/046}{\emph{JHEP} {\bfseries
  03} (2003) 046} [\href{https://arxiv.org/abs/hep-th/0212072}{{\ttfamily
  hep-th/0212072}}].

\bibitem{PhysRevLett.111.125004}
V.~E. Fortov and V.~B. Mintsev, \emph{Quantum bound of the shear viscosity of a
  strongly coupled plasma},
  \href{https://doi.org/10.1103/PhysRevLett.111.125004}{\emph{Phys. Rev. Lett.}
  {\bfseries 111} (2013) 125004}.

\bibitem{Grozdanov:2015qia}
S.~Grozdanov, A.~Lucas, S.~Sachdev and K.~Schalm, \emph{{Absence of
  disorder-driven metal-insulator transitions in simple holographic models}},
  \href{https://doi.org/10.1103/PhysRevLett.115.221601}{\emph{Phys. Rev. Lett.}
  {\bfseries 115} (2015) 221601}
  [\href{https://arxiv.org/abs/1507.00003}{{\ttfamily 1507.00003}}].

\bibitem{Lucas:2017ggp}
A.~Lucas and S.~A. Hartnoll, \emph{{Resistivity bound for hydrodynamic bad
  metals}},  \href{https://arxiv.org/abs/1704.07384}{{\ttfamily 1704.07384}}.

\bibitem{Kadanoff}
L.~P. {Kadanoff} and P.~C. {Martin}, \emph{{Hydrodynamic equations and
  correlation functions}},
  \href{https://doi.org/10.1016/0003-4916(63)90078-2}{\emph{Annals of Physics}
  {\bfseries 24} (1963) 419}.

\bibitem{Cherman:2009tw}
A.~Cherman, T.~D. Cohen and A.~Nellore, \emph{{A Bound on the speed of sound
  from holography}},
  \href{https://doi.org/10.1103/PhysRevD.80.066003}{\emph{Phys. Rev.}
  {\bfseries D80} (2009) 066003}
  [\href{https://arxiv.org/abs/0905.0903}{{\ttfamily 0905.0903}}].

\bibitem{Cherman:2009kf}
A.~Cherman and A.~Nellore, \emph{{Universal relations of transport coefficients
  from holography}},
  \href{https://doi.org/10.1103/PhysRevD.80.066006}{\emph{Phys. Rev.}
  {\bfseries D80} (2009) 066006}
  [\href{https://arxiv.org/abs/0905.2969}{{\ttfamily 0905.2969}}].

\bibitem{Hohler:2009tv}
P.~M. Hohler and M.~A. Stephanov, \emph{{Holography and the speed of sound at
  high temperatures}},
  \href{https://doi.org/10.1103/PhysRevD.80.066002}{\emph{Phys. Rev.}
  {\bfseries D80} (2009) 066002}
  [\href{https://arxiv.org/abs/0905.0900}{{\ttfamily 0905.0900}}].

\bibitem{Hoyos:2016cob}
C.~Hoyos, N.~Jokela, D.~Rodr\'iguez~Fern\'andez and A.~Vuorinen,
  \emph{{Breaking the sound barrier in AdS/CFT}},
  \href{https://doi.org/10.1103/PhysRevD.94.106008}{\emph{Phys. Rev.}
  {\bfseries D94} (2016) 106008}
  [\href{https://arxiv.org/abs/1609.03480}{{\ttfamily 1609.03480}}].

\bibitem{Grozdanov:2014kva}
S.~Grozdanov and A.~O. Starinets, \emph{{On the universal identity in second
  order hydrodynamics}},
  \href{https://doi.org/10.1007/JHEP03(2015)007}{\emph{JHEP} {\bfseries 03}
  (2015) 007} [\href{https://arxiv.org/abs/1412.5685}{{\ttfamily 1412.5685}}].

\bibitem{Haehl:2015pja}
F.~M. Haehl, R.~Loganayagam and M.~Rangamani, \emph{{Adiabatic hydrodynamics:
  The eightfold way to dissipation}},
  \href{https://doi.org/10.1007/JHEP05(2015)060}{\emph{JHEP} {\bfseries 05}
  (2015) 060} [\href{https://arxiv.org/abs/1502.00636}{{\ttfamily
  1502.00636}}].

\bibitem{Edalati:2010pn}
M.~Edalati, J.~I. Jottar and R.~G. Leigh, \emph{{Holography and the sound of
  criticality}}, \href{https://doi.org/10.1007/JHEP10(2010)058}{\emph{JHEP}
  {\bfseries 10} (2010) 058} [\href{https://arxiv.org/abs/1005.4075}{{\ttfamily
  1005.4075}}].

\end{thebibliography}\endgroup
\end{document}